\documentclass[aps,pra,twocolumn,superscriptaddress,floatfix]{revtex4-2}
\usepackage[caption=false]{subfig}
\usepackage{graphicx}
\usepackage{amsmath}
\usepackage{amssymb}
\usepackage{physics}
\usepackage{hyperref}
\usepackage{xcolor}
\usepackage{booktabs}
\usepackage[normalem]{ulem}
\usepackage{multirow} 
\usepackage{changes}

\begin{document}

\title{Temporal steering of entanglement decay with single-shot control}

\author{Saumya Ranjan Behera}
\affiliation{Raman Research Institute, C.V.Raman Avenue, Sadashivanagar, Bengaluru-560080, Karnataka, India}

\author{Kallol Sen}
\affiliation{Raman Research Institute, C.V.Raman Avenue, Sadashivanagar, Bengaluru-560080, Karnataka, India}
\affiliation{QuSyn Technologies, Bengaluru-560094, Karnataka, India}

\author{Animesh Sinha Roy}
\affiliation{Raman Research Institute, C.V.Raman Avenue, Sadashivanagar, Bengaluru-560080, Karnataka, India}

\author{Snigdhadev Ray}
\affiliation{School of Physical Sciences, Indian Association for the Cultivation of Science, Kolkata - 700032, West Bengal, India}
\affiliation{Raman Research Institute, C.V.Raman Avenue, Sadashivanagar, Bengaluru-560080, Karnataka, India}

\author{Ashutosh Singh}
\affiliation{Raman Research Institute, C.V.Raman Avenue, Sadashivanagar, Bengaluru-560080, Karnataka, India}
\affiliation{Department of Physics and Astronomy, University of Calgary, Alberta T2N 1N4, Canada}

\author{A.R.P. Rau}
\affiliation{Department of Physics and Astronomy, Louisiana State University, Baton Rouge, Louisiana 70803}

\author{Urbasi Sinha}
\email{usinha@rri.res.in}
\affiliation{Raman Research Institute, C.V.Raman Avenue, Sadashivanagar, Bengaluru-560080, Karnataka, India}
\affiliation{Department of Physics and Astronomy, University of Calgary, Alberta T2N 1N4, Canada}

\date{\today}

\begin{abstract}
Entanglement in open quantum systems can vanish abruptly through entanglement sudden death (ESD) under dissipative evolution, posing a challenge for quantum technologies. Here, we show that the timing of a single local unitary operation can deterministically steer the trajectory of entanglement decay. This introduces temporal steering of dissipative quantum dynamics as a control resource. We develop a time-dependent open-system framework in which a time-shift operator generates a family of effective conditional maps parametrized by $x \in [0,1]$, connecting independent and correlated amplitude damping. Within this framework, we derive analytic conditions for ESD and show that a single intermediate local $\sigma_x$ operation can avoid, delay, or hasten ESD by redirecting the dissipative trajectory rather than modifying the environment. We further prove that this single-shot protocol is optimal within a broad class of local control strategies. Experimentally, a displaced-Sagnac interferometer realizes the correlated-damping-like regime and, for the first time in one photonic platform, demonstrates avoidance, delay, and hastening of the separability transition. Our results establish temporal steering of open-system dynamics as a practical paradigm for decoherence control, and demonstrate operational advantages in use cases such as teleportation, illustrating that in dissipative quantum systems knowing \textit{when} to act can be as important as knowing \textit{what} operation to perform.
\end{abstract}

\maketitle

\section{Introduction}

Quantum systems inexorably lose coherence through environmental interactions, with entanglement—the quintessential quantum resource—being particularly fragile. This decoherence problem is universal: it limits qubit lifetimes in superconducting processors~\cite{Biercuk2009}, degrades quantum memories~\cite{Du2009}, and constrains communication distances in quantum networks~\cite{Scarani2009}. One particularly striking manifestation is entanglement sudden death (ESD), where entanglement vanishes in finite time rather than decaying asymptotically~\cite{Yu2004, Yu2009}. Unlike gradual decoherence, ESD represents a threshold phenomenon that poses fundamental questions about the nature of open quantum evolution and presents practical challenges for quantum technologies that rely on maintaining entanglement over extended periods. These challenges motivate the search for control strategies that can shape the trajectory of dissipative evolution, raising a central question: in open quantum systems, can knowing \textit{when} to intervene be as important as knowing \textit{what} operation to perform?

Here we show that temporal steering of open-system dynamics via a single local unitary operation provides a deterministic mechanism for controlling entanglement decay, with timing being a crucial element. We consider a continuous family of amplitude-damping channels, introducing timing as a control resource for steering dissipative quantum trajectories. We report three principal advances. 

First, we demonstrate experimentally that a single local NOT operation, applied at a chosen time during amplitude damping, can deterministically achieve three qualitatively distinct outcomes: complete ESD avoidance (avoidance of finite-time ESD), significant delay (extending the damping parameter range by approximately 30\%), or hastening (accelerating ESD onset). The regime is selected purely by the operation timing, characterized by a parameter $p$ representing the strength of damping experienced before the NOT, with no modification to the environmental noise or coupling strength. Second, we introduce a displaced-Sagnac interferometer architecture where intrinsic path delays naturally suppress the cross-components $\ket{HV}$ and $\ket{VH}$ of the two-photon state, H and V denoting polarization of the photons. These delays, which would be detrimental in conventional interferometric implementations, instead produce behavior closely mimicking correlated amplitude damping —a form of collective decoherence where both qubits flip together or not at all, rather than independently. This realization connects idealized theoretical models to realistic photonic systems without requiring the engineering of global system-bath coupling. Third, we develop a time-dependent framework that captures this physics through a single tunable parameter $x \in [0,1]$ arising from the temporal structure of photon detection. This parameter interpolates continuously between standard amplitude damping ($x=1$, where qubits decohere independently) and correlated amplitude damping ($x=0$, where decoherence is collective). Intermediate values of $x$ describe novel damping channels that are neither standard nor fully correlated, revealing a rich landscape of engineerable noise models accessible through simple geometric control of optical path delays.

Current strategies to mitigate decoherence fall into three broad categories, each with distinct operational requirements and fundamental limitations. Dynamical decoupling (DD)~\cite{Viola1999, Biercuk2009, Du2009, Souza2012, Lucamarini2011} applies sequences of control pulses designed to time-average environmental noise to zero. While effective in many contexts, DD requires precise pulse timing that becomes increasingly challenging as system complexity grows. Moreover, each pulse introduces errors that accumulate over the sequence, and the approach becomes perturbative for strong system-environment coupling. The quantum Zeno effect (QZE)~\cite{Maniscalco2008, Oliveira2008, Kondo2016, Long2022} employs frequent measurements that effectively freeze quantum evolution through repeated projection onto a target subspace. However, achieving the Zeno regime demands measurement rates faster than environmental correlation times, which often proves experimentally prohibitive. Quantum weak measurement reversal (QWMR)~\cite{Sun2010, Lee2023, Wang2016, Korotkov2010, Lee2011, Kim2012, Man2012} attempts probabilistic state recovery through a combination of weak measurements and reversal operations but by its nature succeeds only when post-selected on favorable measurement outcomes, making it unsuitable for deterministic protocols.

Despite their different mechanisms, all these approaches share a common feature: they require \emph{multiple} operations—pulse trains, repeated measurements, or statistical averaging—to suppress decoherence. This multiplicity has both practical and fundamental limitations. Practically, each additional operation introduces overhead and potential for error. Fundamentally, the multi-operation requirement suggests that existing approaches primarily fight decoherence through brute force averaging or projection, rather than exploiting the underlying structure of the open-system dynamics.

Here we demonstrate a fundamentally different paradigm based on a simple observation: quantum decay exhibits a fundamental temporal asymmetry. Excited states decay to ground states, never reverse. For a two-qubit entangled state $\ket{\psi} = \alpha\ket{HH} + \beta\ket{VV}$ under amplitude damping, the $\ket{VV}$ component (both qubits excited) preferentially decays toward $\ket{HH}$ (both ground), $\ket{HV}$, and $\ket{VH}$ (single excitations). Now consider inserting a local NOT operation at an intermediate time $t_1$ during this evolution. The NOT swaps the qubit populations: $\ket{H} \leftrightarrow \ket{V}$. Critically, this population inversion means the subsequent evolution from $t_1$ onward follows a completely different trajectory because the system now ``sees" inverted populations---what was stable becomes unstable and vice versa. This raises a striking question: can the \emph{timing} of a single local NOT operation---with no change to the noise itself---completely control decoherence?

Early theoretical work by Rau, Ali, and Alber ~\cite{Rau2008, Ali2008} showed that appropriately timed local operations can manipulate the onset of entanglement sudden death (ESD). These studies revealed that the timing of a local operation during dissipative evolution can influence whether ESD is hastened, delayed, or avoided. However, these analyses were formulated for specific noise models and did not provide a continuous channel framework describing how such control emerges across a broader family of dissipative dynamics. Subsequent work by Singh et al. ~\cite{Singh2017} proposed an optical realization of this idea using cascaded amplitude-damping channels, suitably adapting H and V for polarization instead of energy excitation. In this case, experimental realization faced a fundamental barrier: the proposed scheme required two separate interferometers implementing sequential damping channels, with path-length matching at the nanometer scale over meter-long setups—a requirement far beyond practical feasibility. This motivates the search for an architecture that captures the underlying physics while remaining experimentally feasible. Our work addresses this challenge through a key insight: rather than eliminating the intrinsic path delays that arise in cascaded optical channels, we exploit them. In a displaced-Sagnac interferometer, these delays naturally suppress the cross-components $\ket{HV}$ and $\ket{VH}$, producing behavior closely mimicking correlated amplitude damping.

The mechanism underlying our protocol is conceptually straightforward. Amplitude damping preferentially drives the transition $\ket{V} \to \ket{H}$ but not the reverse. When applied to $\ket{\psi} = \alpha\ket{HH} + \beta\ket{VV}$ with $\alpha < \beta$, damping causes the $\ket{VV}$ component to decay, generating mixed-state components $\ket{HV}$ and $\ket{VH}$ that destroy entanglement. This leads to a redistribution of the original $|VV\rangle$ amplitude among four components in the $|H\rangle\,, |V\rangle$ basis of the two-qubit system. A local NOT operation swaps the state to $\ket{\psi'}=\beta'\ket{HH} + \alpha'\ket{VV} + \text{cross terms}$, where $\beta' < \beta$ and $\alpha' > \alpha$ reflect partial decay that has already occurred during the first damping stage. When population inversion leads to $\alpha' < \beta'$, subsequent damping attacks can prevent the entanglement from ever reaching zero. This is \emph{not} quantum error correction: the protocol does not restore the initial state. 
Rather, we perform trajectory engineering, exploiting decay asymmetry to steer evolution onto paths where entanglement survives longer or dies faster, depending on when we intervene.

This work establishes that temporal control---\emph{when} to apply operations—can be as important as operational control---\emph{what} operations or \emph{how many}---in determining open-system dynamics. The single-shot, non-perturbative nature of our protocol is relevant to any platform where local gates are fast, but gate sequences accumulate errors, such as superconducting qubits with coherence-limited gate fidelities approaching fundamental bounds. The tunable parameter $x$ connecting idealized and realistic noise models suggests routes to engineer custom decoherence channels by controlling temporal correlations, applicable beyond photonics to circuit QED and trapped ions, where bath spectral properties can be tailored. Our Hamiltonian formulation reveals connections to dynamical decoupling through a custom ``parity operator," but operating non-perturbatively rather than in the weak-coupling limit where standard DD applies.

The remainder of this paper is organized as follows. Section~\ref{sec:theory} develops the theoretical framework, introducing standard and correlated amplitude-damping models, deriving the phase diagram showing avoidance, delay, and hastening regimes, presenting the time-dependent formalism with parameter $x$, and proving that local NOT is the unique optimal operation. Section~\ref{sec:hamiltonian} presents the Hamiltonian formulation connecting our protocol to dynamical decoupling. Section~\ref{sec:experiment} describes the experimental implementation using a displaced-Sagnac interferometer. Section~\ref{sec:results} demonstrates complete ESD manipulation across all three regimes. Section~\ref{sec:discussion} discusses the physical mechanism, platform generalization, and error analysis. We conclude in Section~\ref{sec:conclusion} with implications and open questions.\\

\noindent The principal theoretical and experimental results are as follows:
\begin{enumerate}
    \item We develop a time-dependent open-system framework for a continuous family of amplitude-damping-like channel (ADC) and correlated amplitude-damping-like channels (CADC). A physically motivated time-shift operator induces an effective correlation parameter $x \in [0,1]$ that interpolates between independent ADC-like ($x=1$) and a CADC-like limit ($x \approx 0$), and we obtain closed-form expressions for the concurrence and the associated ESD boundary as functions of the damping parameters $(p,P)$ of the first and second damping channels, respectively, and the time delay parameter $x$.
    \item Within this family we analytically classify all possible single-shot control behaviors and prove the optimality of a single local NOT. We show that a single local $\sigma_x$ applied at an intermediate time can avoid, delay, or hasten ESD by steering the dissipative trajectory, and we demonstrate that cascaded $(\mathrm{CADC{-}NOT})^n$ protocols and a broad class of general $SU(2)$ controls cannot outperform this single-shot scheme.
    \item We experimentally implement the framework in a displaced-Sagnac interferometer and realize a CADC-like channel with a tunable $x$, fully characterize both ADC and CADC-like behavior, and perform new control experiments with and without the NOT, thereby observing all three regimes of ESD manipulation in a single platform with quantitative agreement between theory and experimental data.
\end{enumerate}
Taken together, these results present a unified theoretical and experimental demonstration that temporal steering of open-system dynamics via a single local NOT operation can provide deterministic control of entanglement loss within a widely relevant noise family.

\section{Theoretical Framework}
\label{sec:theory}

We consider a two-qubit state $\rho_{in} = \ket{\psi}\bra{\psi}$ with $\ket{\psi} = \alpha\ket{HH} + \beta\ket{VV}$ evolving through two cascaded ADCs with an intermediate local NOT operation to steer the ESD trajectory. The damping channels are characterized by parameters $p$ and $P$. These parameters represent the probability that an excited state $\ket{V}$ decays to the ground state $\ket{H}$ during each damping stage. The evolution, described by Kraus operators, depends crucially on whether a local NOT operation is inserted between the two channels. The initial state $\rho_{in}$ evolves through the damping channel through the action of Kraus operators $\mathbb{K}_i$ to give,
\begin{equation}\label{rho_evolution}
\rho_{out}=\sum_i \mathbb{K}_i\rho_{in}\mathbb{K}_i^\dagger\,, \ \ \text{subject to }\ \ \sum_i \mathbb{K}_i^\dagger\mathbb{K}_i=\mathbb{I}\,.
\end{equation}

One of the most widely used measures of quantum entanglement between two qubits is the concurrence~\cite{WottersConcurrence}. For an arbitrary bipartite quantum state $\rho$, the concurrence is defined as
\begin{eqnarray}
 \label{eqn:concurrence1}
C(\rho) = \text{max}\{ 0, \sqrt{\lambda_{1}}- \sqrt{\lambda_{2}}- \sqrt{\lambda_{3}}- \sqrt{\lambda_{4}}\},
\end{eqnarray}
where the quantities $\lambda_{i}$ are the eigenvalues in decreasing order of the matrix:
\begin{eqnarray*}
R = \rho (\sigma_{y}\otimes\sigma_{y})\rho^{*}(\sigma_{y}\otimes\sigma_{y}).
\end{eqnarray*}
 Here $\rho^{*}$ is the complex conjugate of the matrix representing the input state $\rho$ and $\sigma_{y}$ is the Pauli matrix,
 \begin{eqnarray*}
\label{eqn:concurrence3}
 \sigma_{y} = \begin{pmatrix}
0 & -i \\
i & 0
\end{pmatrix}.
\end{eqnarray*}

The detailed derivation of the Kraus operators from the unitary evolution of the system and environment is given in Appendix \ref{app:kraus}.
For standard ADC, each qubit couples independently to its local environment. The single-photon Kraus operators are $K_1 = \text{diag}(1, \sqrt{1-p})$ and $K_2 = \text{off-diag}(0, \sqrt{p})$, where we use shorthand notation for the $2\times2$ matrices. The two-photon evolution involves tensor products of these operators. Without a NOT operation between channels, the concurrence is
\begin{eqnarray}
C(p,P) &=& 2 \max\Big[0, \big\{|\alpha\beta| - |(p(1-P) + P)\beta^2|\big\}\nonumber\\
&&\times(1-p)(1-P)\Big].
\label{eq:adc_no_not}
\end{eqnarray}

With a local NOT operation $\sigma_x\otimes\sigma_x$ applied after the first channel but before the second, and for our choice of parameters $\alpha$ and $\beta$, the output becomes an X-state \cite{Yu2004, Yu2009, Rau2008, Singh2017} with the concurrence
\begin{eqnarray}
C(p,P) &= 2 \max\Big[0, \big\{ 1-p(1-P)\big\}(1-P)|\alpha\beta|-\nonumber\\
&(1-P)\big[ P-(1-p)\big\{ P-p(1-P)\big\}\beta^{2}\big] \Big],
\label{eq:adc_with_not}
\end{eqnarray}
 The key observation is that NOT fundamentally changes which density matrix elements compete in determining the concurrence.

CADC represents a qualitatively different decoherence process where the two qubits couple to a common environment, leading to collective rather than independent decay. The Kraus operators act on the full two-qubit Hilbert space, with the transformation $\ket{VV} \to \sqrt{p}\ket{HH} + \sqrt{1-p}\ket{VV}$ being correlated—both qubits flip together or not at all. This suppresses the single-excitation states $\ket{HV}$ and $\ket{VH}$. For a CADC first channel followed by a standard ADC second channel, without NOT, the concurrence is
\begin{equation}
C(p,P) = 2 \max\left[0, \{|\alpha\beta| - \sqrt{1-p} P \beta^2\}\sqrt{(1-p)}(1-P)\right].
\label{eq:cadc_no_not}
\end{equation}

With NOT between channels:
\begin{equation}
C(p,P) = 2 \max\left[0, \sqrt{1-p}|\alpha\beta| - P(\alpha^2 + p\beta^2)\right](1-P).
\label{eq:cadc_with_not}
\end{equation}

The ESD line---defined as the set of $(p,P)$ values where $C(p,P) = 0$ for the first time---partitions the parameter space into regions with qualitatively different behavior. For $\rm ADC-NOT-ADC$:
\begin{equation}
[1 - p(1-P)]|\alpha\beta| = [P - (1-p)(P - p(1-P))]\beta^2.
\label{eq:esdline_adc}
\end{equation}

For $\rm CADC-NOT-ADC$:
\begin{equation}
\sqrt{1-p}|\alpha\beta| = P(\alpha^2 + p\beta^2).
\label{eq:esdline_cadc}
\end{equation}

\begin{figure}[t]
\centering
\includegraphics[width=\linewidth]{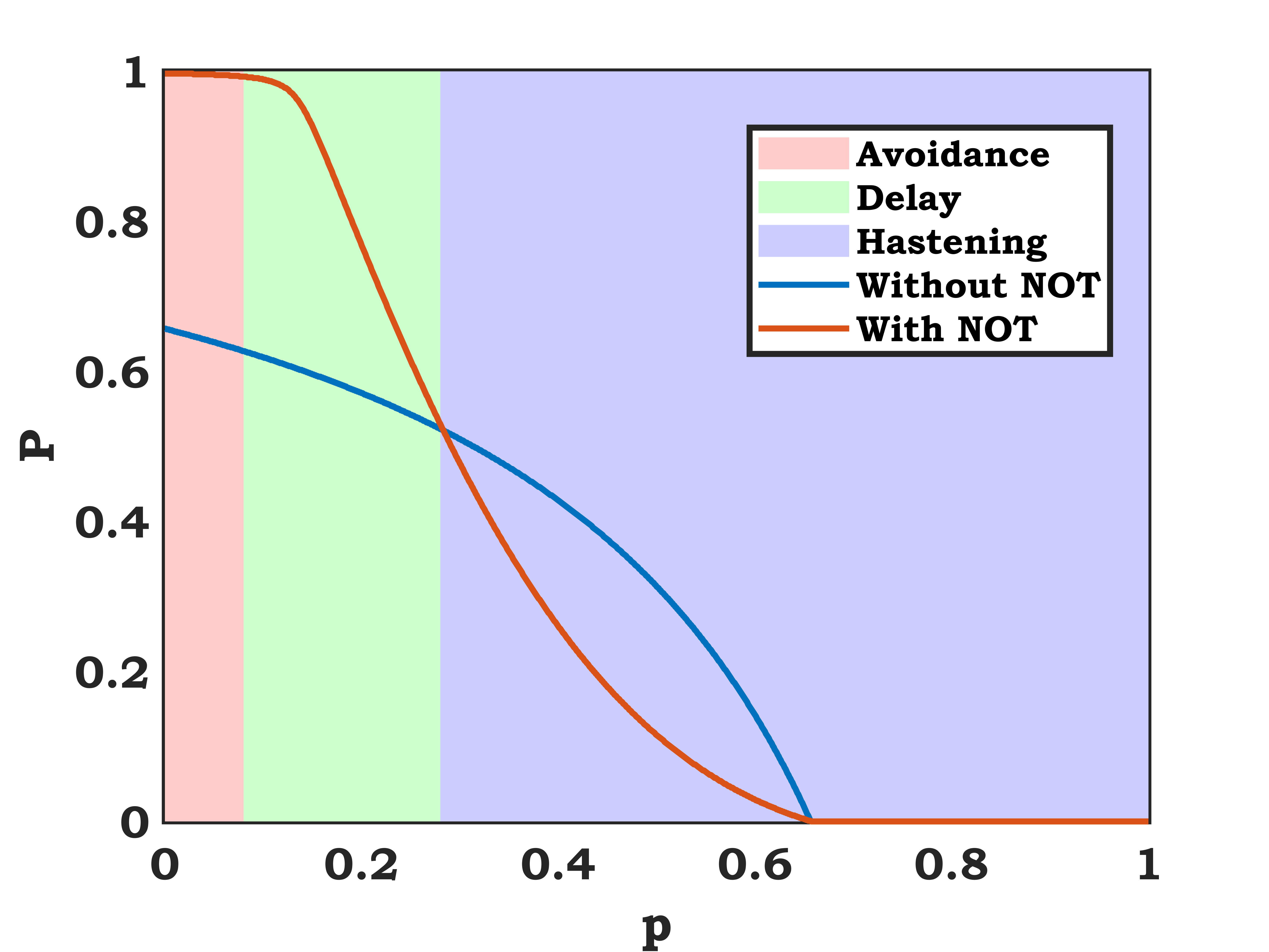}
\caption{A contour plot showing the ESD line $C(p,P)=0$ for $x=1$ (standard ADC) with initial state $|\Psi\rangle=\alpha|HH\rangle+\beta|VV\rangle$, where $\alpha=0.55$. The $(p,P)$ parameter space is divided into three regions: avoidance (red), delay (green), and hastening (blue). The blue and red curves denote the ESD boundaries without and with the NOT operation, respectively.}
\label{fig: adc_adc}
\end{figure}

\begin{figure}[t]
\centering
\includegraphics[width=\linewidth]{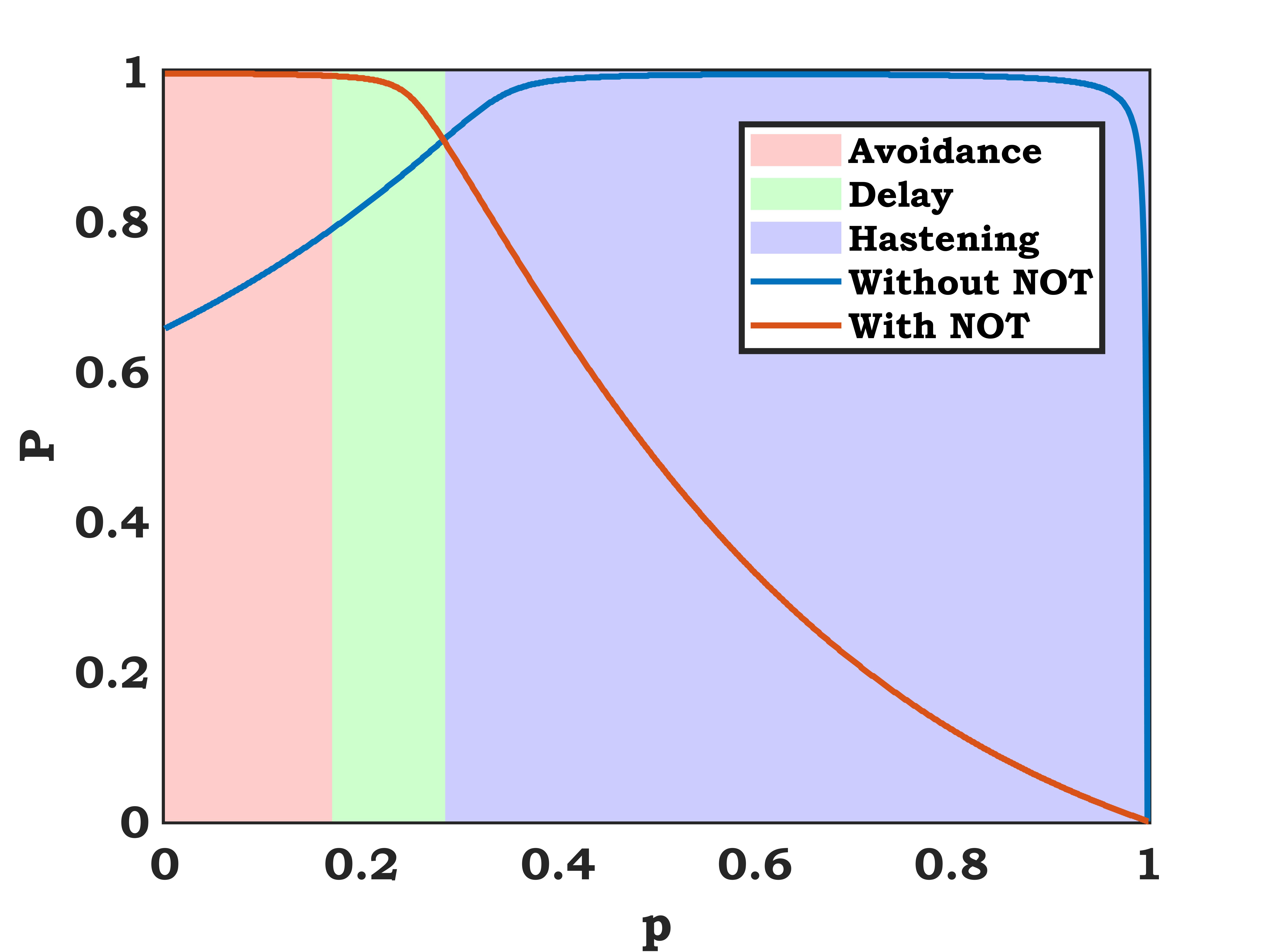}
\caption{ESD lines for $x=0$ ($\rm{CADC}-like$) with $\alpha=0.55$. The three regions retain the same physical interpretation but the ``without NOT" ESD line has a positive slope, opposite to the ADC case.}
\label{fig: cadc_adc}
\end{figure}

These ESD lines depend sensitively on the initial state parameter $\alpha$. Figures~\ref{fig: adc_adc} and \ref{fig: cadc_adc} show the resulting phase diagrams for $\alpha = 0.55$. The $(p,P)$ plane divides into three distinct regions. In the avoidance regime (small $p \lesssim 0.17$), the first damping barely affects the state before the NOT. Population inversion then places the larger coefficient on the stable $\ket{HH}$ state, preventing ESD. The delay regime (intermediate $0.17 < p < 0.28$) represents partial degradation before NOT. The inversion reduces but does not eliminate the asymmetry driving ESD, postponing it by $\Delta P \approx 0.3$. The hastening regime (large $p \gtrsim 0.28$) exhibits counterintuitive behavior where the NOT accelerates or induces ESD because the state is already too degraded. Detailed derivations are done in Appendix \ref{app: ESDLine}.

\begin{table}[t]
\centering
\caption{Summary of ESD manipulation regimes for $\alpha=0.55$.}
\label{tab:regimes}
\begin{tabular}{lcc}
\toprule
Regime & Condition & Effect of NOT \\
\midrule
Avoidance & $p \lesssim 0.17$ & $P_{\text{ESD}} \to 1$ \\
Delay & $0.17 < p < 0.28$ & $\Delta P \approx 0.3$ \\
Hastening & $p \gtrsim 0.28$ & ESD accelerated \\
\bottomrule
\end{tabular}
\end{table}

\begin{figure}
    \centering
    \includegraphics[width=0.9\linewidth]{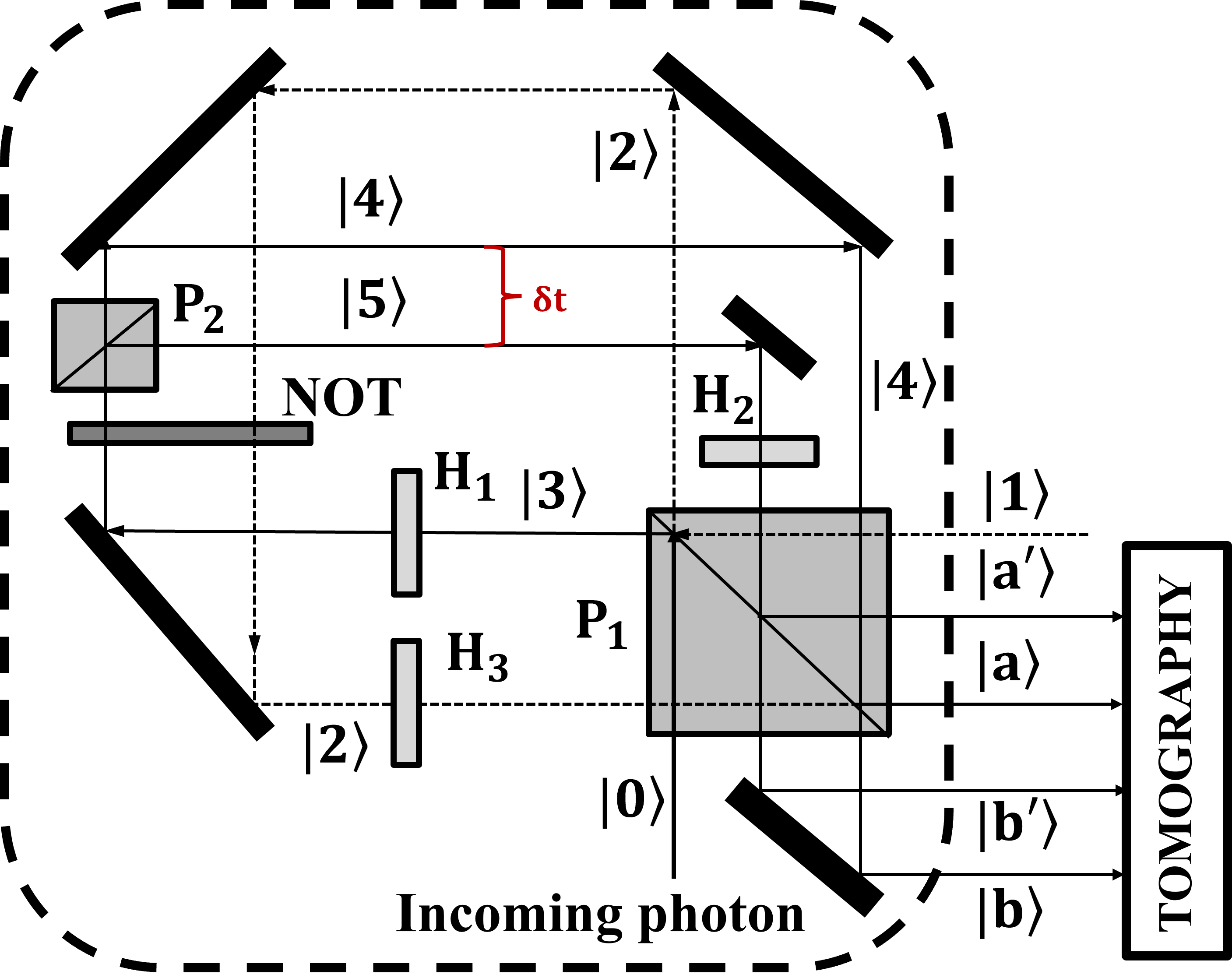}
    \caption{Detailed schematic diagram of each interferometric setup. The time delay between the two paths of the second polarizing beam-splitter $\mathcal{P}_2$ is demarcated by $\delta t$ in red. The delay $\delta t$ between the two paths of the second polarizing beam splitter controls the suppression of the cross components $\ket{HV}$ and $\ket{VH}$ within the coincidence window.}
    \label{fig:setup_dt}
\end{figure}

The experimental challenge is implementing these different damping channels. Previous proposals~\cite{Singh2017} required two separate interferometers with nanometer path matching—experimentally infeasible. Our solution exploits intrinsic path delays. In a displaced-Sagnac interferometer, photon paths have different lengths, creating a time delay $\delta t$, as illustrated in Fig. \ref{fig:setup_dt}.
If $\delta t$ exceeds the coincidence window $\Delta t$, pairs with one photon in each path are suppressed. Crucially, the $\ket{HV}$ and $\ket{VH}$ components populate such path combinations, while $\ket{HH}$ and $\ket{VV}$ do not. This naturally suppresses cross-terms, mimicking CADC.

We model the input state as a time-ensemble:
\begin{equation}
\ket{\psi_{\text{in}}} = \sum_t [a(t)\bar{a}(t)\ket{H_0H_0'} + b(t)\bar{b}(t)\ket{V_0V_0'}],
\label{eq:time_ensemble}
\end{equation}
where $a(t)\,,\bar{a}(t)$ and $b(t)\,,\bar{b}(t)$ are the temporal amplitude envelopes for the $|H\rangle$ and $|V\rangle$ polarized photons along paths $|0\rangle\,, |0'\rangle$ respectively, and define a time-shift operator through
\begin{equation}
 x(\delta t)  = \abs{\frac{\int d\omega\,a(\omega)\bar a(-\omega)e^{-i\omega\delta t}}{\int d\omega\,a(\omega)\bar a(-\omega)}}\,,
\label{eq:x_definition}
\end{equation}
See Appendix~\ref{app:x_explain}  for details. The parameter $x\in[0,1]$ measures the effective temporal overlap of delayed photon wavepackets within the coincidence window: $x=1$ corresponds to the $\rm ADC-like$ limit in which the cross terms survive, whereas $x=0$ corresponds to the $\rm CADC-like$ limit in which the cross terms are suppressed.

Importantly, the maps generated in this way are effective conditional maps associated with the detected coincidence ensemble. At $x=1$ the map reduces to $\rm ADC-like$ which is fully CPTP. However, for $x<1$ the cross components $|HV\rangle$ and $|VH\rangle$ are progressively suppressed by the coincidence window, and the output state is no longer normalized. The normalization factor
\begin{equation}
\eta_{\mathrm{post}}(p,x;\alpha,\beta)=N(p,x;\alpha,\beta)=1-2p(1-p)(1-x^2)|\beta|^2,
\label{eq:eta_post_note}
\end{equation}
tracks precisely how much probability is carried by the suppressed cross components at each value of $x$ and as a function of the damping parameter $p$. The $x=0$ limit, realized in the present experiment, corresponds to complete suppression of the cross components and is directly verified in Fig. \ref{fig: hvvh}. \\ 

\begin{figure}[t]
\centering
\includegraphics[width=0.9\linewidth]{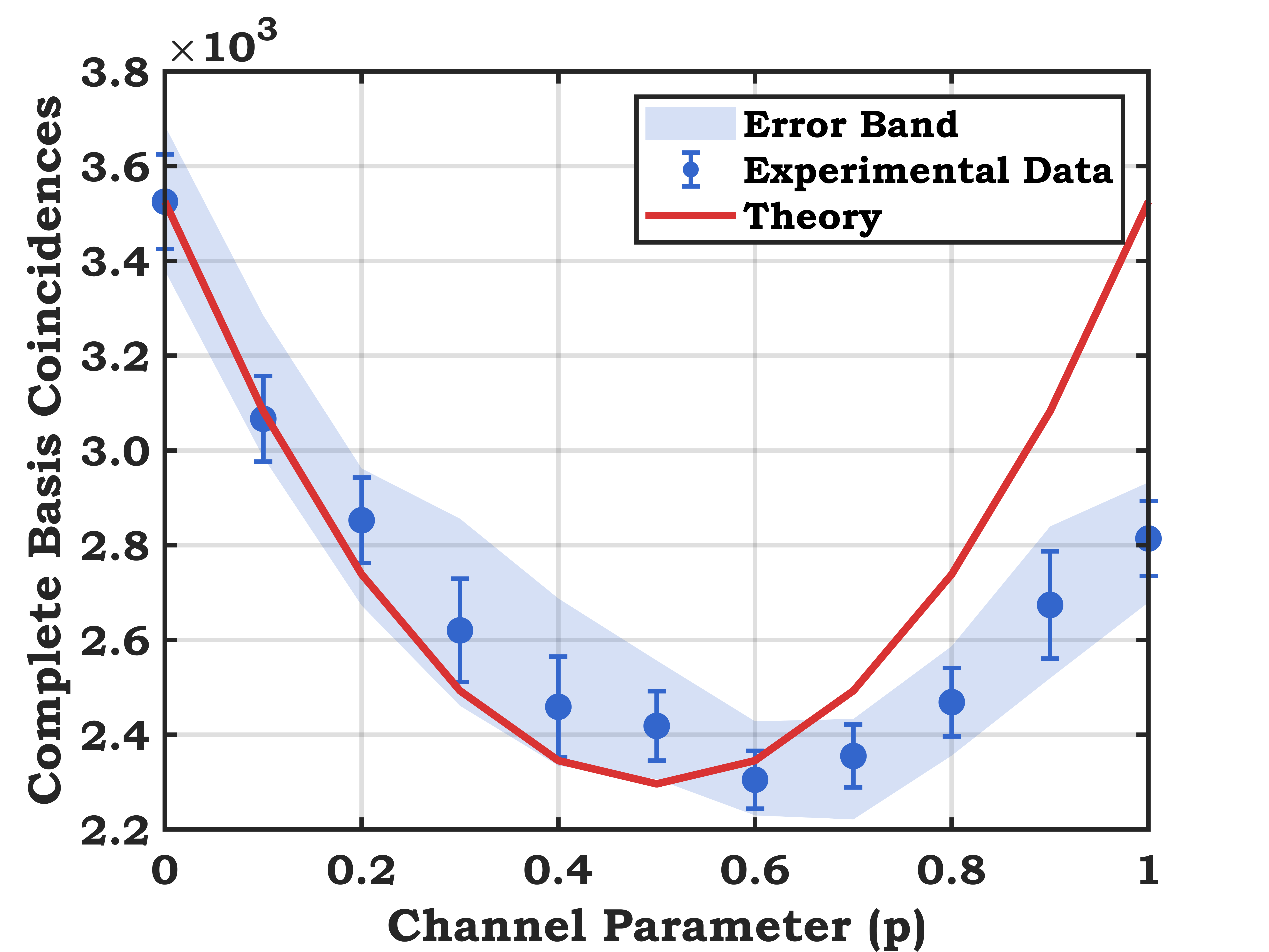}
\caption{Experimental coincidence counts (blue markers) compared against theoretical counts (red line) $\eta_{\rm post}C_{ref}$ as a function of the channel parameter for the first channel $p$. Here \(C_{\rm ref}\) is the initial coincidence count at \(p=0\), and \(\eta_{\rm post}\) is the theoretically admitted-event fraction given by the normalization factor in Eq.~\eqref{eq:eta_post_note}.}
\label{fig:Normalization_verification}
\end{figure}

For the main experimental state, $\alpha=0.55$, and the $\rm CADC-like$ regime \(x \approx 0\), $\eta_{\rm post}$ gives a minimum admitted-branch weight of approximately \(0.65\) at \(p=0.5\). This shows that the detected branch is not a vanishing-probability component of the dynamics. We compare the theoretically expected coincidence counts using the admitted event fraction $\eta_{\rm post}$ multiplied with the initial counts $C_{ref}$ ($p=0$), with the experimentally observed coincidence counts in Fig.~\ref{fig:Normalization_verification}. The difference in the experimental tail from theoretical expectations is attributed to variable coupling efficiencies at different detectors. As $p$ transitions from $p<0.5$ to $p>0.5$, more photons are directed toward the $L_2 R_2$ couplers, which has lower coupling efficiency in comparison to $L_1 R_1$, causing a drop in coincidence counts (see Fig.~\ref{ESDExpt}).

With an intermediate local NOT, the output conditional state takes the form
\begin{equation}
\rho_{\mathrm{out}}=\frac{1}{N}
\begin{pmatrix}
A&0&0&\mathcal{X}\\
0&B&0&0\\
0&0&C&0\\
\mathcal{X}^{*}&0&0&D
\end{pmatrix},
\end{equation}
where
\begin{align}
    \begin{split}
        A &= (1-p)^2\beta^2+P\big[P-(1-p)\big\{ P-p(2x^2-P)\big\}\beta^2\big]\,,\\
        B &=C=(1-P)\big[ P-(1-p)\big\{ P-p(x^2 -P)\big\} |\beta|^2\big]\,,\\
        D &= (1-P)^2 \left(|\alpha|^2+|\beta|^2 p^2\right)\,,\ \mathcal{X} =\alpha^\star  \beta  (1-p)(1-P)\,,\\
        N &= 1-2p(1-p)(1-x^2)|\beta|^2\,.
    \end{split}
\end{align}
 The associated concurrence is
\begin{eqnarray}
C^{\rm }(p,P)&=&\frac{2}{N}\max\Big[0,(1-p)|\alpha\beta|-\Big\{P-(1-p)\nonumber\\
&&\big(P-p(x^2-P)\big)|\beta|^2\Big\}\Big](1-P).
\label{eq:concurrence_timedep}
\end{eqnarray}

\begin{figure}
\centering
\includegraphics[width=0.9\linewidth]{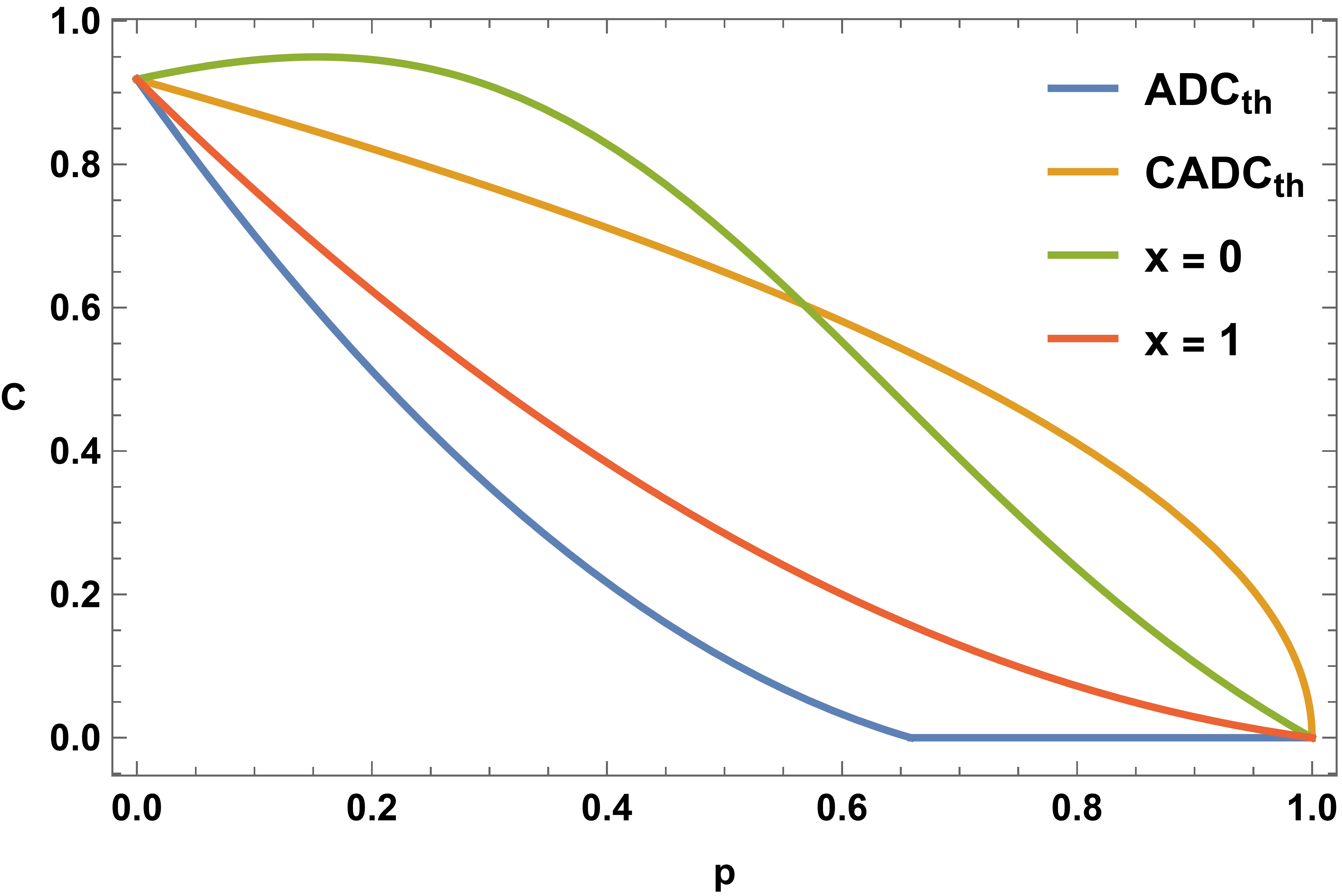}
\caption{Comparison of the effective conditional-map framework with the conventional ADC (blue) and CADC (orange) for $\alpha=0.55$. The time-dependent formalism continuously interpolates between the $\rm ADC-like$ ($x=1)$ (red) and $\rm CADC-like$ ($x=0$) (green) limits, which are different from the conventional damping channels.}
\label{fig:theory_expt_comp}
\end{figure}

Figure~\ref{fig:theory_expt_comp} illustrates the behavior of a conventional ADC (blue curve) and CADC (orange curve) channel. In the same figure, we also plot the limiting cases $x=1$ (red curve) and $x=0$ (green curve) (see Appendix \ref{app: ESDLine} for the derivation of the concurrence for the time-dependent case). This demonstrates the difference between $\rm ADC-like$ and $\rm CADC-like$ limits. \eqref{eq:concurrence_timedep} approaches the apparatus-defined $\rm ADC-like$ at $x=1$ and $\rm CADC-like$ limit at $x=0$. Intermediate values of $x$ correspond physically to partial overlap of delayed photon wavepackets and continuously modulate the suppression of the cross terms $\ket{HV}$ and $\ket{VH}$. This provides a compact description of how geometry and detection timing reshape dissipative entanglement dynamics.

To verify NOT is optimal, we test the general SU(2) operator $\vec{\sigma} = n_0 I + i n_x \sigma_x + i n_y \sigma_y + i n_z \sigma_z$. Analysis (Appendix~\ref{app:su2}) shows NOT ($n_x = 1$, others zero) maximizes the concurrence for delaying ESD. Population inversion is the key physical ingredient, not arbitrary unitary scrambling.

\section{Hamiltonian Formulation and Connection to Dynamical Decoupling}
\label{sec:hamiltonian}
The Hamiltonian perspective reveals why single shot protocol performs better in comparison to traditional approaches such as Dynamical Decoupling \cite{Lucamarini2011}. The total unitary $U = \exp(-iH)$ consists of $H = H_{\text{free}} + H_{\text{int}}$, where $H_{\text{free}}$ produces no system-environment entanglement and $H_{\text{int}}$ creates correlations responsible for decoherence. $H_{\text{free}}=H_A^{free}\otimes H_B^{free}$ where $A\,, B$ are the respective  single qubit subsystems such that,
\begin{align}
\begin{split}
H_{\text{free}}(\alpha |HH\rangle+\beta|VV\rangle)\otimes|00\rangle\rightarrow (\alpha' |HH\rangle+\beta'|VV\rangle)\\ \otimes|kk\rangle\,.
\end{split}
\end{align}
Thus, $H_{\text{int}}$ is solely responsible for system-environment correlations. Dynamical decoupling seeks a parity operator $\Pi$ satisfying $[\Pi, H_{\text{free}}] = 0$ and $\{\Pi, H_{\text{int}}\} = 0$. Inserting the parity kicks between the total unitary $U$, results in
\begin{equation}
    (\Pi\cdot U\cdot \Pi)\cdot U = U_{\rm free}^2\,.
\end{equation}
Hence repeated operations of the iterative kicks ($n-$times) cancel the environmental interaction in the following sense,
\begin{equation}
    {\rm n-terms}\dots\underbrace{\left(\Pi\cdot U\cdot\Pi\right)\cdot U}_{\rm 1st-term} = \exp(-2i n H_{\rm free}) = U_{\rm free}^{2n}\,,
\end{equation}
Note that the parity operator for the system is then given by $\Pi=\Pi_A\otimes \Pi_B$, the tensor product of the operators for each subsystem. In order to determine $\Pi_A$, we write $H_A^{free}$ in its eigen basis, (\cite{SM} See Supplementary Material) and decompose,
\begin{equation}
H_A^{int}=\sqrt{p}\sum_{i\neq j}\mu_{ij}|\lambda_i\rangle\langle \lambda_j|+O(p)\,.
\end{equation}
Therefore, to the first order, we can write the interacting Hamiltonian as
\begin{equation}
H_{\text{int}}=\sqrt{p}\left(H_A^{int}\otimes H_B^{free}+H_A^{free}\otimes H_B^{int}\right)+O(p)\,.
\end{equation}
The ``parity" operator $\Pi_A=\sum_i s_i |\lambda_i\rangle\langle \lambda_i|$ by construction, implies $s_i=\pm1\,,\ \forall\ i$. Of these $s_i=-1$ for $i=6\,,7\,,16\,,17$ while the rest are $1$. The parity operator is thus,
\begin{equation}
\Pi_A = I - 2(\ket{\lambda_6}\bra{\lambda_6} + \ket{\lambda_7}\bra{\lambda_7} + \ket{\lambda_{16}}\bra{\lambda_{16}} + \ket{\lambda_{17}}\bra{\lambda_{17}})\,,
\label{eq:parity_operator}
\end{equation}
where the explicit forms of the relevant eigenvectors are given in Appendix~\ref{app:eigenvectors}. In terms of the computational basis, {\it i.e.,} polarization and path states, we can write $\Pi_A$ as,
\begin{align}\label{total_parity}
\begin{split}
    \Pi_A = &|H\rangle\langle H|\otimes\left(I-|1\rangle\langle1|-|3\rangle\langle3|-|4\rangle\langle4|-|9\rangle\langle9|\right)\\&+|V\rangle\langle V|\otimes\left(I-|0\rangle\langle0|-|3\rangle\langle3|-|5\rangle\langle5|-|8\rangle\langle8|\right)\\&-|H\rangle\langle V|\otimes\left(|1\rangle\langle0|+|3\rangle\langle3|+|4\rangle\langle5|+|9\rangle\langle8|\right)\\&-|V\rangle\langle H|\otimes\left(|0\rangle\langle1|+|3\rangle\langle3|+|5\rangle\langle4|+|8\rangle\langle9|\right)\,.
\end{split}
\end{align}
where $|H\rangle$ and $|V\rangle$ are the polarization states of the photon and $|i\rangle$ denote the path states. $\Pi_B$ is of the exact same form as $\Pi_A$ with appropriate renaming of the path variables for subsystem B. \eqref{total_parity} demonstrates the need for custom set of offset pulses on various paths applied in parallel, to cancel the effective system-environment interaction.

\begin{table}[t]
\centering
\caption{Comparison of existing methods with our protocol.}
\label{tab:final_comparison}
\resizebox{\columnwidth}{!}{
\begin{tabular}{p{1.7cm} p{2cm} p{2cm} p{2cm} p{2.1cm}}
\toprule
Feature & \centering{DD} & \centering{QZE} & \centering{WMR} & This work \\
\midrule
Coupling & \centering{Weak} & \centering{Weak} & \centering{N/A} & Non-perturbative \\

Operational Characteristics & \centering{Pulse sequence} & \centering{Multiple measurements} & \centering{Pre/Post selection} & Single shot \\

Control Sequences & \centering{$n\gg 1$} & \centering{$n\gg 1$} & \centering{weak interaction + post-selection} & One time unitary evolution\\

Timing Precision & \centering{$\omega\tau=\pi$} & \centering{N/A} & \centering{Coarse} & Coarse ($\Delta p\!\sim\!0.05$) \\

State Recovery & \centering{Iterative} & \centering{Iterative and continuum} & \centering{Probabilistic} & Deterministic\\

Hardware Error & \centering{Accumulates iteratively} & \centering{Accumulates iteratively} & \centering{Limited by single weak interaction} & Limited by the single unitary \\
\bottomrule
\end{tabular}
}
\end{table}

Table~\ref{tab:final_comparison} compares our approach to other methods operating in weak coupling that require pulse sequences with $\Omega_{\text{pulse}} \gg \gamma_{\text{decay}}$ and precise timing. Our protocol is non-perturbative, requires only one operation with coarse timing tolerance, and avoids error accumulation. The parity operator projects onto system-environment eigenstates aligned with the NOT-induced population swap. A single application redirects the trajectory rather than averaging over many cycles. This clarifies why single-shot works: we perform steering, not averaging. Appendix~\ref{app:cascade} illustrates that cascading $n>1$ stages degrade performance, confirming that single-shot is optimal.

\section{Experimental Implementation}
\label{sec:experiment}

To realize time-delay-induced $\rm CADC-like$ behavior, we designed an architecture around four requirements: cascaded amplitude damping with tunable $p$ and $P$, intermediate NOT operation without destroying coherence, controllable path delays engineering parameter $x$, and interferometric stability over meter scales. Previous proposals~\cite{Singh2017} envisioned two separate Sagnac interferometers cascaded in series---experimentally infeasible due to nanometer path-matching requirements. Our solution recognizes that path delays, traditionally detrimental, can be exploited. By embedding both damping channels within a single displaced-Sagnac interferometer (DSI), intrinsic path differences naturally suppress $\ket{HV}$ and $\ket{VH}$ components. 

\begin{figure*}[t]
\centering
\includegraphics[width=\textwidth]{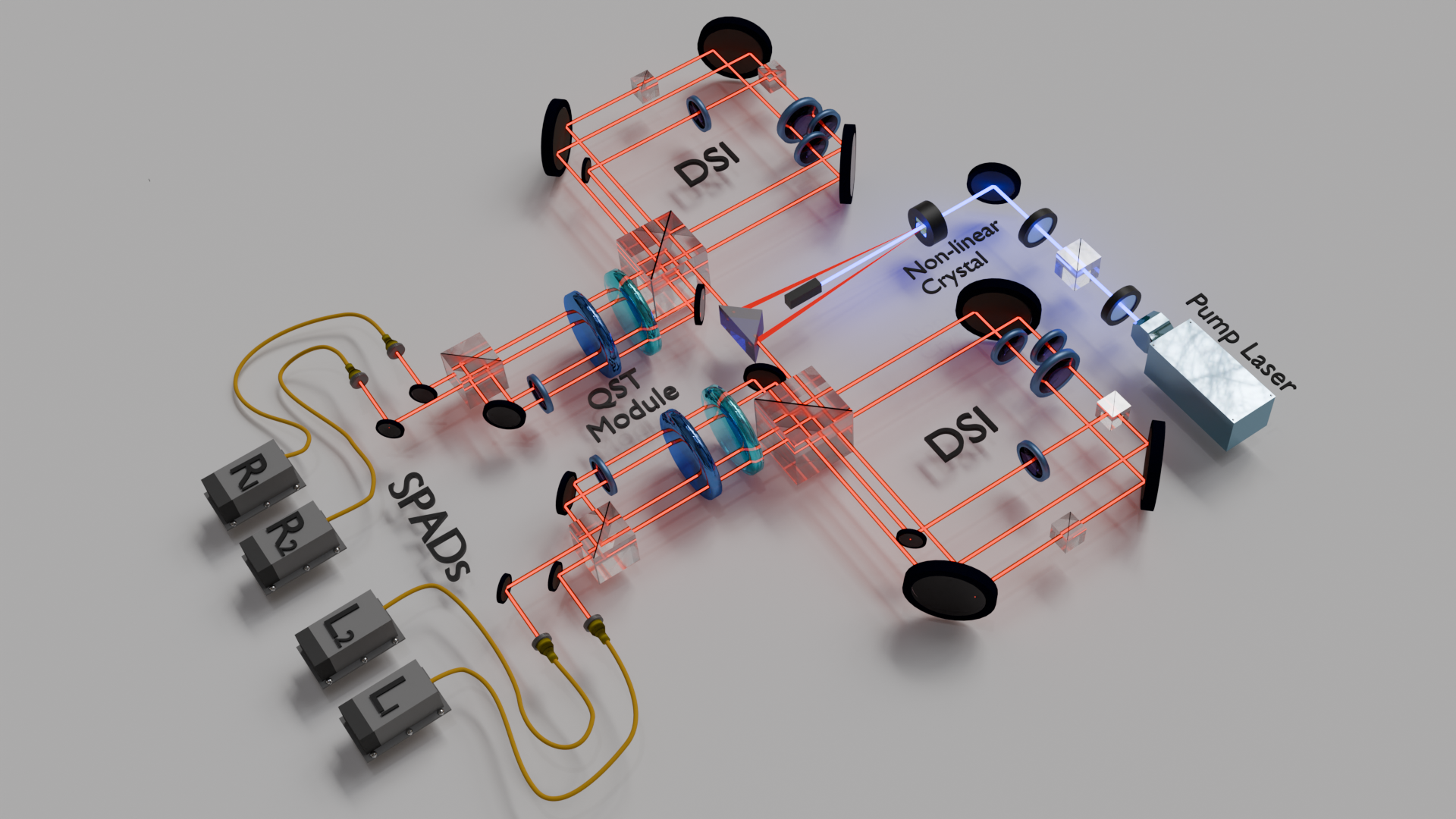}
\caption{Experimental schematic overview showing entangled photon source (left), displaced-Sagnac interferometer implementing cascaded damping protocol (center), and quantum state tomography module (right). DSI: Displaced-Sagnac Interferometer, QST: Quantum State Tomography, SPAD: Single-Photon Avalanche Diode.}
\label{ESDExpt}
\end{figure*}

Figure~\ref{ESDExpt} provides the system overview. The entangled photon source uses Type-I SPDC with two sandwiched $\beta$-BBO (Beta Barium Borate) crystals 
pumped by 405 nm laser (100 mW). Crystals oriented with perpendicular optic axes produce the state $\ket{\Psi} = \cos\alpha_0 \ket{HH} - \sin\alpha_0 \ket{VV}$ where pump polarization is controlled by a combination of polarization beam splitter (PBS) and a half-wave plate (HWP), 
tunes $\alpha = \cos\alpha_0$ from 0
to 1. After down-conversion at 810 nm, photons are spectrally filtered (3 nm bandwidth
) and spatially filtered (
using pinhole apertures) before fiber coupling. The typical single count rates were approximately $4*10^4/s$ per coupler, with the maximum coincidence count rate between different detector pairs reaching the order of $5*10^3/s$. 

\begin{figure}[t]
\centering
\includegraphics[scale=0.125]{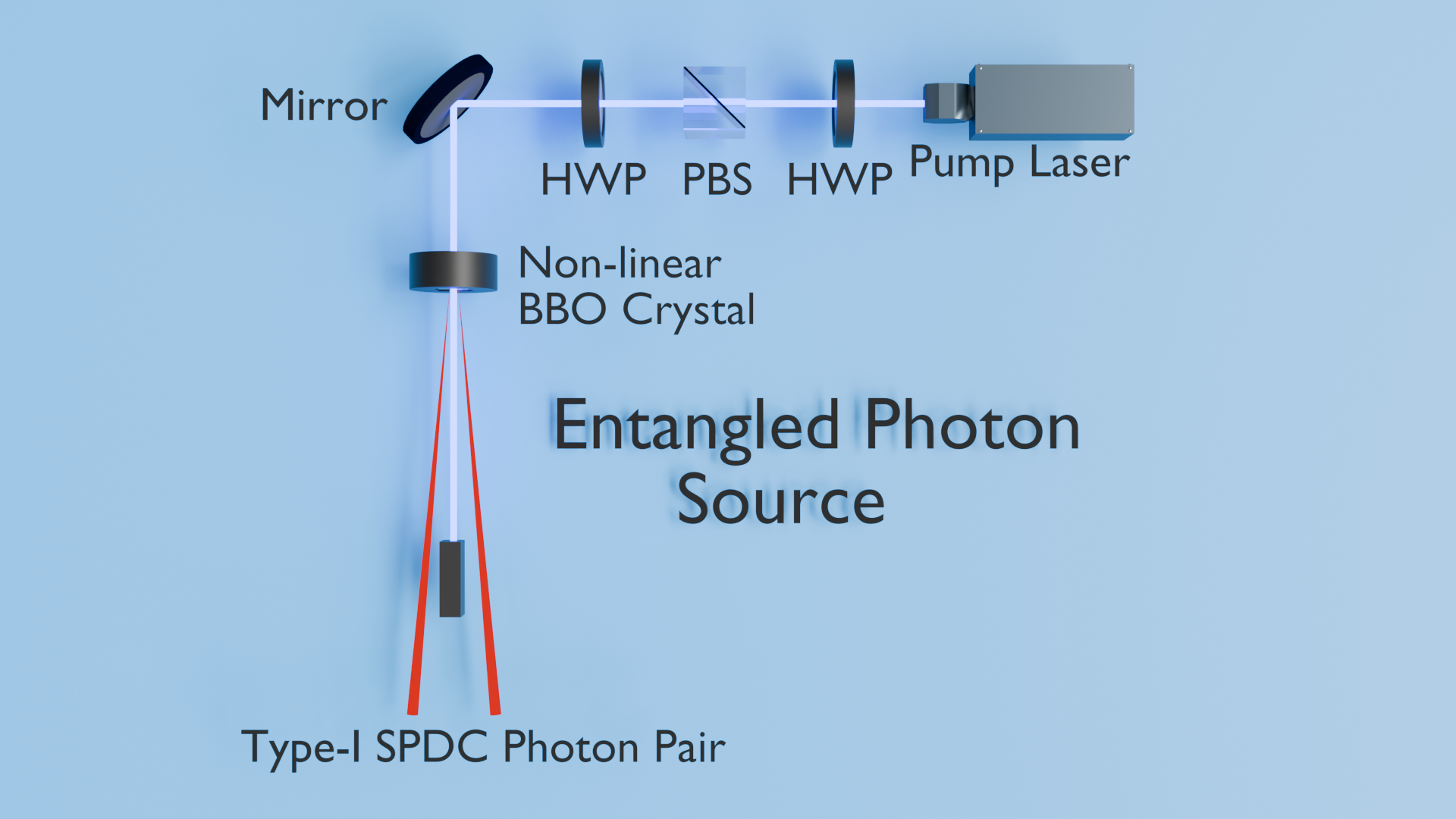}
\caption{Type-I SPDC entangled photon source. HWP H$_0$ controls pump polarization and entanglement parameter $\alpha$. Two 1 mm $\beta$-BBO crystals produce collinear degenerate pairs at 810 nm. DM: dichroic mirrors, BPF: bandpass filters, PH: pinholes, HWP: Half-Wave Plate, PBS: Polarizing Beam Splitter, BBO: Beta-Barium Borate.}
\label{phsource}
\end{figure}

Figure~\ref{phsource} shows source details. The source is aligned 
to maximize coincidence rates and verify high-quality entanglement ($C > 0.95$ for maximally entangled states) before ESD manipulation experiments.

\begin{figure*}[t]
\centering
\includegraphics[scale=0.35]{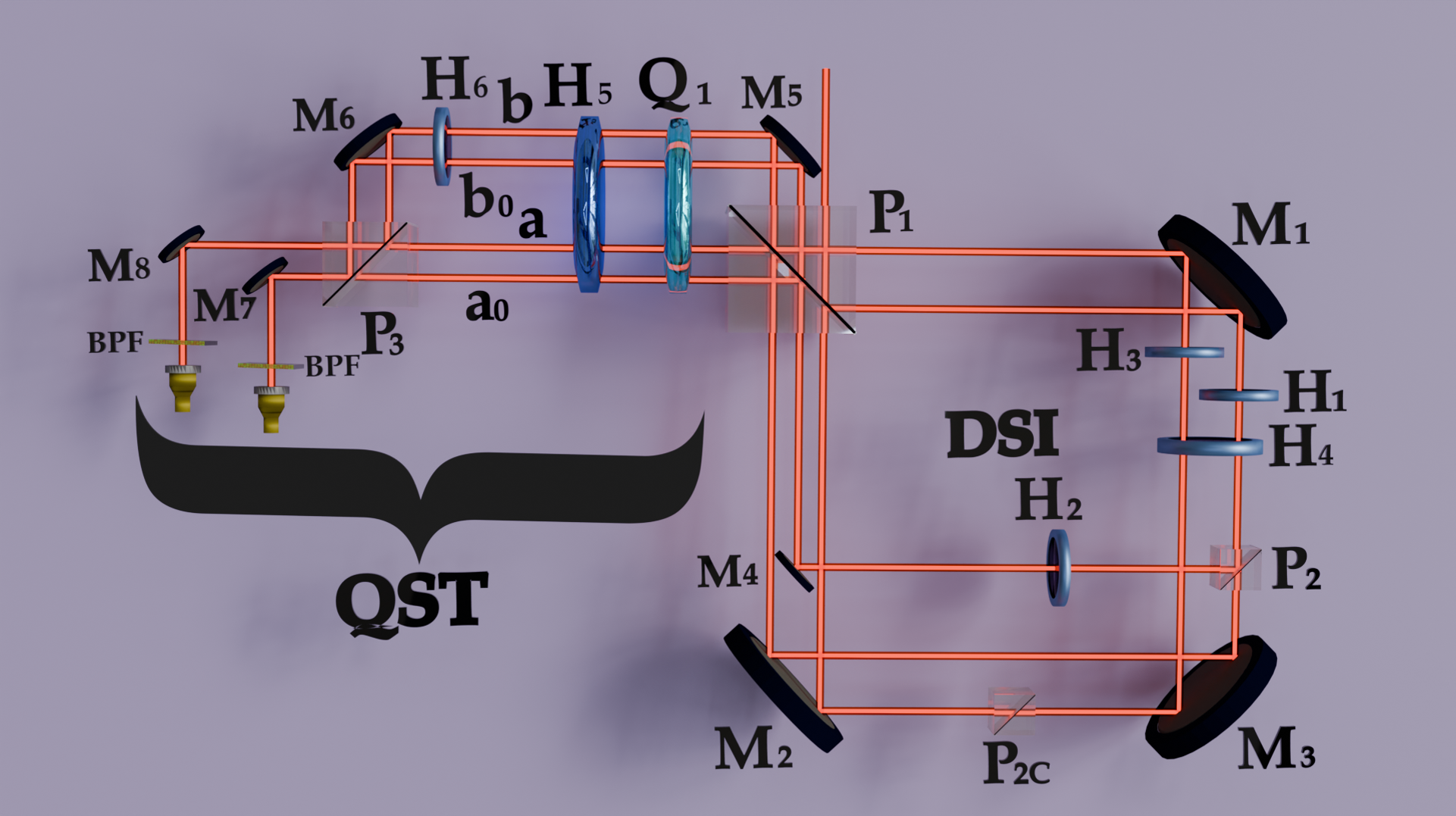}
\caption{Detailed displaced-Sagnac interferometer and QST schematic. PBS P$_1$ splits photons into clockwise (CW, path $\ket{2}$) and counter-clockwise (CCW, path $\ket{3}$). On CCW path, HWP H$_1$ at angle $\theta$ implements first ADC with $p = \sin^2(2\theta)$, then PBS P$_2$ reflects $\ket{V}$ to path $\ket{5}$ and transmits $\ket{H}$ to path $\ket{4}$. CW path bypasses H$_1$ and P$_2$. HWP H$_4$ at 45° implements NOT on both paths. HWPs H$_2$, H$_3$ at angle $\phi$ implement second ADC with $P = \sin^2(2\phi)$. Paths $\ket{4}$ and $\ket{5}$ have different spatial lengths creating time delay $\delta t \approx 1$ ns (for $\Delta L \approx 30$ cm). Combined with 1 ns coincidence window, this produces $x \approx 0$. QST module uses Q$_1$, H$_5$, H$_6$, PBS P$_3$ for 16 projective measurements onto SPADs L1, L2, R1, R2. M$_1$--M$_8$: mirrors, BPF: band-pass filter.}
\label{QST}
\end{figure*}

Figure~\ref{QST} shows the complete DSI architecture. Each photon enters PBS P$_1$, which splits by polarization into clockwise (CW) and counter-clockwise (CCW) paths. On the CCW path $\ket{3}$, HWP H$_1$ at angle $\theta$ implements polarization rotation mimicking spontaneous emission: $\ket{VV} \to \sqrt{p}\ket{HH} + \sqrt{1-p}\ket{VV}$ with $p = \sin^2(2\theta)$. PBS P$_2$ reflects $\ket{V}$ to path $\ket{5}$ and transmits $\ket{H}$ to path $\ket{4}$. The CW path bypasses both H$_1$ and P$_2$. HWP H$_4$ at 45° acts on both paths implementing $\sigma_x$: $\ket{H} \leftrightarrow \ket{V}$. HWPs H$_2$ (on CW) and H$_3$ (on CCW path $\ket{5}$), both at angle $\phi$, implement second ADC with $P = \sin^2(2\phi)$. Paths recombine at PBS P$_1$.

The crucial innovation: PBS P$_2$ creates spatially separated paths with length difference $\Delta L > 30$ cm, corresponding to time delay $\delta t > 1$ ns. This is comparable to coherence time $\tau_c > 1$ ns and larger than coincidence window $\Delta t = 1$ ns. For $\ket{HH}$ and $\ket{VV}$ components, both photons follow similar paths experiencing negligible relative delay. For cross-components $\ket{HV}$ and $\ket{VH}$, photons follow different paths with relative delay $\delta t$. When $\delta t$ becomes comparable to $\Delta t$, these are partially excluded from detected state. This is the physical origin of $x$ from Eq.~\eqref{eq:x_definition}. Our configuration gives $\delta t/\Delta t \ge 1$, yielding $x \approx 0$.

After the DSI, QST requires projective measurements onto 36 basis states. Our QST module uses quarter-wave plate Q$_1$ ($0^\circ$ or $45^\circ$), half-wave plate H$_5$ ($0^\circ$, $22.5^\circ$, $45^\circ$ or $67.5^\circ$), and half-wave plate H$_6$ ($45^\circ$) combined with PBS P$_3$ separating H and V. Detection uses four SPADs with a 1 ns coincidence window. For each projection, we integrate coincidence counts over 10 s, yielding counts that vary with projection efficiency and damping strength. The density matrix $\rho$ is reconstructed via maximum-likelihood estimation with positivity constraint. The concurrence of a two-qubit density matrix $\rho$ is estimated using Eq. \ref{eqn:concurrence1} 
as $C(\rho) = \max\left(0,\; \sqrt{\lambda_1} - \sqrt{\lambda_2} - \sqrt{\lambda_3} - \sqrt{\lambda_4}\right)$. Fidelities with the maximally entangled state are computed to be $\mathcal{F} > 0.95$ for high-quality states ($p, P =0$).

To verify $\rm CADC-like$ behavior of the first channel, we measure diagonal elements $\rho_{|HV\rangle\langle HV|}$ and $\rho_{|VH\rangle\langle VH|}$ of the two-photon output polarization state. Standard ADC predicts coincidences varying as $p(1-p)$ and peaking at $p=0.5$ whereas CADC predicts zero. Our implementation with $\rm CADC-like$ implementation ($x = 0$) should show suppression but not complete elimination in practice.

\begin{figure}[t]
\centering
\includegraphics[width=\linewidth]{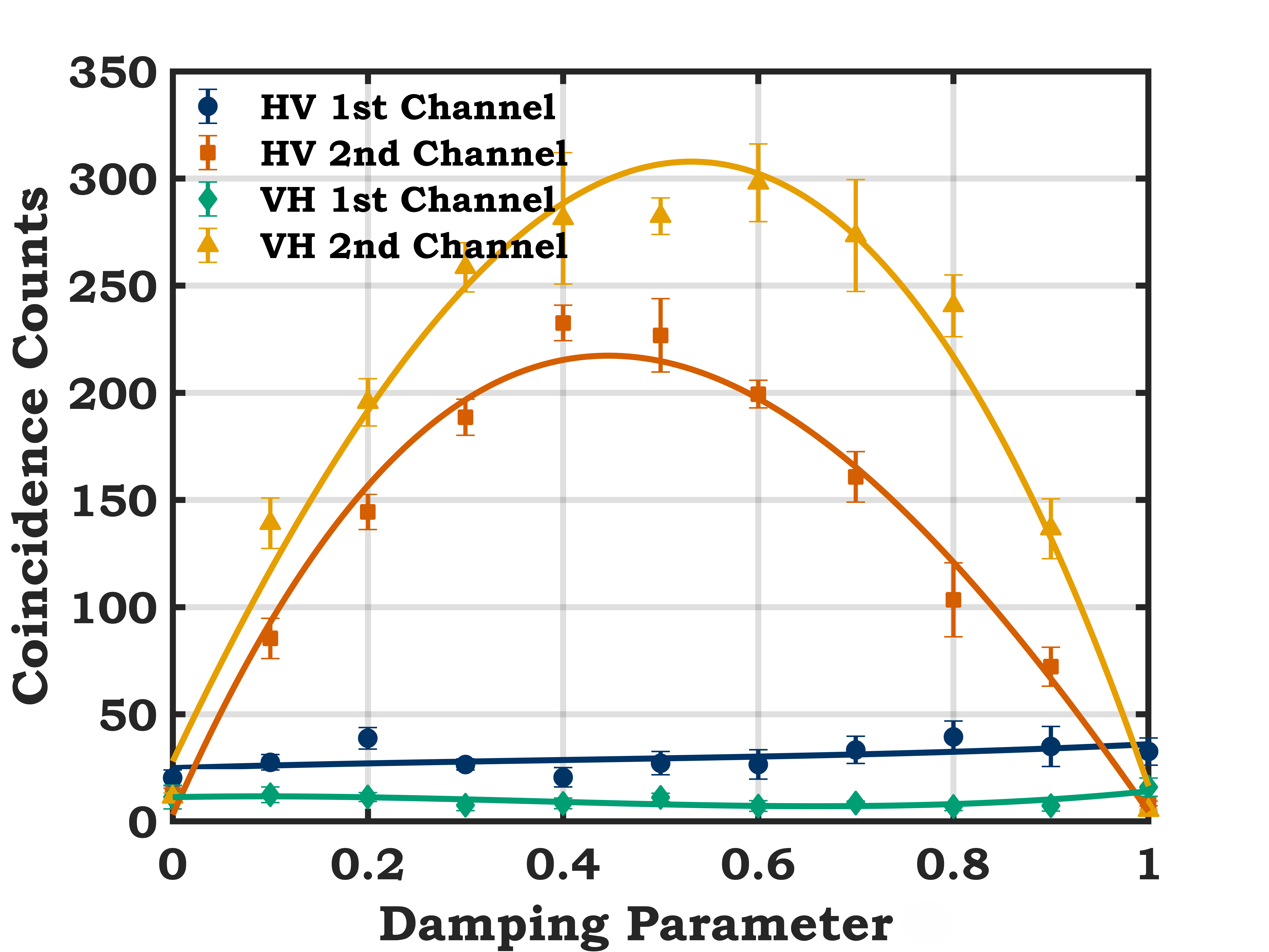}
\caption{Experimental verification of $\rm CADC-like$ behavior. Blue circles and green diamonds with respective error bars: measured coincidence counts in $\ket{HV}$ and $\ket{VH}$ bases versus the first ADC parameter. Orange squares and yellow triangles with respective error bars: measured coincidence counts in $\ket{HV}$ and $\ket{VH}$ bases versus the second ADC parameter. Solid curves fitting with measured data showing at least $85\%$ suppression. Strong deviation from ADC confirms $\rm CADC-like$ characteristics.}
\label{fig: hvvh}
\end{figure}

Figure~\ref{fig: hvvh} shows that measured $HV$ and $VH$ counts for the first channel deviate strongly from the normal ADC prediction, following the $x=0$ characteristics with a minimum of $85\%$ suppression. The observed $15\%$ occurrence of coincidence counts primarily results from accidental coincidences caused by noise in the system. This confirms that our DSI realizes $\rm CADC-like$ physics through path-delay-induced cross-term suppression without requiring engineered global system-bath coupling.

In the following section, for the results pertaining to the ESD manipulation, implementing the without-NOT configuration directly within the displaced-Sagnac interferometer is experimentally challenging. Removal of the NOT operation is impractical as it leads to a complete interferometric realignment for the second ADC realization. A detailed description of the strategy adopted for the implementation of with-NOT and without-NOT configurations is provided in Appendix~\ref{app: WNI_ESD}. Briefly, the post-ADC1/post-NOT state was reproduced at the source by tuning the pump polarization, with the reconstructed state matching the measured with-NOT input Concurrence to within 4$\%$, ensuring the second-channel evolution faithfully represents the without-NOT dissipative trajectory.

\section{Results}
\label{sec:results}

We demonstrate complete ESD manipulation across all three regimes, comparing experimental data to theoretical predictions from Eq.~\eqref{eq:concurrence_timedep} using $x = 0$ implementation and prepared initial state with $\alpha = 0.55$ and $\beta = 0.835$. This state exhibits ESD for both ADC and $\rm CADC-like$ configurations without NOT, making it ideal for testing manipulation.

We first characterize each damping channel independently to validate our model and establish baseline ESD behavior.
\begin{figure}[t]
\centering
\includegraphics[width=\linewidth]{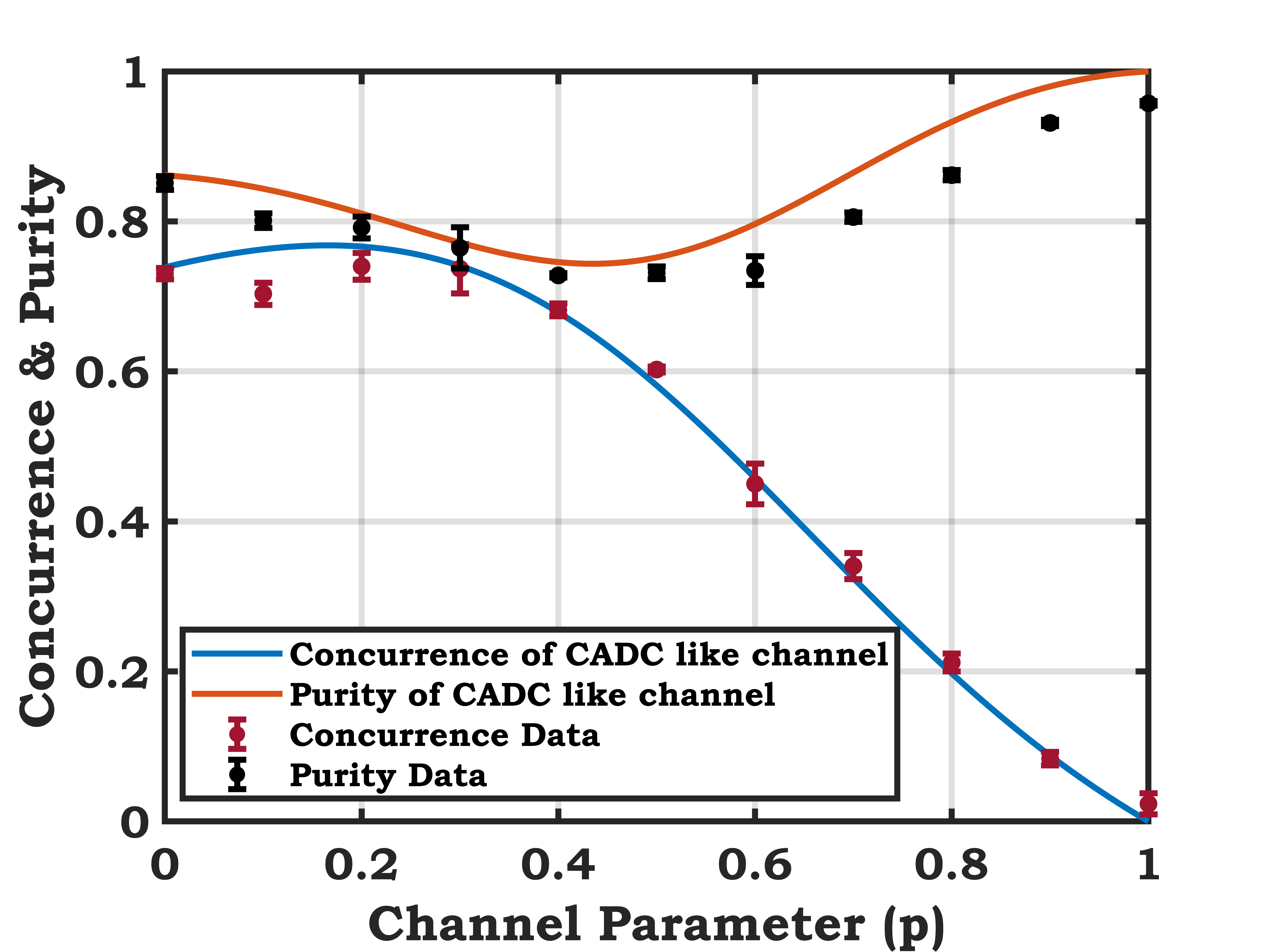}
\caption{First channel characterization. Initial state prepared with $p=0$ (HWP $H_1$ with its fast axis aligned with the vertical direction). Parameter $p$ varied 0 to 1 while keeping second channel disabled ($P = 0$) with NOT active. The red and black circles with error bars represent the measured concurrence and purity data, respectively. Both quantities vary systematically with $p$, consistent with the expected $\rm CADC-like$ behavior as predicted by the blue and red solid lines.}
\label{fig: adc_1}
\end{figure}
\begin{figure}[t]
\centering
\includegraphics[width=\linewidth]{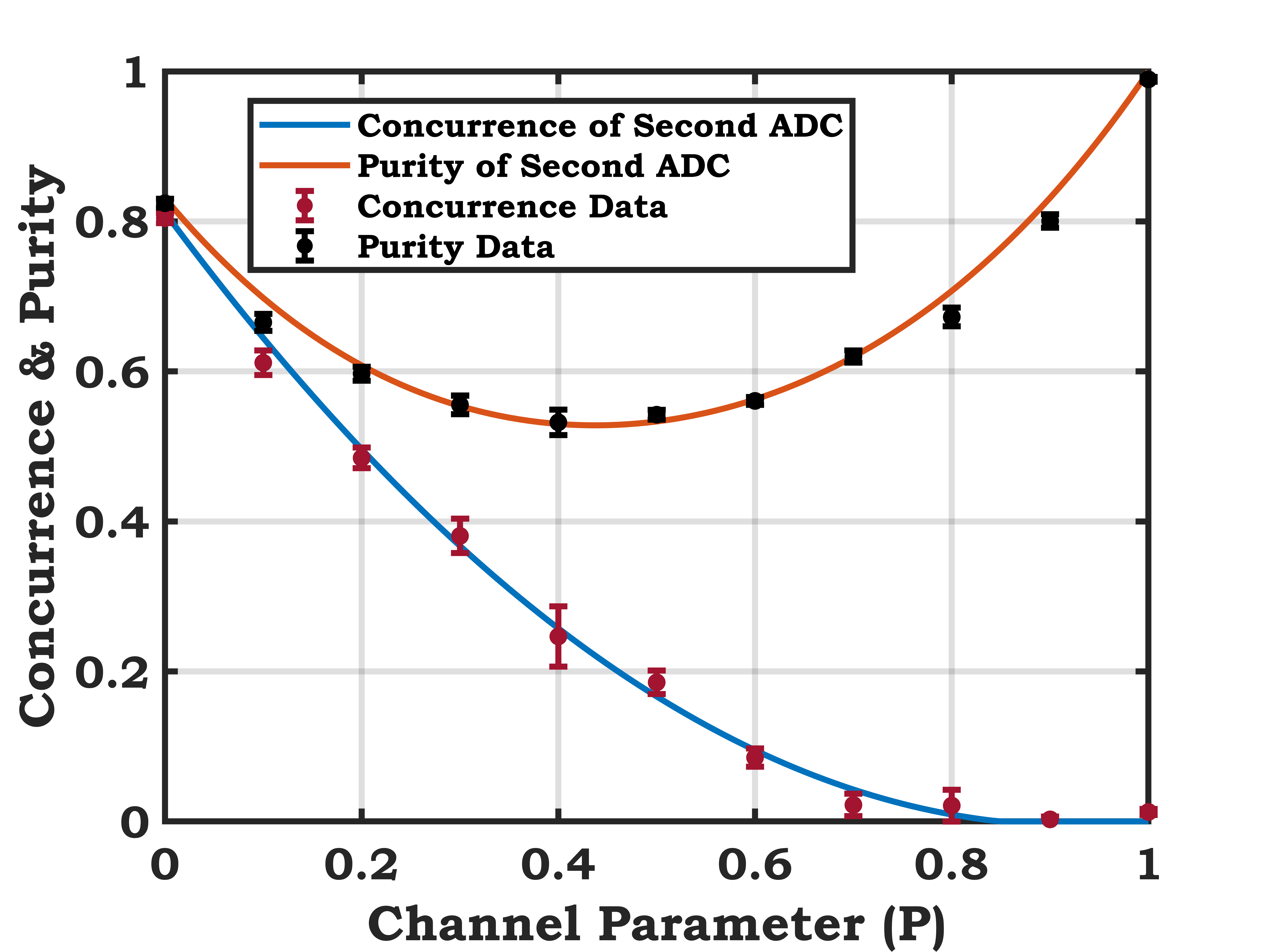}
\caption{Second channel characterization. First channel disabled ($p = 0$) gives initial $C \approx 0.82$, purity $> 80\%$. Parameter $P$ varied from 0 to 1. The experimentally measured concurrence and purity are shown as red and black circles with error bars, respectively, whereas the solid blue and orange curves represent the corresponding theoretical predictions. ESD occurs at $P \approx 0.85$ and the channel behavior agrees with the ADC prediction.}
\label{fig: adc_2}
\end{figure}

In the first run, the second channel is disabled, and the damping parameter $p$ of the first channel is varied while the NOT operation remains active, allowing us to examine the evolution of Concurrence and purity under the first damping channel. In the second run, the first channel is disabled, and the parameter $P$ of the second channel is varied starting from the initially prepared entangled state. Figures~\ref{fig: adc_1} and \ref{fig: adc_2} show channel characterizations confirming that our implementation behaves as modeled.

\begin{figure}[t]
\centering
\includegraphics[width=\linewidth]{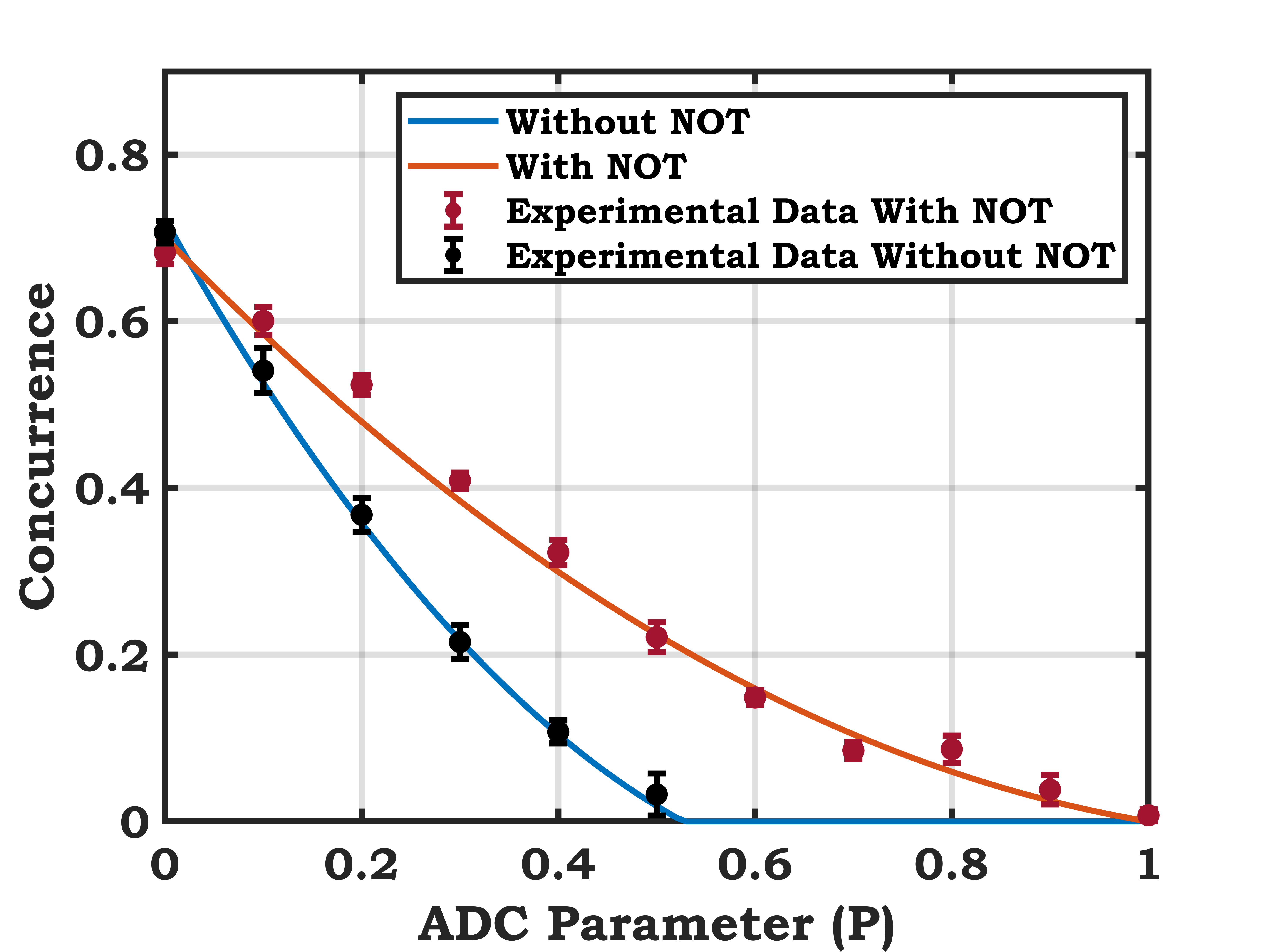}
\caption{ESD avoidance for $p = 0$. Black circles: without NOT, ESD at $P = 0.53$ matches with the theory (blue solid line). With NOT, Concurrence remains nonzero till $P = 1$ (red circles with error bars and the red solid line presenting experimental data and theory, respectively). The avoidance window $\Delta P = 0.47$ demonstrates complete ESD prevention within the accessible range of $P$.}
\label{fig: avoidance}
\end{figure}

Figure~\ref{fig: avoidance} shows the avoidance regime ($p=0$). Without NOT, ESD occurs at $P = 0.53$ agreeing with theory. With NOT, Concurrence remains positive across the entire range to $P = 1$, demonstrating complete ESD avoidance. At $p = 0$, the NOT swaps $0.55\ket{HH} + 0.835\ket{VV} \to 0.835\ket{HH} + 0.55\ket{VV}$, placing the larger coefficient on stable $\ket{HH}$. Subsequent damping depletes the smaller $\ket{VV}$ component, which never dominates, and ESD is prevented.

\begin{figure}[t]
\centering
\includegraphics[width=\linewidth]{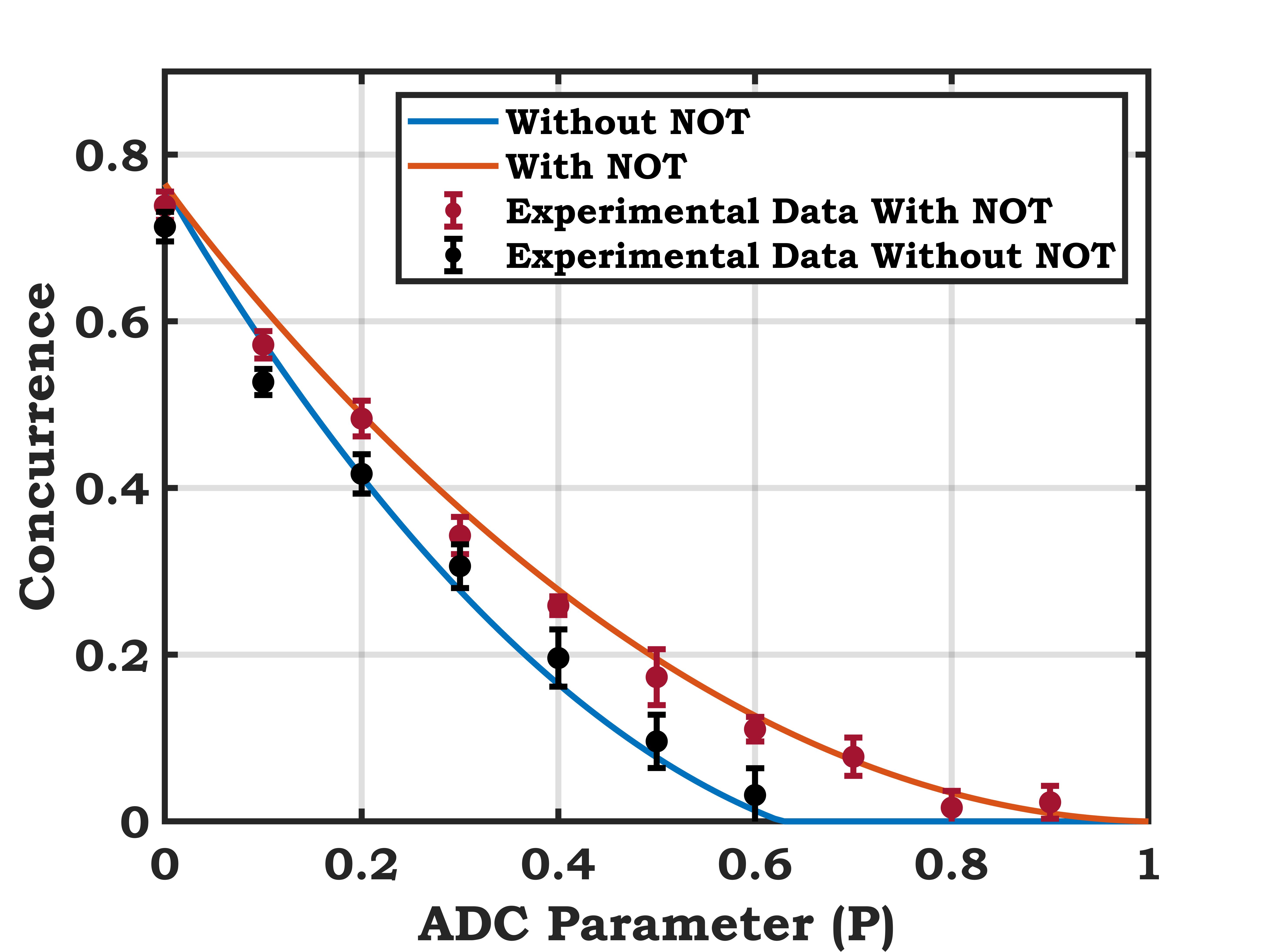}
\caption{Delay of ESD. The initial state $|\Psi\rangle$ was created with $\alpha=0.55$. The first damping channel is kept active with damping strength $p=0.22$. In the absence of the NOT operation, we observe ESD at a damping parameter $P=0.62$ (black circles with error bars represent experimental data, and the blue solid line denotes the theoretical prediction). With NOT applied, we observe ESD delayed to the damping parameter $P=0.93$ (red circles with error bars represent experimental data, and the red solid line corresponds to the theoretical prediction). The observed delay of $\Delta P = 0.31$ represents 50\% extension in the damping parameter range.}
\label{fig: delay}
\end{figure}

Figure~\ref{fig: delay} shows the delay regime ($p=0.22$). Without NOT, ESD at $P = 0.62$; with NOT, postponed to $P = 0.93$. The delay $\Delta P = 0.31$ represents a 50\% extension. This is another practically relevant regime, as the delay mechanism also provides a significant extension of the coherence time.

From an operational perspective, the shift in $P_{\mathrm{ESD}}$ can be translated into the performance of concrete tasks such as teleportation. Using the standard relation between the concurrence and average teleportation fidelity, our delayed-ESD configuration keeps the fidelity above the classical threshold for a longer interval in $P$ (and hence in $t$) than the uncontrolled case (Appendix \ref{app:teleportation}). While we do not optimize protocol parameters for any specific architecture here, teleportation fidelity illustrates how temporal steering of the dissipative trajectory can be directly linked to the useful lifetime of entanglement as a resource.

\begin{figure}[t]
\centering
\includegraphics[width=\linewidth]{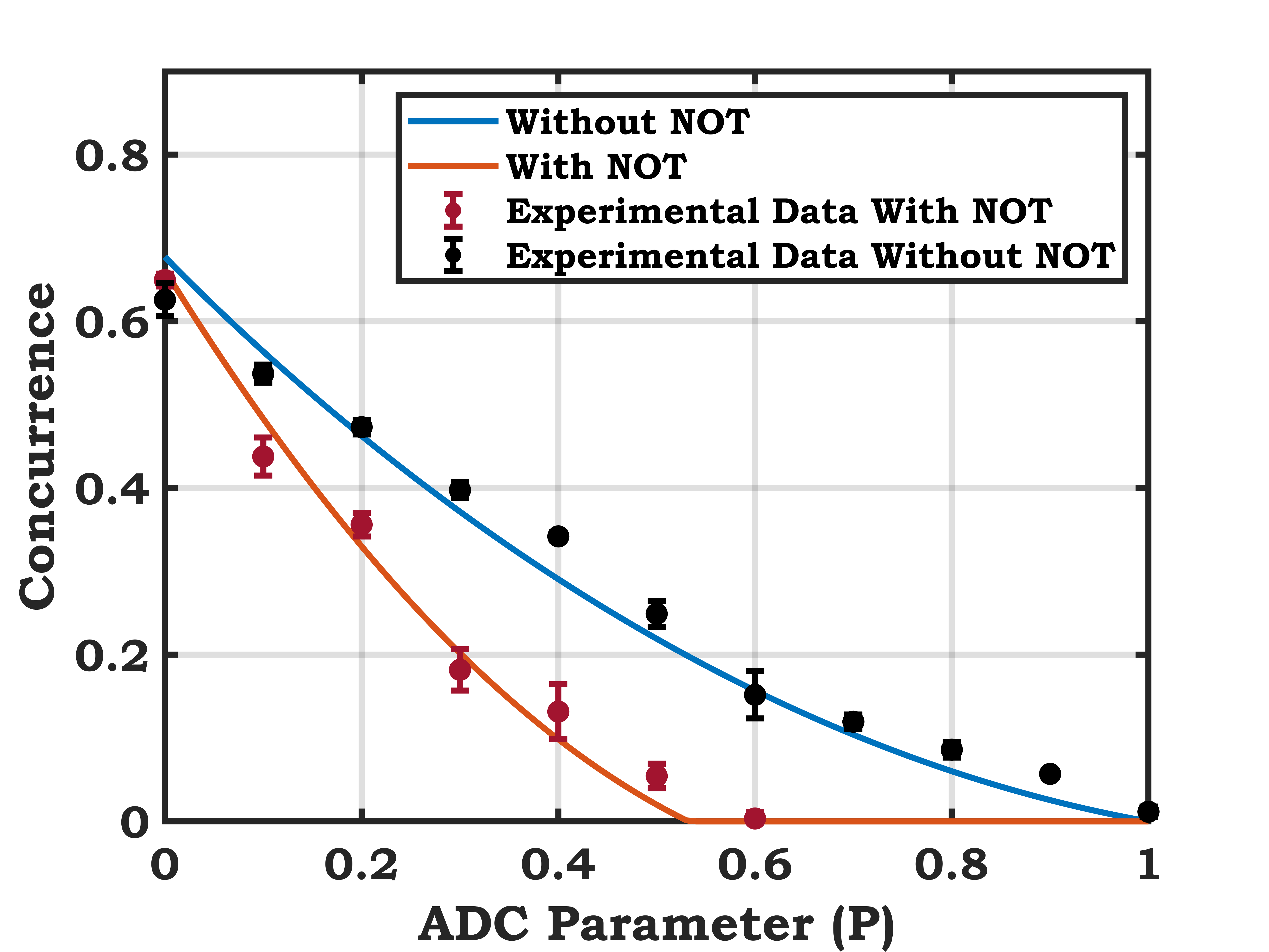}
\caption{Hastening of ESD. The initial state $|\Psi\rangle$ was created with $\alpha=0.55$. The first damping channel parameter is set at $0.43$. Without the NOT operation, we observe an avoidance of ESD (black circles with error bars: represent experimental data, blue solid line: prediction from the theoretical model). With NOT applied, we observe ESD occurring at the damping parameter $P=0.6$ (red circles with error bars: represent experimental data, red solid line: theoretical prediction). Hastening was observed with $\Delta P = 0.4$.}
\label{fig: hastening}
\end{figure}

Figure~\ref{fig: hastening} shows the counterintuitive hastening regime ($p=0.43$). Without NOT, Concurrence decays asymptotically, becoming zero at $P=1$. With NOT, ESD is \emph{induced} at $P \approx 0.6$. After strong first-channel damping, the state is highly mixed (purity $< 0.7$). The NOT introduces additional asymmetry that pushes the system across the separability threshold rather than away from it. This demonstrates that population inversion is not universally beneficial; rather, its effectiveness is strongly dependent on the timing of its implementation. 

\begin{table}[t]
\centering
\caption{Summary of experimental ESD manipulation results for $\alpha=0.55$. First damping channel implemented with $x=0$. 
}
\label{tab:results_summary}
\begin{tabular}{lcccc}
\toprule
Regime & $p$ & \multicolumn{2}{c}{$P_{\text{ESD}}$} & $\Delta P$ \\
\cmidrule(lr){3-4}
 & & w/o NOT & w/ NOT & \\
\midrule
Avoidance & 0 & 0.53 & No ESD ($\to 1$) & $\approx0.47$ \\
Delay & 0.22 & 0.62 & 0.93 & $\approx0.31$ \\
Hastening & 0.43 & No ESD ($\to 1$) & 0.60 & -0.40 \\
\bottomrule
\end{tabular}
\end{table}

Table~\ref{tab:results_summary} summarizes all three regimes. Agreement between experiment and theory is excellent (within 0.02--0.04 for Concurrence), strongly validating our theoretical framework and confirming that our DSI implementation realizes the predicted physics. 

To demonstrate the robustness of our single-shot protocol, we present additional experimental results in the Supplementary Material~\cite{SM} for initial state parameters $\alpha = 0.45$ and $\alpha = 0.5$, complementing the primary experimental run at $\alpha = 0.55$ reported in the main text. These results confirm that our key findings on manipulation of ESD --- are not specific to a single state parameter but persist across a range of input states. Porting the protocol to different initial conditions demonstrates its versatility and validates the theoretical predictions across varying entanglement regimes, further underscoring the robustness of the experimental implementation.

\section{Discussion}
\label{sec:discussion}

The experimental results confirm that appropriately timed single NOT operations can completely control entanglement dynamics by exploiting quantum decay asymmetry. Amplitude damping has directional asymmetry: $\ket{V} \to \ket{H}$ occurs but not reverse. For $\ket{\psi} = \alpha\ket{HH} + \beta\ket{VV}$ with $\alpha < \beta$, the larger $\beta$ on $\ket{VV}$ preferentially decays, populating mixed components $\ket{HV}$ and $\ket{VH}$ that eventually dominate coherence, causing ESD.

The NOT performs population inversion: $\ket{H} \leftrightarrow \ket{V}$. Applied at intermediate time with parameter $p$, it swaps to $\beta'\ket{HH} + \alpha'\ket{VV}+ \text{cross terms}$, where $\beta' < \beta$. If $\alpha' < \beta'$, the population asymmetry is inverted and ESD is avoided. This is not quantum error correction---we do not restore the initial state. We perform trajectory engineering: using decay asymmetry to steer evolution onto paths avoiding the ESD threshold despite ongoing decoherence.

Timing determines the state's condition when the operation acts. Early application (small $p$) catches the state before degradation, allowing effective redirection. Late application (large $p$) finds the state badly damaged; redirection becomes counterproductive, as hastening demonstrates. Comparison to other methods is instructive: for example, DD averages away the interaction Hamiltonian $H_{\text{int}}$ through many operations requiring weak coupling and precise timing. Our approach applies one operation that redirects the trajectory, working for strong coupling with coarse timing tolerance but sacrificing generality.

While demonstrated in photonics, the physics applies to any platform with amplitude-damping-dominated decoherence, fast local gates, and the ability to delay gate application. Superconducting qubits have 20--50 ns gates and 50--150 $\mu$s $T_1$ with dominant energy relaxation. Implementation would engineer one damping stage via tunable resonator couplings and a $\sigma_x$ operation on each qubit. Trapped ions offer 1--10 $\mu$s gates and $> 1$ ms motional lifetimes, with cascaded damping via sequential motional mode coupling. NV centers have 100 ns manipulation and 1--10 ms $T_1$, though engineering independent damping channels is challenging. The tunable parameter $x$ is more platform-specific. In photonics, $x$ arises naturally from path geometry. For other platforms, achieving analogous tuning would require engineering temporal correlations in system-bath coupling through resonator quality factors, multimode cavities, or modulated laser coupling.

Our measured Concurrences fall below theory by approximately 1--14\%. We identify three dominant error sources: PBS extinction ratio ($\sim 10^{-2}-10^{-3}$), HWP angle precision ($\delta\theta$), and path mismatch contributes to the $\Delta C$. A detailed first-order quantitative analysis of systematic imperfections affecting the concurrence measurement is presented in Appendix~\ref{app:errors}. We find the PBS extinction to dominate the error budget.

\section{Conclusion and Outlook}
\label{sec:conclusion}

We have demonstrated complete, deterministic control of entanglement sudden death using a single local NOT operation whose timing determines the outcome. Three regimes---avoidance, delay, and hastening---are achieved by varying the damping parameter \emph{when} the NOT operation is applied, without modifying the noise itself. This establishes temporal control — knowing when to act — as a practical and underexplored resource for shaping open-system dynamics, complementing operational control over what gates to apply or how many.

The single-shot paradigm is relevant to platforms where gate sequences accumulate errors. The tunable parameter $x$ suggests routes to engineer custom decoherence channels by controlling temporal structure, extending beyond photonics to circuit QED and trapped ions. Our approach performs trajectory steering rather than averaging, operating non-perturbatively, unlike DD or QZE.

Open questions include extending the time-delay framework to phase-damping or depolarizing channels, dynamically tuning $x$ for error compensation, and understanding multi-qubit population inversion for graph states. This work opens a research direction: temporal engineering of decoherence. As quantum devices scale, knowing \emph{when} to act may be as important as knowing \emph{what} to do.

\begin{acknowledgments}
US thanks the DST for support in the form of a core research grant for this work as well as partial support from a Canada Excellence Research Chair professorship. US, KS, ASR, and SRB thank the National Quantum Mission of the DST for partial support. SR acknowledges RRI-VSP and DST INSPIRE-Scholarship for Higher Education.
\end{acknowledgments}

\section{Data Availability}
There are no publicly available research data or software supporting this manuscript. Requests for further information or data should be sent to the corresponding author.

\appendix

\section{Kraus Operator Derivation}
\label{app:kraus}

The polarizing beam-splitter $\mathcal{P}_1$ is given by
\begin{equation}
\mathcal{P}_1=|H_2\rangle\langle H_0|+|V_3\rangle\langle V_0|+ \dots,
\end{equation}
where we neglect contributions orthogonal to the input path $|00\rangle$. The half-wave plate on path $|3\rangle$ is
\begin{equation}
\mathcal{H}_1(\theta) = H(\theta)\otimes|3\rangle\langle 3|+ \mathbb{I}\otimes |2\rangle\langle 2|,
\end{equation}
where the HWP matrix is
\begin{align}
\begin{split}
H(\theta)&=-\cos2\theta |H\rangle\langle H|+\cos2\theta |V\rangle\langle V|\\&+\sin2\theta\left(|H\rangle\langle V|+|V\rangle\langle H|\right), \\
\mathbb{I}&=|H\rangle\langle H|+|V\rangle\langle V|.
\end{split}
\end{align}

The NOT operation is
\begin{equation}\label{local_not}
U_{NOT}= \sigma_x\otimes\left(|3\rangle\langle 3|+|2\rangle\langle 2|\right), \ \sigma_x=|H\rangle\langle V|+|V\rangle\langle H|.
\end{equation}

The second PBS with time-delay operator $X$ is
\begin{equation}
\mathcal{P}_2(X)= |H_4\rangle\langle H_3|+ X |V_5\rangle\langle V_3|+ |H_2\rangle\langle H_2|+|V_2\rangle\langle V_2|+\dots,
\end{equation}
and the second HWP is
\begin{equation}
\mathcal{H}_2(\phi)= H(\phi)\otimes\left(|5\rangle\langle 5|+|2\rangle\langle 2|\right)+\mathbb{I}\otimes|4\rangle\langle 4|.
\end{equation}

Finally, the recombination PBS is
\begin{align}
\begin{split}
\mathcal{P}_1^\dagger = &|H_a\rangle\langle H_2|+|V_b\rangle\langle V_2|+|H_b\rangle\langle H_4|\\& \hspace{0.5cm}+|V_{a'}\rangle\langle V_5|+ |H_{b'}\rangle\langle H_5|+|V_a\rangle\langle V_4|.
\end{split}
\end{align}

The total unitary operator can be written as $U^A=|F\rangle\langle H_0|+|G\rangle\langle V_0|+\dots$, and for two photons:
\begin{align}\label{unitary_00}
\begin{split}
U_T &=U^A\otimes U^B = |GG\rangle\langle V_0V_0|+|FF\rangle\langle H_0H_0|\\&+|FG\rangle\langle H_0V_0|+|GF\rangle\langle V_0H_0|+\dots.
\end{split}
\end{align}
Applying to input path $|00\rangle$ gives
\begin{align}\label{ut}
\begin{split}
\widetilde{U} = U|00\rangle &=  |FF\rangle\langle HH|+ |GG\rangle\langle VV|\\&+|GF\rangle\langle VH|+|FG\rangle\langle HV|,
\end{split}
\end{align}
which can be expanded in terms of Kraus operators:
\begin{equation}\label{ukraus}
\widetilde{U}=\sum_{ij}\mathbb{K}_{ij}|ij\rangle, \ \ |ij\rangle\in |\text{output paths}\rangle.
\end{equation}
The explicit forms of the Kraus operators for with and without NOT cases for cascaded damping channels are given in Supplementary Material~\cite{SM}. In the next two subsections, we will consider cascading damping channels in detail with and without the NOT operation and calculate the expressions of the concurrence in each cases.

\section{Details of Phase Diagram: ESD line}\label{app: ESDLine}

\subsection{Time Independent Formalism}
In this case, the Kraus operators are that corresponding to traditional ADC and CADC. We discuss the cascades of these damping channels with and without NOT operator below. 

\subsection*{Cascading ADC}\label{cascade_adc}
We have two cascaded amplitude-damping channels connected either with or without a NOT operator. The single photon Kraus operators are $2\times2$ matrices given by
\begin{equation}
    K_{1} = \begin{pmatrix}
        1 & 0 \\
        0 & \sqrt{1-p}
    \end{pmatrix}
    ~~\text{and}~~
    K_{2} = \begin{pmatrix}
        0 & \sqrt{p} \\
        0 & 0
    \end{pmatrix}\,.
\end{equation}
The Kraus operators for the two-photon state are given by
\begin{equation}
\mathbb{K} = \left(K_1\otimes K_1\,, K_1\otimes K_2\,, K_2\otimes K_1\,, K_2\otimes K_2\right)\,.
\end{equation}
For the cascaded ADCs on the initial state in \eqref{rho_evolution}, the final output state is obtained by applying \eqref{rho_evolution} in succession first without the NOT operation so that,
\begin{equation}
\rho_{out}=\sum_j \mathbb{K}_j(P) \left(\sum_i\mathbb{K}_i(p)\rho_{in}\mathbb{K}_i(p)^\dagger\right)\mathbb{K}_j(P)^\dagger\,,
\end{equation}
where $p$ and $P$ are the parameters characterizing the first and second ADC. $\rho_{out}$ is the form of an $X-$state given by
\begin{equation}
   \rho_{out} = \begin{pmatrix}
  \rho_{11}   & 0 & 0 &  \rho_{14}\\
    0 & \rho_{22} & 0 & 0 \\
    0 & 0 & \rho_{33} & 0 \\
    \rho_{41} & 0 & 0 & \rho_{44}
    \end{pmatrix}\,,
\end{equation}
where 
\begin{eqnarray}
    \rho_{11} &=& \alpha^{2}+(p+P-pP)^{2}\beta^{2}\nonumber\,,\\
    \rho_{14} &=& \rho_{41} = (1-p)(1-P)\alpha\beta \nonumber\,,\\
    \rho_{22} &=& \rho_{33} = (1-p)(1-P)\big(p(1-P)+P\big)\beta^{2}\nonumber\,,\\
    \rho_{44} &=& (1-p)^{2}(1-P)^{2}\beta^{2}\,.
\end{eqnarray}
The concurrence of the output state is \cite{Rau2008},
\begin{equation}\label{adaadc}
    C(p\,,P) = 2\ \text{max}\Big[0,\{|\alpha\beta |-|\big(p(1-P)+P\big)\beta^{2 }|\} (1-p)(1-P)\Big] \,.
\end{equation}
Including the NOT operation in between the two damping channels gives rise to a different scenario. In this case, starting from the total unitary operator for the cascade, given by
\begin{equation}
U_{TOT}=U_{ADC}(P)U_{NOT}U_{ADC}(p)\,,
\end{equation}
we get a total of 9 Kraus operators. The output state is
 \begin{equation}
     \rho_{out} = \sum_{i=1}^{9}\mathbb{K}_{i}\rho_{in}\mathbb{K}_{i}^{\dagger} = \begin{pmatrix}
    \rho_{11} & 0 & 0 &  \rho_{14} \\
    0 & \rho_{22} & \rho_{23} & 0 \\
    0 & \rho_{32} & \rho_{33} & 0 \\
    \rho_{41} & 0 & 0 & \rho_{44}
    \end{pmatrix}\,,
 \end{equation}
 where,
 \begin{eqnarray}
    \rho_{11} &=& P^{2}+(1-p)(1-P)\Big\{ 1+P-p(1-P)\Big\} \beta^{2}\nonumber\,,\\
    \rho_{14} &=& \rho_{41} = \Big\{ 1-p(1-P)\Big\}(1-P)\alpha\beta \nonumber\,,\\
    \rho_{22} &=& \rho_{33} = (1-P)\Big[ P-(1-p)\big\{ P-p(1-P)\big\}\beta^{2}   \Big]\nonumber\,,\\
    \rho_{23} &= & \rho_{32} = 2\sqrt{p(1-p)P}(1-P)\alpha\beta\nonumber\,,\\
    \rho_{44} &= & (1-P)^{2}(\alpha^{2}+p^2\beta^{2})\,,
 \end{eqnarray}
with Concurrence
\begin{equation}
C(p\,,P) = 2\ \text{max}\left(0, |\rho_{23}| - \sqrt{\rho_{11}\rho_{44}} , |\rho_{14}| - \sqrt{\rho_{22}\rho_{33}} \right)\,.
\end{equation}
Since the quantity $|\rho_{23}| - \sqrt{\rho_{11}\rho_{44}} $ is negative for the entire range of state and channel parameters, it does not contribute to the concurrence. So,
\begin{eqnarray}\label{adcnotadc}
C(p\,,P) &=& 2\ \text{max}\left(0, |\rho_{14}| - \sqrt{\rho_{22}\rho_{33}} \right)\,.\nonumber\\
&=& 2\max\!\Big[0,\left(1-p(1-P)\right)(1-P)|\alpha\beta|\nonumber\\
&-&(1-P)\left(P-(1-p)\{P-p(1-P)\}\beta^2\right)\Big].\nonumber\\
\end{eqnarray}

\subsection*{Cascading CADC with ADC}\label{cascade_cadc}
Here we consider a CADC channel with channel parameter $p$ and with or without local NOT operation on each qubit and finally an ADC channel with channel parameter $q$. First we consider the CADC + ADC channel scenario.  The CADC channel acts globally on both photonic polarization states. The Kraus operators are
\begin{equation}
    K_{1} = \begin{pmatrix}
        1 & 0 & 0 & 0 \\
        0 & 1 & 0 & 0 \\
        0 & 0 & 1 & 0 \\
        0 & 0 & 0 & \sqrt{1-p}
    \end{pmatrix}
    ~~\text{and}~~
    K_{2} = \begin{pmatrix}
        0 & 0 & 0 & \sqrt{p} \\
        0 & 0 & 0 & 0 \\
        0 & 0 & 0 & 0 \\
        0 & 0 & 0 & 0
    \end{pmatrix}\,.
\end{equation}
After evolution under the CADC channel, the state evolves under an ADC channel with channel parameter $q$ and the final output state becomes:
\begin{equation}
   \rho_{out} = \begin{pmatrix}
    \rho_{11} & 0 & 0 &  \rho_{14} \\
    0 & \rho_{22} & 0 & 0 \\
    0 & 0 & \rho_{33} & 0 \\
    \rho_{41} & 0 & 0 & \rho_{44}
    \end{pmatrix}\,,
\end{equation}
where,
 \begin{eqnarray}
    \rho_{11} &=& \alpha^{2}+\{p+(1-p)P^{2}\}\beta^{2}\nonumber\,,\\
    \rho_{14} &=& \sqrt{(1-p)}(1-P)\alpha\beta \nonumber\,,\\
    \rho_{22} &=& \rho_{33} = (1-p)(1-P)P\beta^{2}\nonumber\,,\\
    \rho_{44} &= & (1-p)(1-P)^{2}\beta^{2}\,,
 \end{eqnarray}
The concurrence of the output state is
\begin{equation}\label{cadcadc}
    C(p\,,P) = 2\ \text{max}\Big[0,\{|\alpha\beta |-\sqrt{1-p}\ P\beta^{2 }|\} \sqrt{(1-p)}(1-P)\Big] \,.
\end{equation}

Finally, for a CADC followed by a NOT and then an ADC, we get a total of 7 nonzero Kraus operators. The output state is
 \begin{equation}
     \rho_{out} = \sum_{i=1}^{7}\mathbb{K}_{i}\rho_{in}\mathbb{K}_{i}^{\dagger} = \begin{pmatrix}
    \rho_{11} & 0 & 0 &  \rho_{14} \\
    0 & \rho_{22} & 0 & 0 \\
    0 & 0 & \rho_{33} & 0 \\
    \rho_{41} & 0 & 0 & \rho_{44}
    \end{pmatrix}\,,
 \end{equation}
 where
\begin{align}
\begin{split}
    \rho_{11} &=\alpha^{2}P^{2}+\big\{1-p(1-P^{2}\big\}\beta^{2}\,,\\
    \rho_{14} &= \rho_{41} = \sqrt{(1-p)}~(1-P)\alpha\beta\,,\\
    \rho_{22} &= \rho_{33} = (1-P)P\big(\alpha^{2} +p\beta^{2}\big)\,, \\
    \rho_{44} &= (1-P)^{2}\big(\alpha^{2} +p\beta^{2}\big)\,,
\end{split}
\end{align}
with Concurrence
\begin{align}\label{cadcnotadc}
\begin{split}
C(p\,,P) &= 2\ \text{max}\big[0, |\rho_{14}| - \sqrt{\rho_{22}\rho_{33}} \big]\\
&=2\ \text{max}\big[0, \big\{ \sqrt{1-p}\ |\alpha\beta|-P(\alpha^{2}+p\beta^{2})\big\} (1-P)\big]\,.
\end{split}
\end{align}

\subsection{Time dependent formalism}
In this case, the tunable parameter $x$ controls whether the channel is ADC or CADC like. Hence in this case, we keep the $x$ general and group the observation under the categories of whether a local NOT operation is considered or not in the cascade of our custom damping channels. For both cases, the Kraus operators are derived in Appendix \ref{app:kraus}.

\subsection*{Setup With NOT}
The output density matrix is given by,
\begin{equation}
\rho_{out}=\sum_{ij}\mathbb{K}_{ij} \rho_{in}\mathbb{K}_{ij}^\dagger=\frac{1}{N}
\begin{pmatrix}
 A & 0 & 0 & \mathcal{X}\\ 0 & B & 0 & 0\\0 & 0 & C & 0\\ \mathcal{X}^\star & 0 & 0 & D\\
\end{pmatrix}\,,
\end{equation}
where
\begin{align}
    \begin{split}
        A &= (1-p)^2\beta^2+P\big[P-(1-p)\big\{ P-p(2x^2-P)\big\}\beta^2\big]\,,\\
        B &=C=(1-P)\big[ P-(1-p)\big\{ P-p(x^2 -P)\big\} |\beta|^2\big]\,,\\
        D &= (1-P)^2 \left(|\alpha|^2+|\beta|^2 p^2\right)\,,\\
        \mathcal{X} &=\alpha^\star  \beta  (1-p)(1-P)\,,\\
        N &= 1-2p(1-p)(1-x^2)|\beta|^2\,.
    \end{split}
\end{align}
The parameters $p =\sin^22\theta$ associated with HWPs $\mathcal{H}_1$ and $P=\sin^22\phi$ with $\mathcal{H}_2\,, \mathcal{H}_3$ for the first and second damping channel, respectively, are real and satisfy $0\leq p\,, q\leq 1$. The concurrence of the output state is

\begin{eqnarray}\label{expnot}
    C(p\,,P) &=& \frac{2}{N}\ \text{max}\Big[0,\big[ (1-p)|\alpha\beta |-\big\{ P-(1-p)\nonumber\\
    &\times &\big(P-p(x^2-P)\big)|\beta|^{2 }\big\}\big](1-P)\Big]\,. 
\end{eqnarray}

The corresponding ESD line is given by
\begin{equation}\label{esd_line_n}
 (1-p)|\alpha\beta |-\Big\{ P-(1-p)\big(P-p(x^2-P)\big)|\beta|^{2 }\Big\}=0\,.
\end{equation}

\subsection*{Setup Without NOT}
In this case, the unitary operator $U^A$ acting on each photonic state is given by
\begin{equation}\label{ua_not}
U^A = \mathcal{P}_1^\dagger\ \mathcal{H}_2(\phi)\ \mathcal{P}_2\ \mathcal{H}_1(\theta)\ \mathcal{P}_1\,.
\end{equation}
The output state becomes
\begin{equation}
\rho_{out}=\sum_{ij}\mathbb{K}_{ij} \rho_{in}\mathbb{K}_{ij}^\dagger=\frac{1}{N}
\begin{pmatrix}
 \rho_{11} & 0 & 0 & \rho_{14}\\ 0 & \rho_{22} & 0 & 0\\0 & 0 & \rho_{33} & 0\\ \rho_{14}^\star & 0 & 0 & \rho_{44}\\
\end{pmatrix}\,,
\end{equation}
where
\begin{align}
    \begin{split}
        \rho_{11} &= P^2+p\big[2P(x^2-P)+p(1-2Px^2+P^2)\big]|\beta|^2\,,\\
        \rho_{22} &=\rho_{33}=(1-P)\big[ P+p\big\{x^2(1-p)-P(2-p)\big\} |\beta|^2\big]\,,\\
        \rho_{44} &= (1-P)^2 \left(1-p(2-p)|\beta|^2\right)\,,\\
        \rho_{14} &=p(1-P)\alpha^\star  \beta \,,\\
        N &= 1-2p(1-p)(1-x^2)|\beta|^{2}\,.
    \end{split}
\end{align}
The concurrence of the output state is
\begin{eqnarray}\label{expwithoutnot}
    C(p\,,P) &=& \frac{2}{N}\ \text{max}\Big[0,\big[p|\alpha\beta |-\big\{ P+p\big(x^2(1-p)\nonumber\\
    &-& P(2-p)\big)|\beta|^{2 }\big\}\big](1-P)\Big] \,.
\end{eqnarray}
Consequently, the ESD line will be
\begin{equation}\label{esd_line_wn}
p|\alpha\beta |-\big\{ P+p\big(x^2(1-p)-P(2-p)\big)|\beta|^{2 }\big\}=0\,.
\end{equation}

\section{The interpretation of $X$}
\label{app:x_explain}
In order to understand the action of $X$ on $a(t)$ note that,
\begin{equation}
\mathcal{P}_2(X)\mathcal{P}_2(X)^\dagger=\mathbb{I}\,,
\end{equation}
implies that $X X^\dagger=1$. This forces the form of the operator $X=\exp(-i\chi)$. The operator $X$ acting on a function $f(t)$ induces a time delay of $\delta t$. Mathematically,
\begin{equation}\label{ftransform}
Xf(t)X^\dagger=f(t+\delta t)\,,
\end{equation}
which enforces the condition,
\begin{equation}
[\chi,f]=i\ \delta t\ \partial_t f\,.
\end{equation}
Now we analyze the action of $\mathcal{P}_2(X)$ and/or $\mathcal{P}_2(\bar{X})$ on a product $\sum_t a(t)\bar{a}(t)$. To preserve generality, we will take distinct $\chi_1\equiv i \delta t_1\partial_t$ and $\chi_2\equiv i\delta t_2\partial_t$. The entire action of $\mathcal{P}_2(X)\otimes \mathcal{P}_2(\bar{X})$ boils down to,
\begin{eqnarray}\label{aatransform}
X\bar{X}\sum_t a(t)\bar{a}(t) &=& \sum_t \left(X a(t) X^\dagger\right)\otimes \left(\bar{X}\bar{a}(t)\bar{X}^\dagger\right)\,, \nonumber\\  X\bar{X}|\text{pol}_1,\text{pol}_2\rangle &=&|\text{pol}_1,\text{pol}_2\rangle\,.
\end{eqnarray}
Next, we take the mode decomposition of $a(t)$,
\begin{equation}
a(t)=\frac{1}{2\pi}\int d\omega\ a(\omega) e^{-i \omega t}\,,
\end{equation}

and similarly for $\bar{a}(t)$. Using \eqref{ftransform}, we can write,
\begin{eqnarray}
X a(t)X^\dagger &=&\frac{1}{2\pi}\int d\omega\ a(\omega) X e^{-i\omega t}X^\dagger \nonumber\\ &=& \frac{1}{2\pi}\int d\omega\ a(\omega) e^{-i\omega(t+\delta t_1)}\,,
\end{eqnarray}
and similarly for $\bar{a}(t)$. Thus \eqref{aatransform} becomes,
\begin{eqnarray}
&& X\bar{X}\sum_t a(t)\bar{a}(t) \nonumber\\
&&=\sum_t \int d\omega_1 d\omega_2\ a(\omega_1)\bar{a}(\omega_2)e^{-i(\omega_1+\omega_2)t}e^{-i\omega_1\delta t_1-i\omega_2\delta t_2}\,.\nonumber
\end{eqnarray}
Approximating \cite{footnote},
\begin{equation}
\sum_t e^{-i(\omega_1+\omega_2)t}\approx 2\pi \delta(\omega_1+\omega_2)\,,
\end{equation}
we can write, 
\begin{equation}
X\bar{X}\sum_t a(t)\bar{a}(t)=\int d\omega\ a(\omega)\bar{a}(-\omega)e^{-i\omega\tau}=I(\tau)\,, 
\end{equation}
where $\tau=\delta t_1-\delta t_2$ is the relative optical time delay. We assume that $a(t)$ and $\bar{a}(t)$ are Gaussian real amplitudes, so that $f(\omega)=a(\omega)\bar{a}(-\omega)$ is real, positive and integrable. The phase factor $\exp(-i\omega\tau)$ oscillates with a period of $2\pi/\tau$ in $\omega$ and the oscillations increase with increasing $\tau$. Hence the positive and negative contributions progressively cancel. To see this more clearly, we integrate by parts, the integrand with respect to $\omega$,
\begin{equation}
    I(\tau)=-\frac{i}{\tau}\int d\omega\ f'(\omega)e^{-i\omega\tau}\,,
\end{equation}
where the total integral vanishes outside the domain of $f(\omega)$. Taking the modulus and applying the integral bound $|\int d\omega g(\omega)|\leq \int d\omega|g(\omega)|$\,,
\begin{equation}
    |I(\tau)|\leq \frac{1}{\tau}\int d\omega\ |f'(\omega)| = \frac{C}{\tau}\,,
\end{equation}
where $C$ is a finite number, independent of $\tau$. Thus $|I(\tau)|\rightarrow 0$ as $\tau\rightarrow\infty$. As an example, for Gaussian profile centered around $\omega=\omega_0$ with width $\sigma$, $f(\omega)=\exp(-(\omega-\omega_0)^2/\sigma^2)$ and
\begin{equation}
    I(\tau)\sim \exp(-\tau^2\sigma^2/4)\exp(-i\omega_0\tau)\,.
\end{equation}
We can thus define, 
\begin{equation}
    x(\tau) = \abs{\frac{I(\tau)}{I(0)}}\,,
\end{equation}
so that $x(\tau)$ is real, positive and bounded between $[0,1]$ where $x(0)=1$ and $x(\tau\rightarrow\infty)\rightarrow0$. We can write the general form of the action as,
\begin{equation}
X\bar{X}\left(\sum_t a(t)\bar{a}(t)\right)= x(\tau) \sum_t a(t)\bar{a}(t)\,, \ x(\tau)\in[0,1]\,,
\end{equation}
Note however, that in a practical setup, the coincidence window ($\Delta t$) behaves as a hard-cutoff such that $x(\tau>\Delta t)=0$. The ability to distinguish $\tau\leq \Delta t$ or $\tau > \Delta t$ depends on the spectral profile of the photon wave packets. $x(\tau)$ quantifies the fraction of the joint spectral distribution of the two photons, separated by $\tau$, that is captured within the coincidence window $\Delta t$. 

Gaussian profiles for photons give soft cutoffs {\it i.e.,} there is no clear boundary where $x(\tau)=0$. In order to get the hard cutoff $\tau_c$, such that $x(\tau>\tau_c)=0$, we start with $a(t)=\bar{a}(t)=\Theta(T-|t|)$ which gives,
\begin{equation}
    f(\omega)=4T^2{\rm sinc}^2(\omega T)\,, \ \ \text{where}\ \ {\rm sinc}(x)=\frac{\sin x}{x}\,.
\end{equation}
Correspondingly,
\begin{equation}
    I(\tau)=2\pi\ {\rm max}(0,2T-\tau)\,, \ \ \text{with}\ \ I(0)=4\pi T\,.
\end{equation}
Consequently,
\begin{equation}
    x(\tau)=\begin{cases}
        1-\tau/(2T)\,, \tau\leq 2T\\ 0\,, \tau>2T
    \end{cases}\,.
\end{equation}
$2T=\Delta t$ is the natural cutoff {\it i.e.,} the coincidence window connected with the detector. The particular functional form of $x(\tau)$ used here corresponds to the chosen input wave-packet profile and is used as a representative model satisfying $x(0)=1$, $x(1)=0$, and monotonic decay with increasing delay. The broader framework does not depend on this specific profile: different spectral envelopes would modify the detailed form of $x(\tau)$ while preserving its role as an effective temporal-overlap parameter.

\section{SU(2) Generalization Analysis}
\label{app:su2}

To determine whether the NOT operation is uniquely optimal for ESD manipulation, we analyze the most general local SU(2) operation. The general single-qubit SU(2) operator can be parameterized as
\begin{equation}
U_{SU(2)} = n_0 I + i n_x \sigma_x + i n_y \sigma_y + i n_z \sigma_z,
\end{equation}
with normalization $|n_0|^2 + |n_x|^2 + |n_y|^2 + |n_z|^2 = 1$ ensuring unitarity.

We reparameterize using $n_\pm = n_x \pm i n_y$, $\bar{n}_0 = n_0 + i n_z$, and $\bar{n}_3 = n_0 - i n_z$. Applying this general operation between the two damping channels and computing the output concurrence yields
\begin{equation}
C = 2 \max(0, |D| - |B|),
\end{equation}
where
\begin{align}
D &= \bar{n}_3(1-P)^2[\alpha^2 + p^2\beta^2] + n_+n_-(1-P)^2p(1-p)\beta^2,
\end{align}
and
\begin{widetext}
\begin{align}
    \begin{split}
   B = &\frac{1}{4} \bigg[\beta^{2}P (1-P)  \big(\bar{n}_0
   \left(\bar{n}_3+2i n_+ \sqrt{p(1-p)}\right)+n_- \left(n_+-2i \bar{n}_3
   \sqrt{p(1-p)}\right)+\left(\bar{n}_0 \bar{n}_3-n_- n_+\right)  (1-2p)\big){}^2\\ &-4 \alpha ^2 \bigg(-n_+ \left(n_- \sqrt{P(1-P)}+i (2
   P-1) \bar{n}_0\right) +\sqrt{P(1-P)} \left(\bar{n}_0^2-\bar{n}_3
   \bar{n}_0+1\right)+n_+^2 \sqrt{P(1-P)}\bigg)\\&\times \Big(-i n_- (2 P-1)
   \bar{n}_3+\sqrt{P(1-P)} \left(-\bar{n}_3^2+\bar{n}_0 \bar{n}_3-1\right)-n_-^2
   \sqrt{P(1-P)}+n_+ n_- \sqrt{P(1-P)}\Big)\\ & +\beta^{2} (1-P) x \Big(i \left(n_+ \bar{n}_0-n_- \bar{n}_3\right) (1-2p)-i n_+ \bar{n}_0-i n_- \bar{n}_3-2\sqrt{p(1-p)}+4 n_- n_+
   \sqrt{p(1-p)}\Big)\\&\times \Big(i \left(n_+ \bar{n}_0-n_-
   \bar{n}_3\right)(1-2p)+i n_+ \bar{n}_0+i n_- \bar{n}_3-2\sqrt{p(1-p)} +4 n_- n_+ \sqrt{p(1-p)}\Big)\bigg]\,.
    \end{split}
\end{align}
\end{widetext}

Analyzing these expressions reveals:

Case 1: NOT operation ($n_x = 1$, others $= 0$): This gives $n_+ = 1$, $n_- = 1$, $\bar{n}_3 = 0$, maximizing $|D| - |B|$ for all parameter values $(p, P, \alpha)$. Note however, there is no difference in concurrence behavior for $\sigma_x\leftrightarrow\sigma_y$. Our expressions for concurrence for $n_x=1$ and $n_y=1$ are identical reflecting this behavior. The reason is attributed to the indifference of amplitude damping channels to relative phases.

Case 2: Identity ($n_0 = 1$, others $= 0$): This gives $n_\pm = 0$, yielding the baseline "without NOT" case. For this case as well, $n_0=1$ or $n_3=1$ gives expressions identical to the ``without-NOT" case. 

Case 3: Hadamard-type operations:
For operations like $(\sigma_x + \sigma_z)/\sqrt{2}$, the mixed coefficients lead to smaller values of $|D| - |B|$ and can even make the concurrence negative (forced to zero by the max operation), effectively destroying entanglement.
 Hadamard-type operations:

The mathematical analysis confirms that among all local SU(2) operations, the NOT operation uniquely maximizes the concurrence for delaying or avoiding ESD. The physical reason is clear: NOT performs population inversion, directly addressing the decay asymmetry ($\ket{V} \to \ket{H}$ but not reverse) that drives ESD. Other unitaries create superpositions, rotate phases, or distribute populations in ways that do not exploit this asymmetry and often worsen entanglement dynamics.

\section{Hamiltonian Eigenvectors}
\label{app:eigenvectors}
The parity operator given in Eq.~\eqref{eq:parity_operator} involves specific eigenvectors $|\lambda_6\rangle$, $|\lambda_7\rangle$, $|\lambda_{16}\rangle$, and $|\lambda_{17}\rangle$ among the set of 24 eigenvectors for $H_{\text{free}}$ denoted by $|\lambda_i\rangle$ for $i=1\,,2\,,\dots\,,24$. These eigenvectors are linear combination of the tensor product of polarization and path states {\it viz.} $|\lambda_i\rangle=|H\rangle\otimes\sum a^i_m|m\rangle+|V\rangle\otimes \sum b^i_n|n\rangle$ where $|m\rangle\,,|n\rangle$ are the path states in the computational basis. These correspond to system-environment states where the photon polarization (system) is entangled with the spatial path (environment), creating correlations that drive decoherence. The generic form of the $|\lambda_i\rangle$ where $i=6,7,16,17$ is given by,
\begin{equation}
    |\lambda_i\rangle= |H\rangle\otimes |P_i\rangle+|V\rangle\otimes |Q_i\rangle\,,
\end{equation}
where the path-states $|P_i\rangle$ and $|Q_i\rangle$ are given by,
\begin{align}
    \begin{split}
        |P_i\rangle &= a_1|1\rangle+p \frac{i}{2\sqrt{2}}|3\rangle-a_1^\star|4\rangle+\frac{1}{2\sqrt{2}}|9\rangle\,,\\
        |Q_i\rangle &= a_1|0\rangle+p \frac{i}{2\sqrt{2}}|3\rangle-a_1^\star|5\rangle+\frac{1}{2\sqrt{2}}|8\rangle\,,\\
    \end{split}
\end{align}
where the corresponding coefficients $(a_1\,, p)$ for the relevant eigenvectors are given in Table \ref{tab:Eigenvectors}.
\begin{table}[]
    \centering
    \begin{tabular}{|c|c|}
    \hline
    \text{Eigenvector} & \text{Parameter $(a_1,p)$}  \\
    \hline
    $\lambda_6$ & $\left((1-i)/4,+\right)$\\
    \hline
    $\lambda_7$ & $\left((1+i)/4,-\right)$\\
    \hline
    $\lambda_{16}$ & $\left(-(1+i)/4,-\right)$\\
    \hline
    $\lambda_{17}$ & $\left(-(1-i)/4,+\right)$\\
    \hline
    \end{tabular}
    \caption{Values of the parameters $(a_1,p)$ for eigenvectors $|\lambda_6\rangle\,,|\lambda_7\rangle\,,|\lambda_{16}\rangle$ and $|\lambda_{17}\rangle$ from the full set $\{|\lambda_i\rangle\}$ of eigenvectors for $i=1\dots 24$ for $H_{\text{free}}$, contributing to the parity operator in \eqref{eq:parity_operator}.}
    \label{tab:Eigenvectors}
\end{table}
The parity operator $\Pi$ in Eq.~\eqref{eq:parity_operator} projects onto the subspace orthogonal to these four states. Physically, this means $\Pi$ removes the system-environment correlations that cause decoherence. A single application of $\Pi$ (via the NOT operation) redirects the state onto eigenvectors where system and environment remain factorizable, preventing further entanglement loss between system and bath.

This explains the connection to dynamical decoupling: DD repeatedly applies $\Pi$ to average away $H_{\text{int}}$, while our protocol applies it once at an optimal time to steer onto a favorable eigenspace trajectory. The full set of eigenvectors for $H_{\text{free}}$ is given in Supplementary Material~\cite{SM}.

\section{Cascaded (CADC-NOT)$^n$ Channels}
\label{app:cascade}
\begin{figure}
    \centering
    \includegraphics[width=0.8\linewidth]{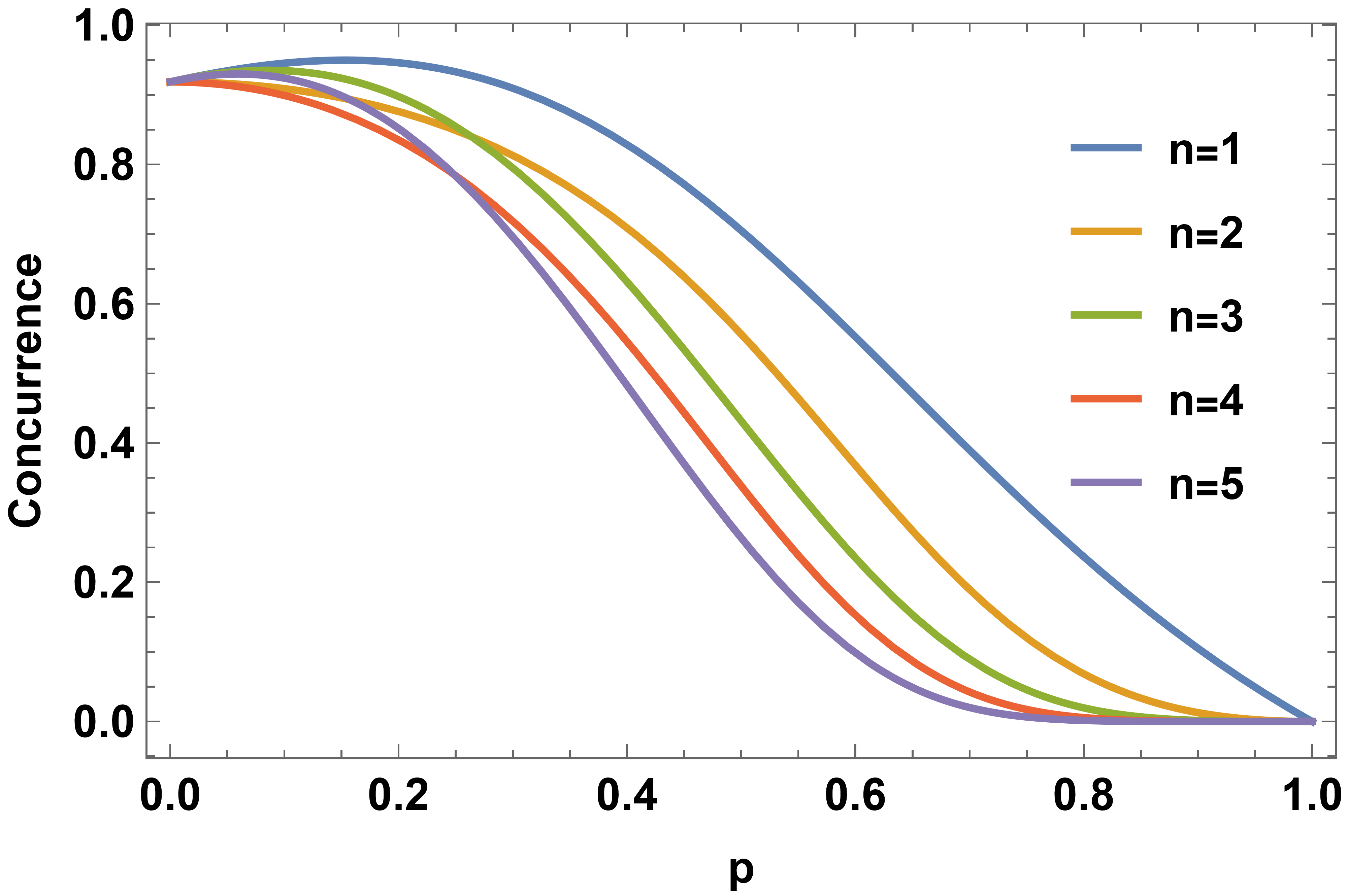}
    \caption{We show the effect of cascading identical $\rm (CADC-like)-NOT$ blocks for $n=1\,,2\,,3\,,4\,, 5$. The blue curve is for $n=1$ which has the highest concurrence.}
    \label{fig:cnmax}
\end{figure}

We examined repeated cascades of $\rm CADC-like$ blocks interspersed with NOT operations. For an $n$-cascade, the output state is
\begin{equation}
\rho_{out}=\sum_{i,j,k\dots}\left(\dots\left(K_k\left(K_j\left(K_i\rho_{in}K_i^\dagger\right)K_j^\dagger\right)K_k^\dagger\right)\dots\right),
\end{equation}
where $K_i$ are the Kraus operators for individual blocks and $\rho_{in}=|\psi\rangle\langle\psi|$ with $|\psi\rangle=\alpha |HH\rangle+\beta|VV\rangle$. The maximum value for the concurrence $C_1(\alpha)$ for $n=1$ and $0<p_1<1$, $0<\alpha\leq 1/\sqrt{2}$ is
\begin{equation}
    C_1^{max}(\alpha) =\frac{\alpha}{\sqrt{2}-\sqrt{1-\alpha^2}}\,, \ \text{for}\  p_1^\star = 1-\frac{1}{\sqrt{2(1-\alpha^2)}}\,.
    \label{eq:cascade1}
\end{equation}
For $\alpha=0.55$, $C_1^{max}=0.949$ which occurs at $p_1^\star=0.153$. For $n=2$ and higher identical cascades, the attainable concurrence remains below $C^{\max}_1(\alpha)$, as shown in Table~\ref{tab:ncascade}. Although the values are not strictly monotonic for the fixed choice $p_i=p_1^\star$, none of the cascaded protocols exceeds the single-shot value. This behavior is illustrated in Fig.~\ref{fig:cnmax}. Note that we have kept the channel strength identical for all the cascades $p_i=p_1$ to compare the effects as we apply the cascades in sequence.
\begin{table}[]
    \centering
    \caption{Concurrence for $n>1$ cascade for $\alpha=0.55$ and $p_i=p_1^\star$ in comparison to $C_1^{max}(\alpha)$.}
    \label{tab:ncascade}
    \begin{tabular}{cccccccc}
    \toprule
         $n$ & 1 & 2 & 3 & 4 & 5 & 6 & 7\\   
         \midrule
        Concurrence & 0.949 & 0.895 & 0.923 & 0.871 & 0.896 & 0.848 & 0.8711 \\
        \bottomrule
    \end{tabular}
\end{table}

\textit{Lemma: For any fixed initial state parameter $\alpha \in (0,1/\sqrt2)$, the concurrence achievable by an $n$-fold cascade of identical ($\rm (CADC-like)$--NOT) blocks satisfy $C_n(\alpha) \le C_1^{max}(\alpha)$ for all $n \ge 1$ with equality for $n=1$. In particular, within the family of cascaded ($\rm (CADC-like)$-NOT)$^n$ protocols considered here, a single-shot NOT is optimal for maximizing the concurrence and delaying ESD.}
\vspace{0.2cm}
\\
The single-shot protocol is therefore not merely economical in implementation; it is optimal within the analyzed family of local strategies. 

\section{With and Without NOT Implementation for ESD control}
\label{app: WNI_ESD}

In our experiment, we consider the bipartite entangled state $|\Psi\rangle = \alpha |HH\rangle + \beta |VV\rangle,$ with $\alpha = 0.55$, which evolves through a custom ADC of strength $p$, followed by a local NOT operation and a second ADC of strength $P$. By tuning the first damping parameter $p$, we realize the three distinct regimes of ESD control: Avoidance, Delay, and Hastening.

\subsection*{Experimental Strategy for the Without-NOT Configuration}

Within our displaced-Sagnac interferometric architecture, the NOT gate is intrinsically coupled with the second PBS and the waveplates that implement the second ADC. Therefore, direct physical removal of the NOT operation is impractical, as it would necessitate removal of multiple optical components and a complete realignment of the interferometer before switching back to the ``with-NOT'' configuration. To preserve interferometric stability and avoid systematic misalignment errors, we adopted an emulation strategy for the ``without-NOT'' case.

Instead of physically removing the NOT gate, we reproduced at the source the quantum state that would exist immediately after the first ADC and the NOT operation. For each initial state corresponding to $P=0$ in Figs.~\ref{fig: avoidance}, \ref{fig: delay}, and \ref{fig: hastening}, we first extracted the parameters $(\alpha,\beta)$ from the experimentally measured ``with-NOT'' input state, which already incorporates the effects of the first damping ($p$) and the NOT operation.

Using these extracted parameters, we tuned the pump polarization to $(\alpha |H\rangle + \beta |V\rangle)$ via the pump waveplates. This produced, through SPDC at the crystal, the entangled state
\[
(\beta |HH\rangle + \alpha |VV\rangle).
\]
This prepared state faithfully mirrors the post-ADC1, post-NOT condition. The concurrence of this reconstructed state matched that of the measured ``with-NOT'' input state within a $4\%$ deviation, confirming the accuracy of the emulation procedure. The fidelities were computed by applying an additional NOT operation numerically to the measured initial state corresponding to the without-NOT case and comparing it to the measured with-NOT input state.

Once prepared, this state was propagated through the second ADC with all interferometer settings unchanged. The resulting evolution corresponds effectively to the ``without-NOT'' configuration. The measured data for this case are represented as black markers with corresponding error bars in the Figs.~\ref{fig: avoidance}, \ref{fig: delay}, and \ref{fig: hastening}.

\subsection*{Input-State Characterization}

To validate the correctness of state preparation in both configurations, we measured the concurrence of the input states for the with-NOT and without-NOT cases over five independent experimental runs in each ESD regime. The mean concurrence values and their standard deviations are summarized in Table~\ref{MeanStdConcurrenceFidelity}.

\begin{table}[h!]
\centering
\caption{Comparison of measured Concurrence between the input states for with-NOT and without-NOT configurations across the three ESD control regimes.}
\label{MeanStdConcurrenceFidelity}
\renewcommand{\arraystretch}{1.5}
\setlength{\tabcolsep}{8pt}
\resizebox{\columnwidth}{!}{
\begin{tabular}{lcccc}
\hline
\textbf{Regime} & \textbf{Configuration} & \textbf{Mean Concurrence} & \textbf{Std. Dev.} \\
\hline
\multirow{2}{*}{\textit{Avoidance}} 
 & With-NOT & 0.68 & 0.01 \\
 & Without-NOT & 0.71 & 0.01  \\
\hline
\multirow{2}{*}{\textit{Delay}} 
 & With-NOT & 0.74 & 0.02 \\
 & Without-NOT & 0.71 & 0.02 \\
\hline
\multirow{2}{*}{\textit{Hastening}} 
 & With-NOT & 0.65 & 0.01 \\
 & Without-NOT & 0.64 & 0.02 \\
\hline
\end{tabular}
}
\end{table}

The close agreement between the concurrence values for the two configurations confirms that the prepared states accurately represent the intended theoretical conditions. The small standard deviations demonstrate good reproducibility across independent runs. Because the fidelities and the concurrence values are well-matched in both scenarios, we were able to demonstrate complete and controlled ESD manipulation experimentally across all three regimes. 

\section{Temporal steering extends the operational lifetime of entanglement}
\label{app:teleportation}
To connect temporal steering to an operational quantum-information task, we evaluate the average teleportation fidelity of the evolving two-qubit
state. The relevant benchmark is the classical threshold for the average fidelity $\mathcal{F}_{avg \rm -cl}=2/3$. Starting point is the output state,
\begin{equation}
\rho = \mathcal{N}^{-1}\begin{pmatrix} a &0 &0& d\\ 0 &b &c &0\\ 0 &c &b &0\\ d &0 &0 &f\end{pmatrix}\,, \ \text{with}\ \mathcal{N}=a+2b+f\,.
\end{equation}
This form of $\rho$ is an $X$-state. To compute the maximum attainable average teleportation fidelity for the $X-$state, we refer to Horodecki's formula,\cite{HORODECKI199621,Nandi2018}
\begin{equation}
\label{tf_max}
\mathcal{F}_{avg}=\frac{1}{2}\left(1+\frac{1}{3} \text{tr}\sqrt{T^\dagger T}\right)\,,
\end{equation}
where $\text{tr}\sqrt{T^\dagger T}$ is the trace norm of $T =\{t_{mn}\}$ and $t_{mn}=\text{tr}\left(\rho\ (\sigma_m\otimes \sigma_n)\right)$ where $\sigma_i$ are Pauli matrices for $(m,n)=1,2,3$. For the coincidence-conditioned implementation, the appropriate rate-aware quantity $\eta_{\rm post}\mathcal{F}_{avg}$ is not a new fidelity but a yield-weighted operational advantage. For the $\rm (ADC-like)-ADC$ case, $\eta_{\rm post}=1$, and the conditional teleportation fidelity itself determines the useful region. For the $\rm (CADC-like)-ADC$ case, the coincidence-conditioned map has a nontrivial admitted-branch weight, so we also consider a yield-weighted operational figure of merit. We define
\begin{equation}\label{eq:mod_tel}
\mathcal{U}_{\rm tel}(p,P)=\eta_{\rm post}(p)\max\!\left[0,\,\mathcal{F}_{avg}(p,P)-\frac{2}{3}\right]\,,
\end{equation}
which weights the teleportation advantage above the classical threshold by the admitted-event fraction of the conditional map. Superscripts indicate whether the local NOT operation is applied. Equation~\eqref{eq:mod_tel} should not be interpreted as defining a new teleportation fidelity; rather, it measures the rate-weighted usefulness of the entangled resource within the coincidence-detected ensemble.

\begin{figure}[t]
\centering
\subfloat[Yield weighted Teleportation score.\label{fig:tf_adc}]{\includegraphics[width=0.47\linewidth]{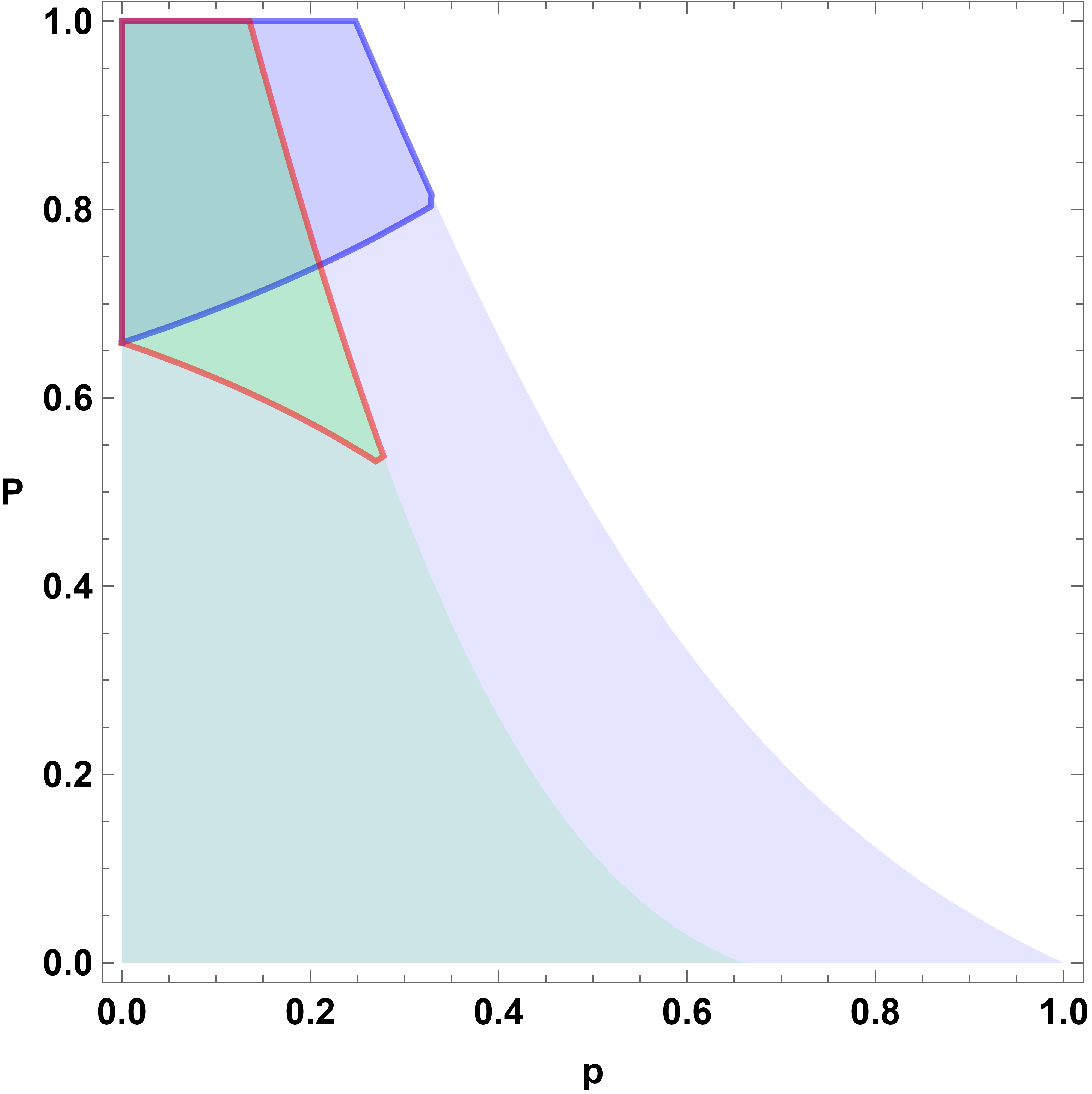}}\hfill
\subfloat[$\Delta P_{\rm tel}$ versus $p$.\label{fig:tf_cadc}]{\includegraphics[width=0.47\linewidth]{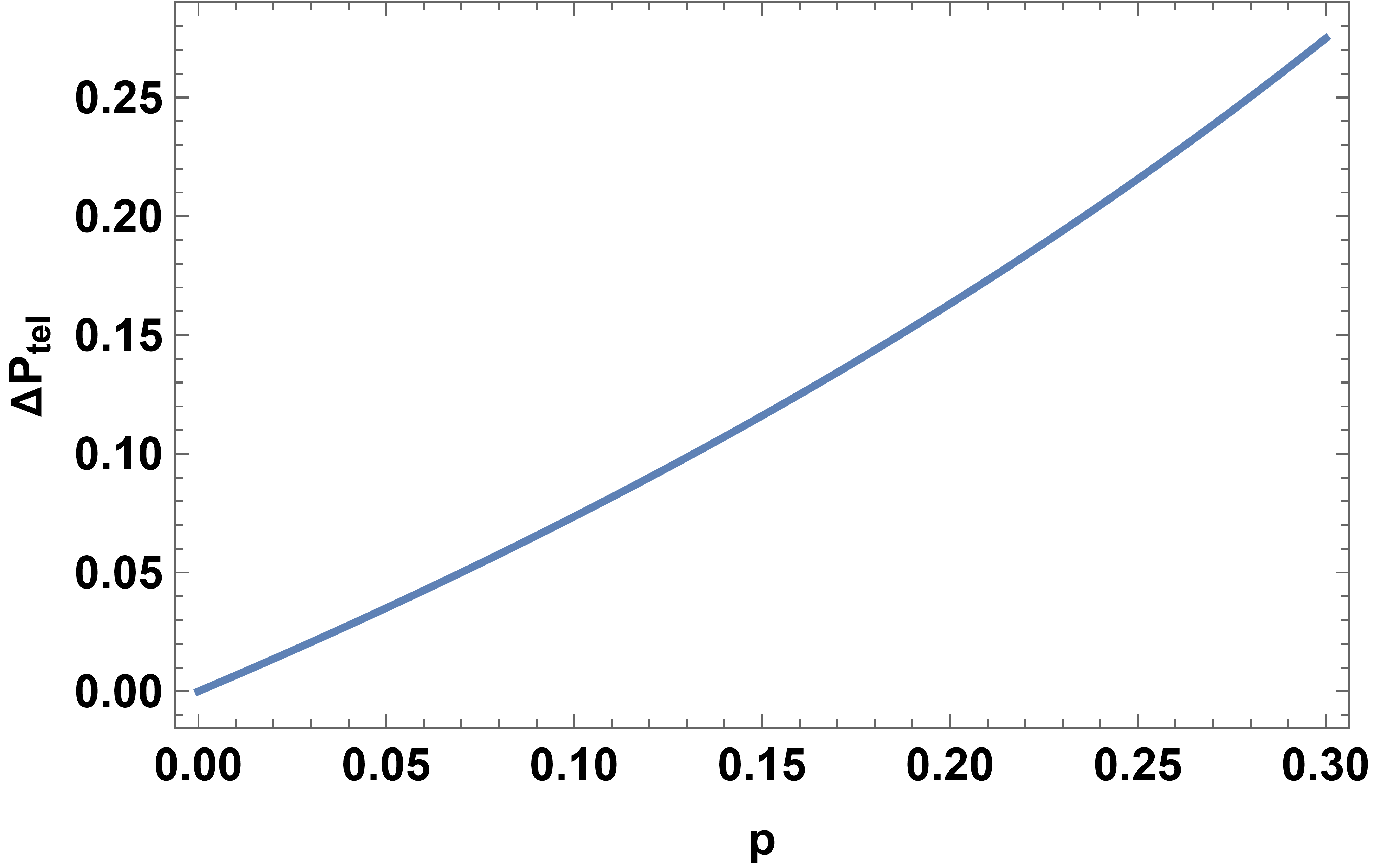}}
\caption{Operational consequence of temporal steering corresponding to input state $\alpha=0.55$. Fig.~\ref{fig:tf_adc} shows the yield-weighted score where $\mathcal{U}_{\rm tel}(p,P)>0$ for the $\rm (ADC-like)-ADC$ cascade (green region) and $\rm (CADC-like)-ADC$ cascade (blue region). The regions bounded by red and blue boundaries mark the advantage of the local NOT operation for the $\rm ADC-like$ and $\rm CADC-like$ cascades respectively. Fig.~\ref{fig:tf_cadc} plots $\Delta P_{\rm tel}$ as a function of $p$ at the boundary of the ``with-NOT" configuration for both the cascades. As $p$ increases till the critical value $p_\star\sim 0.3$, $\Delta P_{\rm tel}$ also increases showing the advantage of the $\rm (CADC-like)-ADC$ cascade over the $\rm (ADC-like)-ADC$ cascade.}
\label{fig:teleportation}
\end{figure}
As shown in Fig.~\ref{fig:tf_adc} for input state corresponding to $\alpha=0.55$, the $\rm (CADC-like)-ADC$ cascade and the $\rm (ADC-like)-ADC$ cascade marked by blue and green regions admit the net yield-weighted score $\mathcal{U}_{\rm tel}(p,P)>0$. The regions bounded by the red and blue boundaries respectively denote the advantage of the local NOT operation for the $\rm (ADC-like)-ADC$ and $\rm (CADC-like)-ADC$ cascade. A global measure is the integrated teleportation advantage given by the total area enclosed by the condition $\eta_{\rm post}(p){\rm max}\left[0,\mathcal{F}_{avg}(p,P)-2/3\right]$.

For the coincidence-conditioned implementation, the appropriate rate-aware quantity is given by,
\begin{equation}
\mathcal{A}_{\rm cond}=\int_{\mathbf{R}} dP\, dp\,\ \left(\mathcal{U}_{\rm tel}^{\rm NOT}(p,P)-\mathcal{U}_{\rm tel}(p,P)\right)\,,
\label{eq:Acond}
\end{equation}
which quantifies the yield-aware area above the classical threshold. The area enclosed by the regions bounded by the red and blue boundaries in Fig.~\ref{fig:tf_adc} is where ``with NOT" dominates ``without-NOT". Comparing
\begin{equation}\label{acond}
    \mathcal{A}_{\rm cond}^{\rm ADC-like}=0.0746\,, \ \mathcal{A}_{\rm cond}^{\rm CADC-like}=0.0704\,.
\end{equation}
we can see that the $\rm (ADC-like)-ADC$ cascade appears to offer slightly more maneuverability. Note however that the global measure is not sufficient to quantify the usefulness of teleportation fidelity. This is because the area for $\rm (CADC-like)-ADC$ channel is suppressed by $\eta_{\rm post}(p)$. In order to further compare the usefulness of the cascades, we quantify the extension of the useful damping range by
\begin{equation}
\Delta P_{\rm tel} = P_{\rm crit}^{\rm NOT} - P_{\rm crit}^{\rm no\,NOT},
\label{eq:DeltaPtel}
\end{equation}
where \(P_{\rm crit}\) is defined by the curve \(\mathcal{F}_{avg}(p,P_{\rm crit})=2/3\). We plot the $\Delta P_{\rm tel}$ as a function of $p$ in Fig.~\ref{fig:tf_cadc} to demonstrate that $\Delta P_{\rm tel}$ increases with $p$ upto the critical $p=p_\star\,, P=P_\star$ such that,
\begin{equation}
    \mathcal{F}_{avg}^{\rm NOT}(p_\star,P_\star)=2/3=\mathcal{F}_{avg}(p_\star,P_\star)\,,
\end{equation}
A positive \(\Delta P_{\rm tel}\), therefore, indicates that the NOT-assisted trajectory preserves teleportation-useful entanglement over a larger range of the second damping parameter for the $\rm (CADC-like)-ADC$ cascade over its counterpart.

Note that since the $p_\star$ for both the cascades are slightly apart, we take the smaller $p_\star$ for the $\rm (ADC-like)-ADC$ cascade for meaningful comparison, since after that the advantage of the $\rm (ADC-like)-ADC$ cascade ceases, as is clearly visible from the Fig.~\ref{fig:tf_adc}. $\Delta P_{\rm tel}$ increases with $p$ and hits a ceiling of $\Delta P_{\rm tel}\approx 0.3$ beyond the $p_\star$ for $\rm ADC-like$ cascade. The findings of \eqref{acond} and \eqref{eq:DeltaPtel} are not conflicting since we must remember the suppression factor in the global measure. These two findings together imply that the total events where we see teleportation fidelity above the classical threshold may be lower in the case of the $\rm CADC-like$ cascade, however the subset of such events will sustain the fidelity for a longer duration in comparison to the $\rm ADC-like$ cascade for non-zero damping.

This preliminary analysis and collective illustration from Fig.~\ref{fig:teleportation} shows that teleportation fidelity can be used as an operational figure of merit for optimizing temporal steering protocols. In particular, it suggests a hardware-agnostic optimization problem: choosing the intervention time and the effective damping structure so as to maximize the teleportation-useful lifetime of entanglement, while accounting for the admitted-event fraction of the conditional map.

\section{Systematic Error Analysis and First-Order Correction}
\label{app:errors}

We analyze systematic errors arising from imperfect optical components affecting the concurrence measurement. The total unitary implemented in the experiment can be written as
\begin{equation}
U_T = U^A \otimes U^B,
\end{equation}
where
\begin{equation}
U^A = P_1^\dagger \cdot H_2(\phi) \cdot P_2 \cdot U_{NOT} \cdot H_1(\theta) \cdot P_1.
\end{equation}
Here $P_i$ denote polarizing beam splitters, $H_i$ half-wave plates, and $U_{NOT}$ is the Pauli-$X$ operation implemented using an HWP at $45^\circ$.

The experimentally observed reduction in the concurrence arises from three dominant sources: (i) state-preparation errors, (ii) PBS leakage and extinction limitations, and (iii) finite angular precision in waveplate rotations implementing amplitude damping and NOT operations.

Error in State Preparation: We prepare the bipartite entangled state
\begin{equation}
|\psi\rangle = \alpha |HH\rangle + \beta |VV\rangle.
\end{equation}

The pump polarization is set using a half-wave plate at angle $\phi$, producing
\begin{equation}
|\psi_{\text{pump}}\rangle = \alpha |V\rangle + \beta |H\rangle.
\end{equation}

Due to the finite least count $\delta\phi$ of the rotation mount (typically $2^\circ$), the parameter $\alpha$ acquires an uncertainty
\begin{equation}
\delta\alpha = 2 \cos(2\phi)\, \delta\phi.
\end{equation}

Thus the prepared state amplitudes become $(\alpha \pm \delta\alpha)$, introducing a small deviation in the initial Concurrence.

Imperfect PBSs: PBS leakage is the dominant source of systematic error. An ideal PBS transmits $|H\rangle$ and reflects $|V\rangle$ perfectly. Real devices exhibit finite extinction ratios, leading to cross-coupling between orthogonal polarizations.

We model the imperfect PBS as
\begin{equation}
P_{\text{real}} =
\begin{pmatrix}
1-\delta & 0 & \delta & 0 \\
0 & \delta' & 0 & 1-\delta' \\
\delta_1 & 0 & 1-\delta_1 & 0 \\
0 & 1-\delta'_1 & 0 & \delta'_1
\end{pmatrix}.
\end{equation}

For high-quality PBSs under standard operation, $\delta \sim 10^{-3}$. However, in our displaced-Sagnac geometry, PBS $P_1$ is driven from three different input ports, and calibration reveals extinction ratios spanning
\[
\delta_i \sim 10^{-3} \text{ to } 10^{-2}.
\]

Waveplate Rotation Errors in ADC and NOT Operations: The amplitude damping parameter is related to HWP angle $\theta$ via
\begin{equation}
p = \sin^2(2\theta).
\end{equation}

The angular uncertainty $\delta\theta$ produces
\begin{equation}
\delta p = 2\sin(4\theta)\, \delta\theta.
\end{equation}

The NOT gate implemented with $\theta = \pi/4$ yields a first-order error
\begin{equation}
\text{Err}_{NOT}
=
\left.
\frac{\partial}{\partial\theta}
\begin{pmatrix}
\cos(2\theta) & \sin(2\theta) \\
\sin(2\theta) & -\cos(2\theta)
\end{pmatrix}
\right|_{\theta=\pi/4}
\delta\theta
=
- 2\sigma_z\, \delta\theta.
\end{equation}

The faulty NOT operator therefore becomes
\begin{equation}
U_{NOT}
=
\left(\sigma_x - 2\mu \sigma_z \right)
\otimes
\left(|1\rangle\langle1| + |2\rangle\langle2|\right),
\quad
\mu = \delta\theta.
\end{equation}

Similar corrections are incorporated for all HWP rotations implementing amplitude damping.

First-Order Kraus Expansion: Including all imperfections, the system evolves as
\begin{equation}
\rho_{\text{out}} = \sum_i \mathbb{K}_i \rho_{\text{in}} \mathbb{K}_i^\dagger,
\quad
\sum_i \mathbb{K}_i^\dagger \mathbb{K}_i = \mathbb{I}.
\end{equation}

Expanding
\[
\mathbb{K}_i = \mathbb{K}_i^0 + \delta\mathbb{K}_i,
\]
we obtain
\begin{align}
\rho_{\text{out}}
=
\rho_{\text{out}}^0
+
\sum_i
\left(
\mathbb{K}_i^0 \rho_{\text{in}} \delta\mathbb{K}_i^\dagger
+
\delta\mathbb{K}_i \rho_{\text{in}} \mathbb{K}_i^{0\dagger}
\right).
\end{align}

Concurrence is computed from the eigenvalues of
\[
\widetilde{\rho}_{\text{out}}
=
\rho_{\text{out}}
(\sigma_y \otimes \sigma_y)
\rho_{\text{out}}^*
(\sigma_y \otimes \sigma_y).
\]

The first-order correction to eigenvalues $\lambda_i$ is
\[
\Delta\lambda_i
=
\langle V_i | \widetilde{\delta\rho}_{\text{out}} | V_i \rangle.
\]

\begin{figure}[htbp]
    \centering
    \includegraphics[scale=0.23]{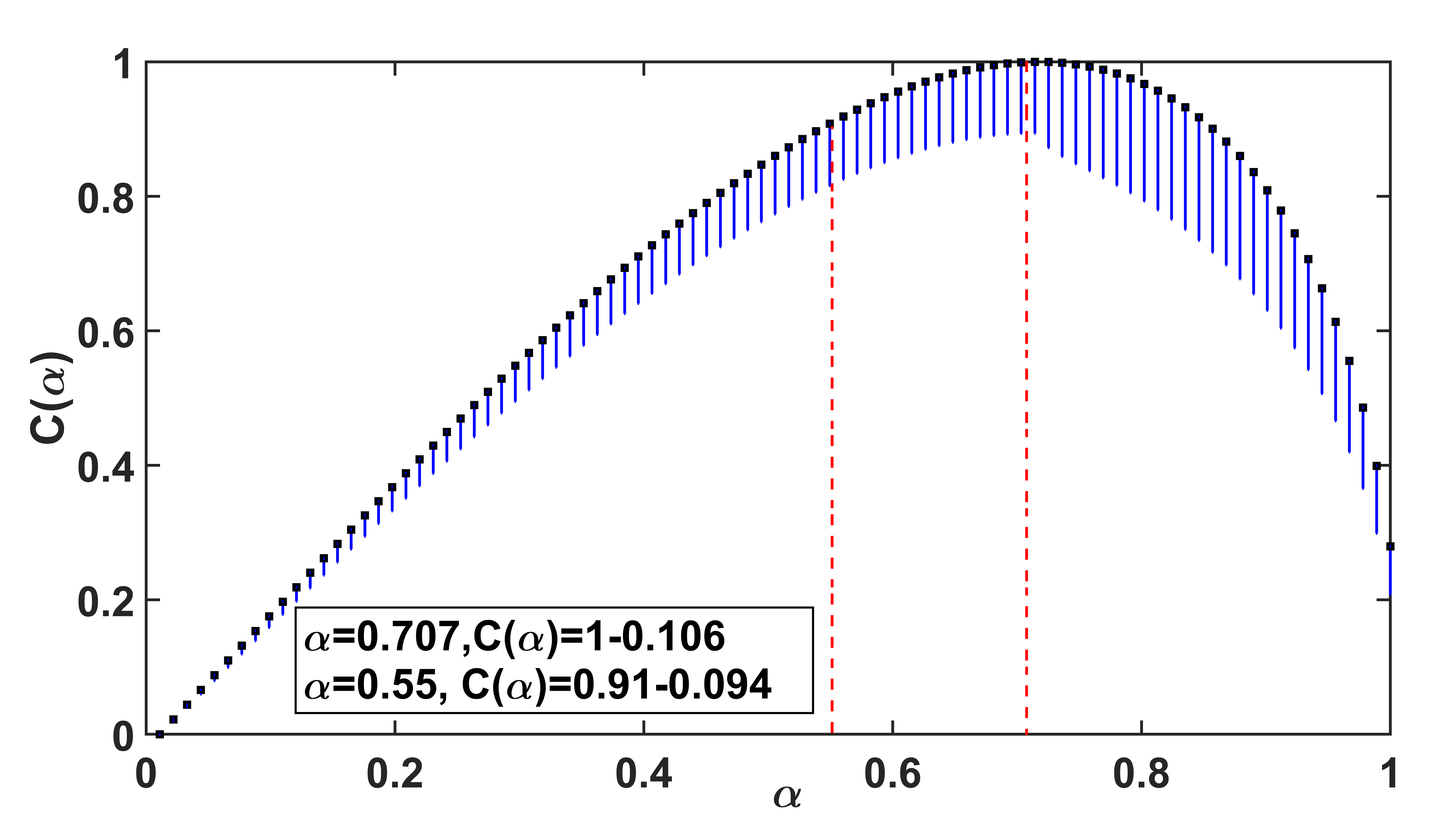}
    \caption{Systematic errors affecting the concurrence measurement. Two representative points corresponding to $\alpha = 0.55$ and $\alpha = 1/\sqrt{2} \approx 0.707$—common in experimental runs—are highlighted. The plot shows the maximum measurable concurrence for these values of $\alpha$, illustrating the impact of systematic errors.}
    \label{SysError}
\end{figure}

Thus, the first-order correction to the concurrence is
\begin{align}
\Delta C
=
\frac{1}{2}
\left(
\frac{\Delta\lambda_1}{\sqrt{\lambda_1}}
-
\frac{\Delta\lambda_2}{\sqrt{\lambda_2}}
-
\frac{\Delta\lambda_3}{\sqrt{\lambda_3}}
-
\frac{\Delta\lambda_4}{\sqrt{\lambda_4}}
\right).
\end{align}
In the error analysis leading to Fig.~\ref{SysError}, the least count of the HWP implementing the NOT operation is denoted by $\mu$, while the uncertainties in the damping parameters $p$ and $P$, arising from the finite angular resolution of the HWPs, are denoted by $\delta p$ and $\delta P$, respectively. These quantities are given by
\[
\mu = \frac{\pi}{180},
\quad
\delta p = \sqrt{p(1-p)}\frac{\pi}{90},
\quad
\delta P = \sqrt{P(1-P)}\frac{\pi}{90},
\]
and $\delta_i \sim 10^{-3}$--$10^{-2}$, we obtain
\begin{align}
\delta_i \sim 10^{-3} &:\quad \Delta C = 1.57\%\pm 0.91\%, \\
\delta_i \sim 10^{-2} &:\quad \Delta C = 9.8\%\pm 4.5\%.
\label{Error}
\end{align}

For $\alpha = 1/\sqrt{2}$, the total concurrence error lies in the range
\[
\Delta C \sim 1\% \text{--} 9\%.
\]

The experimentally observed $1$--$14\%$ deviation from the ideal concurrence is quantitatively explained by the first-order error model. Systematic errors discussed above lead to lower entanglement than theoretically expected. Fig. \ref{SysError} incorporates all such experimentally measured sources of error and highlights representative values of the concurrence for selected values of $\alpha$ used during data acquisition. We conclude, however, that the dominant limiting error arises from PBS imperfections. Further details are given in the Supplementary Material~\cite{SM}.


\begin{thebibliography}{99}
\bibitem{footnote}
Strictly speaking, the equality holds only when $t\in[-\infty,\infty]$. We have assumed that the run-time of the experiment is long enough to disregard finite domain corrections.
\end{thebibliography}
\end{document}


\title{Supplementary Material for\\``Temporal steering of entanglement decay with single-shot control''}

\author{Saumya Ranjan Behera}
\affiliation{Raman Research Institute, C.V.Raman Avenue, Sadashivanagar, Bengaluru-560080, Karnataka, India}

\author{Kallol Sen}
\affiliation{Raman Research Institute, C.V.Raman Avenue, Sadashivanagar, Bengaluru-560080, Karnataka, India}
\affiliation{QuSyn Technologies, Bengaluru-560094, Karnataka, India}

\author{Animesh Sinha Roy}
\affiliation{Raman Research Institute, C.V.Raman Avenue, Sadashivanagar, Bengaluru-560080, Karnataka, India}

\author{Snigdhadev Ray}
\affiliation{School of Physical Sciences, Indian Association for the Cultivation of Science, Kolkata - 700032, West Bengal, India}
\affiliation{Raman Research Institute, C.V.Raman Avenue, Sadashivanagar, Bengaluru-560080, Karnataka, India}

\author{Ashutosh Singh}
\affiliation{Raman Research Institute, C.V.Raman Avenue, Sadashivanagar, Bengaluru-560080, Karnataka, India}
\affiliation{Department of Physics and Astronomy, University of Calgary, Alberta T2N 1N4, Canada}

\author{A.R.P. Rau}
\affiliation{Department of Physics and Astronomy, Louisiana State University, Baton Rouge, Louisiana 70803}

\author{Urbasi Sinha}
\email{usinha@rri.res.in}
\affiliation{Raman Research Institute, C.V.Raman Avenue, Sadashivanagar, Bengaluru-560080, Karnataka, India}
\affiliation{Department of Physics and Astronomy, University of Calgary, Alberta T2N 1N4, Canada}

\date{\today}
\maketitle

This supplementary material provides additional information on the theoretical computations and experimental details for our study. The full derivation of Kraus operators, error analysis and details of cascade analysis will be discussed in the following sections.

\section{Kraus Operator Derivation}
\label{Sup:kraus}
While the detailed derivation for the Kraus operators from the unitary describing the system and environment evolution, is given in the appendix of the paper, the explicit expressions for the Kraus operators obtained therein, for with and without NOT scenarios are given below.

\subsection*{With NOT}
For the case with NOT operation between channels:
\begin{align}
\begin{split}
|F\rangle&=\cos2\phi |V_b\rangle+\sin2\phi|H_a\rangle, \
|G\rangle =x\sin2\theta \left(\cos2\phi|V_{a'}\rangle+\sin2\phi|H_{b'}\rangle\right)+\cos2\theta|H_b\rangle.
\end{split}
\end{align}
The Kraus operators are:
\begin{align}\label{kraus_operators}
\begin{split}
\mathbb{K}_1 &= \sin^22\phi |HH\rangle\langle HH|,
\mathbb{K}_2 = \frac{\sin 4\phi}{2} |HV\rangle\langle HH| +\cos2\theta\sin2\phi |HH\rangle\langle HV|,
\mathbb{K}_3 = x\sin2\theta\frac{\sin4\phi}{2} |HV \rangle\langle HV|,\\ 
\mathbb{K}_4 &= x\sin2\theta\sin^22\phi |HH \rangle\langle HV|,
\mathbb{K}_5 = \frac{\sin 4\phi}{2} |VH\rangle\langle HH| +\cos2\theta\sin2\phi |HH\rangle\langle VH|,\\
\mathbb{K}_6 &= \cos^22\phi |VV\rangle\langle\ HH|+\cos^22\theta |HH\rangle\langle VV|+\cos2\theta\cos2\phi (|VH \rangle\langle HV|+|HV\rangle\langle VH|),\\
\mathbb{K}_7 &= x\frac{\sin 4\theta}{2}\cos2\phi |HV\rangle\langle VV| + x\sin2\theta\cos^22\phi |VV \rangle\langle HV|,
\mathbb{K}_8 = x\frac{\sin 4\theta}{2}\sin2\phi |HH\rangle\langle VV| + x\sin2\theta\frac{\sin4\phi}{2} |VH \rangle\langle HV|,\\
\mathbb{K}_9 &= x\sin2\theta\frac{\sin4\phi}{2} |VH \rangle\langle VH|,  
\mathbb{K}_{10} = x\frac{\sin 4\theta}{2}\cos2\phi |VH\rangle\langle VV| +x\sin2\theta\cos^22\phi |VV\rangle\langle VH|,\\
\mathbb{K}_{11} &= \sin^22\theta\cos^22\phi |VV\rangle\langle VV|,  
\mathbb{K}_{12} =\sin^22\theta \frac{\sin 4\phi}{2} |VH\rangle\langle VV|, 
\mathbb{K}_{13} = x\sin2\theta\sin^22\phi |HH\rangle\langle VH|,\\
\mathbb{K}_{14}&= x\frac{\sin 4\theta}{2}\sin2\phi |HH\rangle\langle VV| +x\sin2\theta\frac{\sin4\phi}{2} |HV\rangle\langle VH|,
\mathbb{K}_{15} = \sin^22\theta \frac{\sin 4\phi}{2} |HV\rangle\langle VV|, 
\mathbb{K}_{16} =\sin^22\theta\sin^22\phi |HH\rangle\langle VV|.
\end{split}
\end{align}

\subsection*{Without NOT}
Without the NOT operator:
\begin{align}
\begin{split}
|F\rangle&=-\sin2\phi|H_a\rangle+\cos2\phi|V_b\rangle\,, \
|G\rangle =x\cos2\theta\left(\cos2\phi|V_{a'}\rangle+\sin2\phi|H_{b'}\rangle\right)+\sin2\theta|H_b\rangle.
\end{split}
\end{align}

The Kraus operators become:
\begin{align}\label{kraus_op_without_not}
\begin{split}
\mathbb{K}_1 & = \sin^22\phi|HH\rangle\langle HH|, \ \mathbb{K}_2= -x\cos2\theta\cos^22\phi|HV\rangle\langle HV|,
\mathbb{K}_3 = -\cos2\phi\left(\sin2\theta|HH\rangle\langle HV|+\sin2\phi|HV\rangle\langle HH|\right), \\ 
\mathbb{K}_4&=-\frac{x}{2}\cos2\theta\sin4\phi |HH\rangle\langle HV|,
\mathbb{K}_5 =-x\cos2\theta\cos^22\phi|VH\rangle\langle VH|,  
\mathbb{K}_6=\cos^22\theta\cos^22\phi|VV\rangle\langle VV|,\\
\mathbb{K}_7&=x\cos2\theta\cos2\phi\left(\sin2\theta|VH\rangle\langle VV|+\sin2\phi|VV\rangle\langle VH|\right),  
\mathbb{K}_8=\frac{1}{2}\cos^22\theta\sin4\phi|VH\rangle\langle VV|,\\
\mathbb{K}_9 &= -\cos2\phi\left(\sin2\theta|HH\rangle\langle VH|+\sin2\phi|VH\rangle\langle HH|\right),
\mathbb{K}_{10}=x\cos2\theta\cos2\phi\left(\sin2\theta|HV\rangle\langle VV|+\sin2\phi|VV\rangle\langle HV|\right),\\
\mathbb{K}_{11}&=\cos^22\phi|VV\rangle\langle HH|+\sin^22\theta |HH\rangle\langle VV|+\sin2\theta\sin2\phi\left(|HV\rangle\langle VH|+|VH\rangle\langle HV|\right),\\
\mathbb{K}_{12}&=x\cos2\theta\sin2\phi\left(\sin2\theta|HH\rangle\langle VV|+\sin2\phi|VH\rangle\langle HV|\right),
\mathbb{K}_{13}=-\frac{x}{2}\cos2\theta\sin4\phi|HH\rangle\langle VH|,\\
\mathbb{K}_{14}&=\frac{1}{2}\cos^22\theta\sin4\phi|HV\rangle\langle VV|,
\mathbb{K}_{15}=x\cos2\theta\sin2\phi\left(\sin2\theta|HH\rangle\langle VV|+\sin2\phi|HV\rangle\langle VH|\right), 
\mathbb{K}_{16}=\cos^22\theta\sin^22\phi|HH\rangle\langle VV|.
\end{split}
\end{align}

\section{Hamiltonian Eigenvectors}
\label{sup:eigenvectors}
In the appendix of the paper, we have quoted the explicit expressions for the relevant eigenvectors for the construction of the parity operator. In this section we give the explicit expressions for the complete set of eigenvectors for $H_A^{free}$:
\begin{eqnarray}
    \lambda_{1} &=& \left\{0,0,0,0,0,0,0,0,0,0,0,0,0,0,0,0,0,0,0,0,0,0,-\frac{1}{\sqrt{2}},\frac{1}{\sqrt{2}}\right\}\,,\nonumber\\
\lambda_{2} &=& \left\{0,\frac{1}{2 \sqrt{2}},0,-\frac{1}{2 \sqrt{2}},\frac{1}{2 \sqrt{2}},0,0,0,0,-\frac{1}{2 \sqrt{2}},0,0,-\frac{1}{2 \sqrt{2}},0,0,\frac{1}{2 \sqrt{2}},0,-\frac{1}{2 \sqrt{2}},0,0,\frac{1}{2 \sqrt{2}},0,0,0\right\}\,,\nonumber\\
\lambda_{3} &=& \left\{0,0,0,0,0,0,0,0,-\frac{1}{\sqrt{2}},0,0,\frac{1}{\sqrt{2}},0,0,0,0,0,0,0,0,0,0,0,0\right\}\,,\nonumber\\
\lambda_{4} &=& \left\{\frac{\sqrt{3}-i}{2 \sqrt{6}},0,\frac{-1+i \sqrt{3}}{2 \sqrt{6}},0,0,0,0,-\frac{i}{\sqrt{6}},0,0,0,0,0,\frac{i
   \left(\sqrt{3}-i\right)}{2 \sqrt{6}},\frac{-\sqrt{3}-i}{2 \sqrt{6}},0,0,0,0,0,0,\frac{1}{\sqrt{6}},0,0\right\}\,,\nonumber\\
   \lambda_{5} &=& \left\{\frac{\sqrt{3}+i}{2 \sqrt{6}},0,\frac{-1-i \sqrt{3}}{2 \sqrt{6}},0,0,0,0,\frac{i}{\sqrt{6}},0,0,0,0,0,-\frac{i
   \left(\sqrt{3}+i\right)}{2 \sqrt{6}},\frac{-\sqrt{3}+i}{2 \sqrt{6}},0,0,0,0,0,0,\frac{1}{\sqrt{6}},0,0\right\}\,,\nonumber\\
\lambda_{6} &=& \left\{0,\frac{1}{4}-\frac{i}{4},0,\frac{i}{2 \sqrt{2}},-\frac{1}{4}-\frac{i}{4},0,0,0,0,\frac{1}{2 \sqrt{2}},0,0,\frac{1}{4}-\frac{i}{4},0,0,\frac{i}{2 \sqrt{2}},0,-\frac{1}{4}-\frac{i}{4},0,0,\frac{1}{2 \sqrt{2}},0,0,0\right\}\,, \nonumber\\
\end{eqnarray}
\begin{eqnarray}
\lambda_{7} &=& \left\{0,\frac{1}{4}+\frac{i}{4},0,-\frac{i}{2 \sqrt{2}},-\frac{1}{4}+\frac{i}{4},0,0,0,0,\frac{1}{2 \sqrt{2}},0,0,\frac{1}{4}+\frac{i}{4},0,0,-\frac{i}{2 \sqrt{2}},0,-\frac{1}{4}+\frac{i}{4},0,0,\frac{1}{2 \sqrt{2}},0,0,0\right\}\,,\nonumber\\
\lambda_{8} &=& \left\{0,0,0,0,0,0,0,0,0,0,0,0,0,0,0,0,\frac{1}{6} \left(-\sqrt{3}-3 i\right),0,\frac{1}{6} \left(-\sqrt{3}+3 i\right),\frac{1}{\sqrt{3}},0,0,0,0\right\}\,,\nonumber\\
\lambda_{9} &=& \left\{0,0,0,0,0,\frac{1}{6} \left(-\sqrt{3}-3 i\right),\frac{1}{6} \left(-\sqrt{3}+3 i\right),0,0,0,\frac{1}{\sqrt{3}},0,0,0,0,0,0,0,0,0,0,0,0,0\right\}\,,\nonumber\\
\lambda_{10} &=& \left\{0,0,0,0,0,0,0,0,0,0,0,0,0,0,0,0,\frac{1}{6} \left(-\sqrt{3}+3 i\right),0,\frac{1}{6} \left(-\sqrt{3}-3 i\right),\frac{1}{\sqrt{3}},0,0,0,0\right\}\,,\nonumber\\
\lambda_{11} &=& \left\{0,0,0,0,0,\frac{1}{6} \left(-\sqrt{3}+3 i\right),\frac{1}{6} \left(-\sqrt{3}-3 i\right),0,0,0,\frac{1}{\sqrt{3}},0,0,0,0,0,0,0,0,0,0,0,0,0\right\}\,,\nonumber\\
\lambda_{12} &=& \left\{-\frac{i}{\sqrt{6}},0,\frac{1}{\sqrt{6}},0,0,0,0,\frac{i}{\sqrt{6}},0,0,0,0,0,-\frac{1}{\sqrt{6}},-\frac{i}{\sqrt{6}},0,0,0,0,0,0,\frac{1}{\sqrt{6}},0,0\right\}\,,\nonumber\\
\lambda_{13} &=& \left\{0,-\frac{i}{2 \sqrt{2}},0,\frac{1}{2 \sqrt{2}},\frac{i}{2 \sqrt{2}},0,0,0,0,-\frac{1}{2 \sqrt{2}},0,0,\frac{i}{2 \sqrt{2}},0,0,-\frac{1}{2 \sqrt{2}},0,-\frac{i}{2 \sqrt{2}},0,0,\frac{1}{2 \sqrt{2}},0,0,0\right\}\,,\nonumber\\
\lambda_{14} &=& \left\{\frac{i}{\sqrt{6}},0,\frac{1}{\sqrt{6}},0,0,0,0,-\frac{i}{\sqrt{6}},0,0,0,0,0,-\frac{1}{\sqrt{6}},\frac{i}{\sqrt{6}},0,0,0,0,0,0,\frac{1}{\sqrt{6}},0,0\right\}\,,\nonumber\\
\lambda_{15} &=& \left\{0,\frac{i}{2 \sqrt{2}},0,\frac{1}{2 \sqrt{2}},-\frac{i}{2 \sqrt{2}},0,0,0,0,-\frac{1}{2 \sqrt{2}},0,0,-\frac{i}{2 \sqrt{2}},0,0,-\frac{1}{2 \sqrt{2}},0,\frac{i}{2 \sqrt{2}},0,0,\frac{1}{2 \sqrt{2}},0,0,0\right\}\,,\nonumber\\
\lambda_{16} &=& \left\{0,-\frac{1}{4}-\frac{i}{4},0,-\frac{i}{2 \sqrt{2}},\frac{1}{4}-\frac{i}{4},0,0,0,0,\frac{1}{2 \sqrt{2}},0,0,-\frac{1}{4}-\frac{i}{4},0,0,-\frac{i}{2 \sqrt{2}},0,\frac{1}{4}-\frac{i}{4},0,0,\frac{1}{2 \sqrt{2}},0,0,0\right\}\,,\nonumber\\
\lambda_{17} &=& \left\{0,-\frac{1}{4}+\frac{i}{4},0,\frac{i}{2 \sqrt{2}},\frac{1}{4}+\frac{i}{4},0,0,0,0,\frac{1}{2 \sqrt{2}},0,0,-\frac{1}{4}+\frac{i}{4},0,0,\frac{i}{2 \sqrt{2}},0,\frac{1}{4}+\frac{i}{4},0,0,\frac{1}{2 \sqrt{2}},0,0,0\right\}\,,\nonumber\\
\lambda_{18} &=& \left\{\frac{-\sqrt{3}-i}{2 \sqrt{6}},0,\frac{-1-i \sqrt{3}}{2 \sqrt{6}},0,0,0,0,-\frac{i}{\sqrt{6}},0,0,0,0,0,-\frac{i
   \left(\sqrt{3}+i\right)}{2 \sqrt{6}},\frac{\sqrt{3}-i}{2 \sqrt{6}},0,0,0,0,0,0,\frac{1}{\sqrt{6}},0,0\right\}\,,\nonumber
\end{eqnarray}
\begin{eqnarray}
\lambda_{19} &=& \left\{\frac{-\sqrt{3}+i}{2 \sqrt{6}},0,\frac{-1+i \sqrt{3}}{2 \sqrt{6}},0,0,0,0,\frac{i}{\sqrt{6}},0,0,0,0,0,\frac{i
   \left(\sqrt{3}-i\right)}{2 \sqrt{6}},\frac{\sqrt{3}+i}{2 \sqrt{6}},0,0,0,0,0,0,\frac{1}{\sqrt{6}},0,0\right\}\,,\nonumber\\
\lambda_{20} &=& \left\{0,0,0,0,0,0,0,0,0,0,0,0,0,0,0,0,0,0,0,0,0,0,\frac{1}{\sqrt{2}},\frac{1}{\sqrt{2}}\right\}\,,\nonumber\\
\lambda_{21} &=& \left\{0,-\frac{1}{2 \sqrt{2}},0,-\frac{1}{2 \sqrt{2}},-\frac{1}{2 \sqrt{2}},0,0,0,0,-\frac{1}{2 \sqrt{2}},0,0,\frac{1}{2 \sqrt{2}},0,0,\frac{1}{2 \sqrt{2}},0,\frac{1}{2 \sqrt{2}},0,0,\frac{1}{2 \sqrt{2}},0,0,0\right\}\,,\nonumber\\
\lambda_{22} &=& \left\{0,0,0,0,0,0,0,0,0,0,0,0,0,0,0,0,\frac{1}{\sqrt{3}},0,\frac{1}{\sqrt{3}},\frac{1}{\sqrt{3}},0,0,0,0\right\}\,,\nonumber\\
\lambda_{23} &=& \left\{0,0,0,0,0,0,0,0,\frac{1}{\sqrt{2}},0,0,\frac{1}{\sqrt{2}},0,0,0,0,0,0,0,0,0,0,0,0\right\}\,,\nonumber\\
\lambda_{24} &=& \left\{0,0,0,0,0,\frac{1}{\sqrt{3}},\frac{1}{\sqrt{3}},0,0,0,\frac{1}{\sqrt{3}},0,0,0,0,0,0,0,0,0,0,0,0,0\right\}\,.\nonumber
\end{eqnarray}
The eigenvectors $\lambda_6\,, \lambda_7\,, \lambda_{16}$ and $\lambda_{17}$ are relevant for the construction of the parity operator given in the main text.

\section{Entanglement Sudden Death (ESD) manipulation using local unitary operator}
The main text presents experimental methods and results for the initial state parameter $\alpha = 0.55$, demonstrating the controlled manipulation of ESD via a local NOT operation. To establish the robustness of our protocol and validate its portability under different initial conditions, we present here additional experimental data for $\alpha = 0.45$ and $\alpha = 0.5$. These measurements validate the independent operation of the protocol and confirm that the controlled manipulation of ESD -- is reproducible across varying entanglement regimes. We further discuss details of individual channels involved in the protocol, in this section. 

\begin{figure*}[htbp]
\centering
\includegraphics[width=0.9\textwidth]{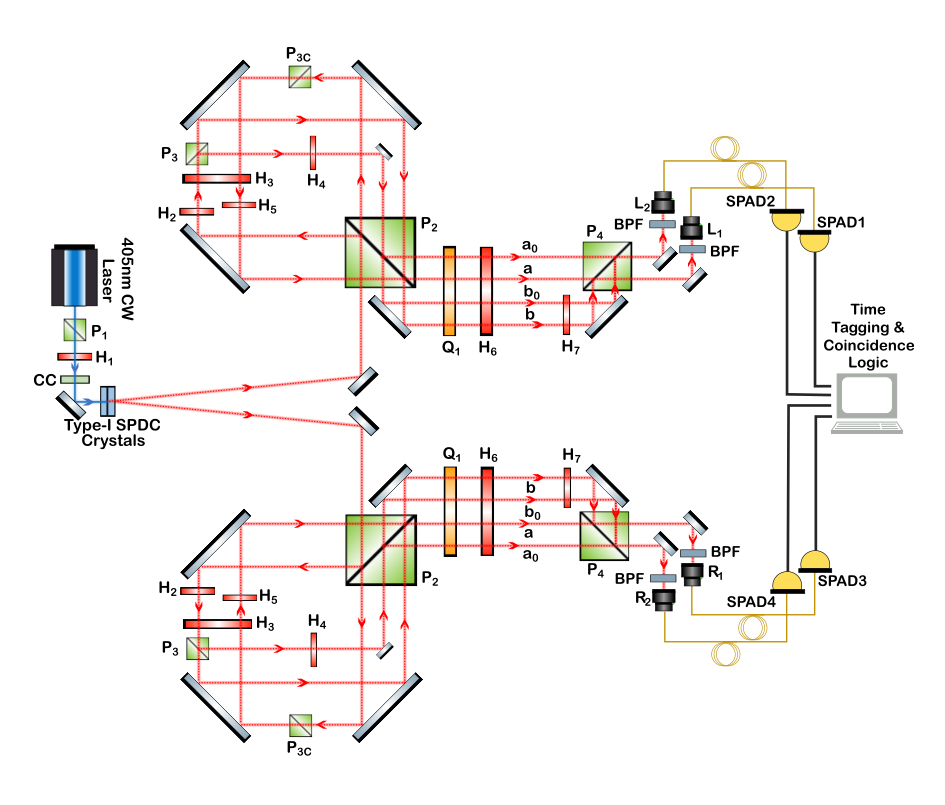}
\caption{Experimental set-up. EPS stands for Entangled Photon Source. $H_1$, $H_2$, $H_3$, $H_4$, $H_5$, $H_6$, $H_7$ are half-wave plates, $Q_1$ is a quarter wave plate, $P_1$, $P_2$, $P_3$, $P_{3C}$ and $P_4$ are polarization beam splitters, $L_1$, $L_2$, $R_1$ and $R_2$ are single photon coupling modules, SMF: Single Mode Fiber, SPAD: Single Photon Avalanche Diode detectors.}
\label{ESDExptScheme}
\end{figure*}

\begin{figure}
    \centering
    \includegraphics[width=0.5\linewidth]{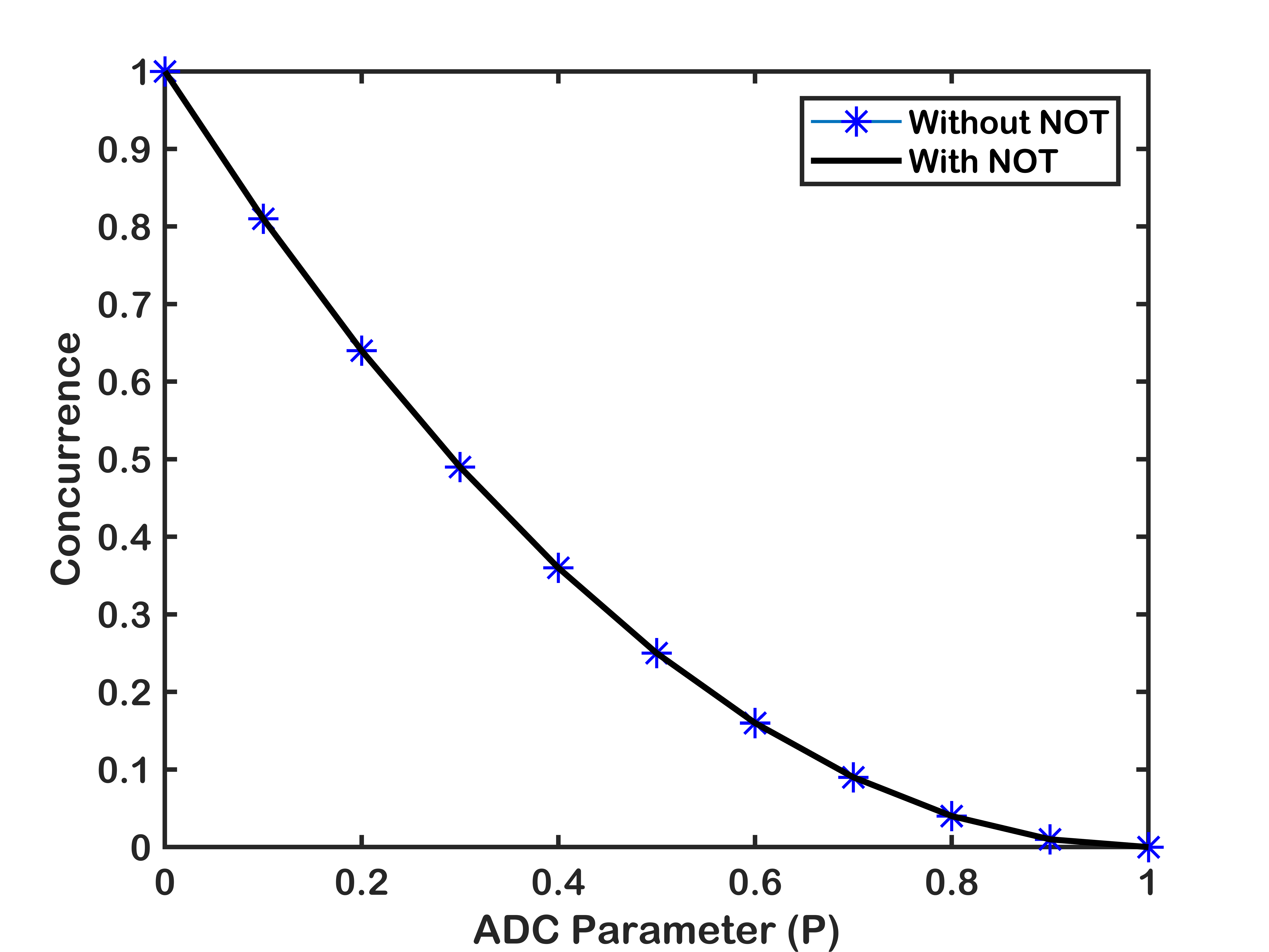}
    \caption{A maximally entangled state evolved through our protocol for ESD manipulation. Both with and without NOT operation we observe asymptotic decay of entanglement.}
    \label{fig:ADE_Ideal}
\end{figure}

\begin{figure}
    \centering
    \includegraphics[width=0.5\linewidth]{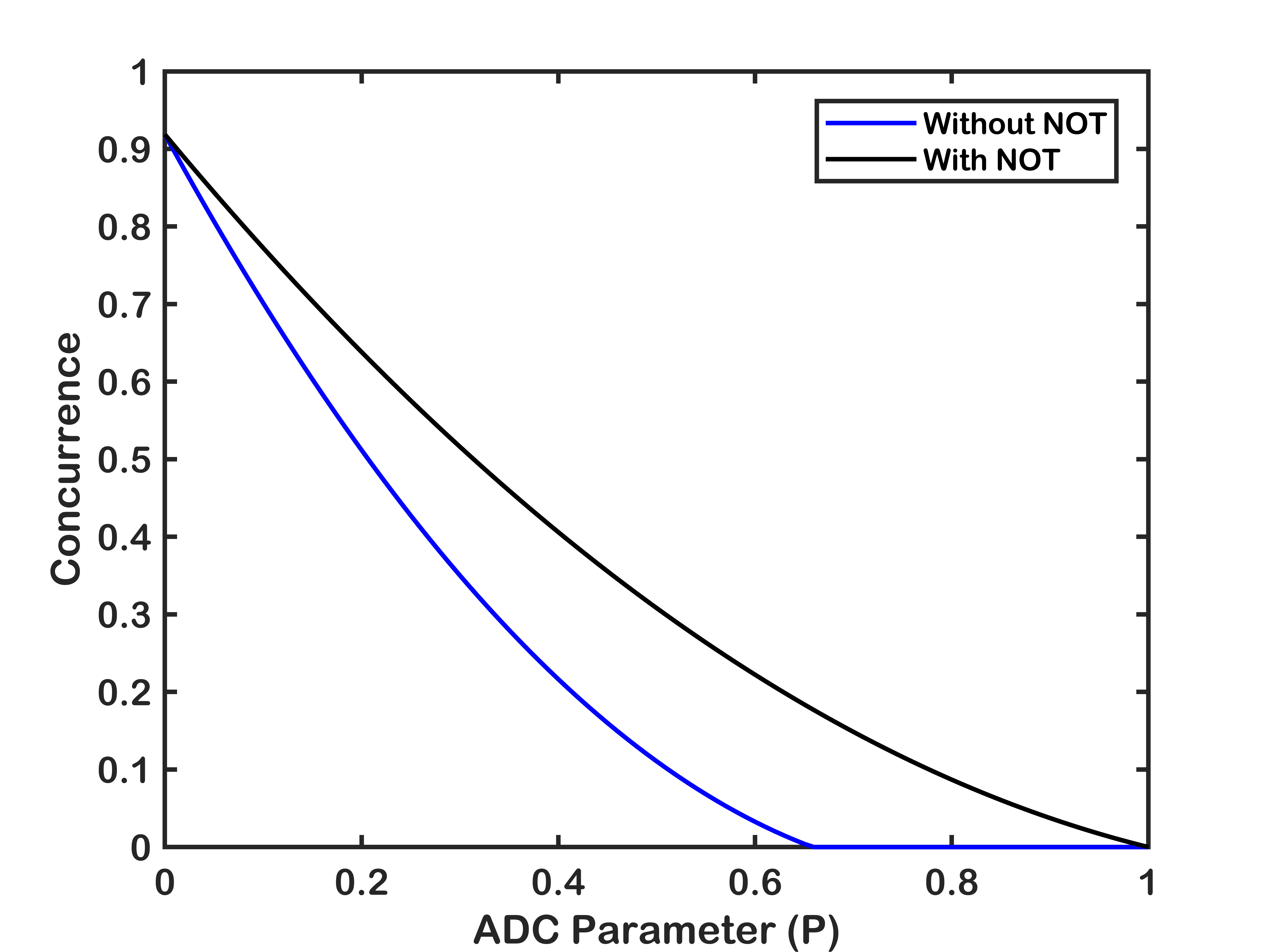}
    \caption{State parameter chosen as $\alpha=0.55$. With the first damping parameter set at zero, evolution through ADC leads to sudden death at 0.64, whereas with NOT operations applied on both qubits, avoidance of ESD is shown.}
    \label{fig:Avoidance_Ideal}
\end{figure}

\begin{figure}
    \centering
    \includegraphics[width=0.5\linewidth]{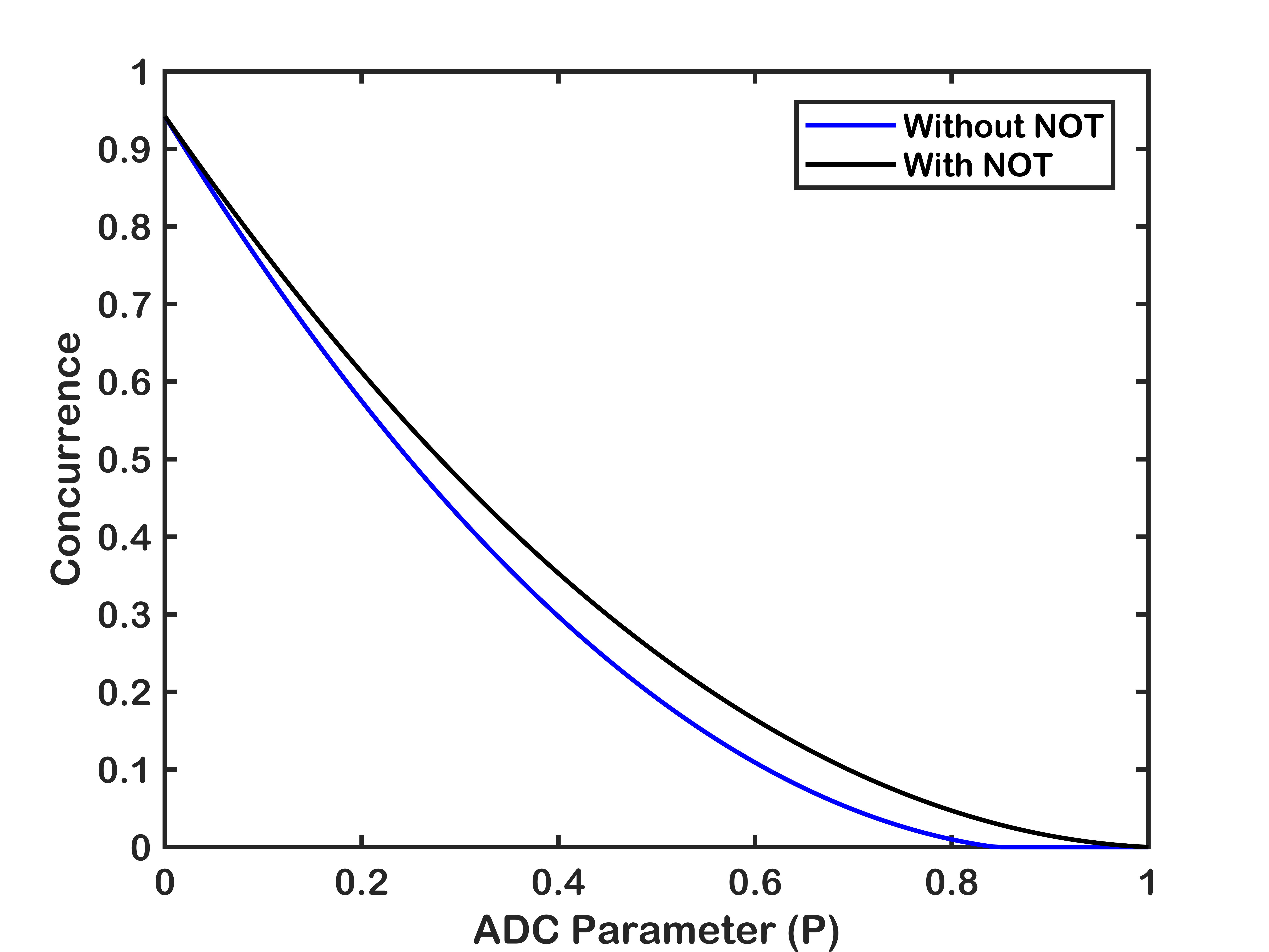}
    \caption{State parameter chosen as $\alpha=0.55$. With first damping parameter was set to 0.22, causing an initial Concurrence of 0.93. Subsequent evolution through ADC leads to sudden death at 0.81, whereas with NOT operations applied on both qubits, the delay of ESD is shown happening at a damping parameter of 0.95.}
    \label{fig:Delay_Ideal}
\end{figure}

\begin{figure}
    \centering
    \includegraphics[width=0.5\linewidth]{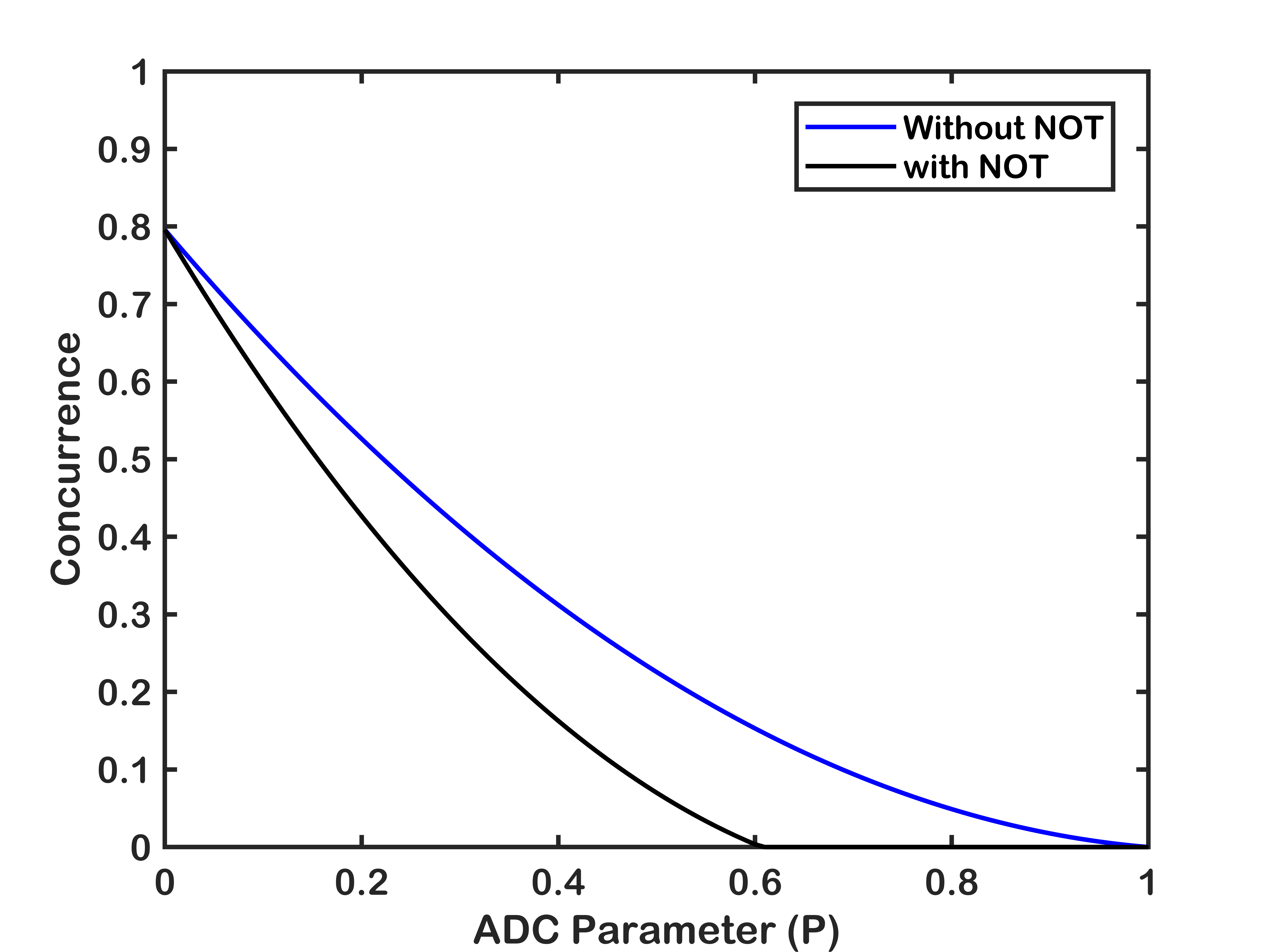}
    \caption{State parameter chosen as $\alpha=0.55$. With first damping parameter was set to 0.43, causing an initial Concurrence of 0.8. Subsequent evolution through ADC leads to sudden death at 0.96, whereas with NOT operations applied on both qubits, the hastening of ESD is shown to be happening at a damping parameter of 0.6.}
    \label{fig:Hastening_Ideal}
\end{figure}

Our protocol involves controlled state evolution through a novel amplitude-damping channel, followed by local unitary NOT operations and finally a normal amplitude-damping channel. The phenomenon of ESD is highly state-dependent\cite{Yu2004,Almeida2007}. Depending on the input state parameter, the evolution through ADC will lead to either asymptotic decay or ESD. Using local unitary operators, we manipulate the entanglement decay trajectory \cite{Rau2008,Singh2017}. Our detailed experimental schematic is shown in Fig. \ref{ESDExptScheme}.
We'll explain the details of the different ADCs we have implemented in the later sections. Using the theoretical framework developed in our paper, for our experimental design, we simulate the evolution of any entangled state in the presence or absence of local unitary operators. Theoretically, we prove the possibility of ESD manipulation as shown in Figures \ref{fig:ADE_Ideal},\ref{fig:Avoidance_Ideal},\ref{fig:Delay_Ideal},\ref{fig:Hastening_Ideal}.
For a state $|\Psi\rangle=\alpha|gg\rangle+\beta|ee\rangle$, where $|g\rangle$ and $|e\rangle$ are ground and excited states, respectively, $\alpha\geq\beta$ leads to asymptotic decay or avoidance of ESD. These characteristics of entanglement dynamics are shown in Fig. \ref{fig:ADE_Ideal}. Similarly, $\alpha < \beta$ leads to ESD during evolution through ADC. In such cases, with careful implementation of the local unitary NOT operation during the evolution can result in hastening, delaying, or the complete avoidance of the onset of ESD. The theoretical simulations for all these manipulation conditions are shown in Fig. \ref{fig:Avoidance_Ideal},\ref{fig:Delay_Ideal},\ref{fig:Hastening_Ideal}.

\subsection{First damping channel}\label{Supp_1stADC}
\begin{figure}
    \centering
    \includegraphics[width=1\linewidth,trim=10 200 250 10,clip]{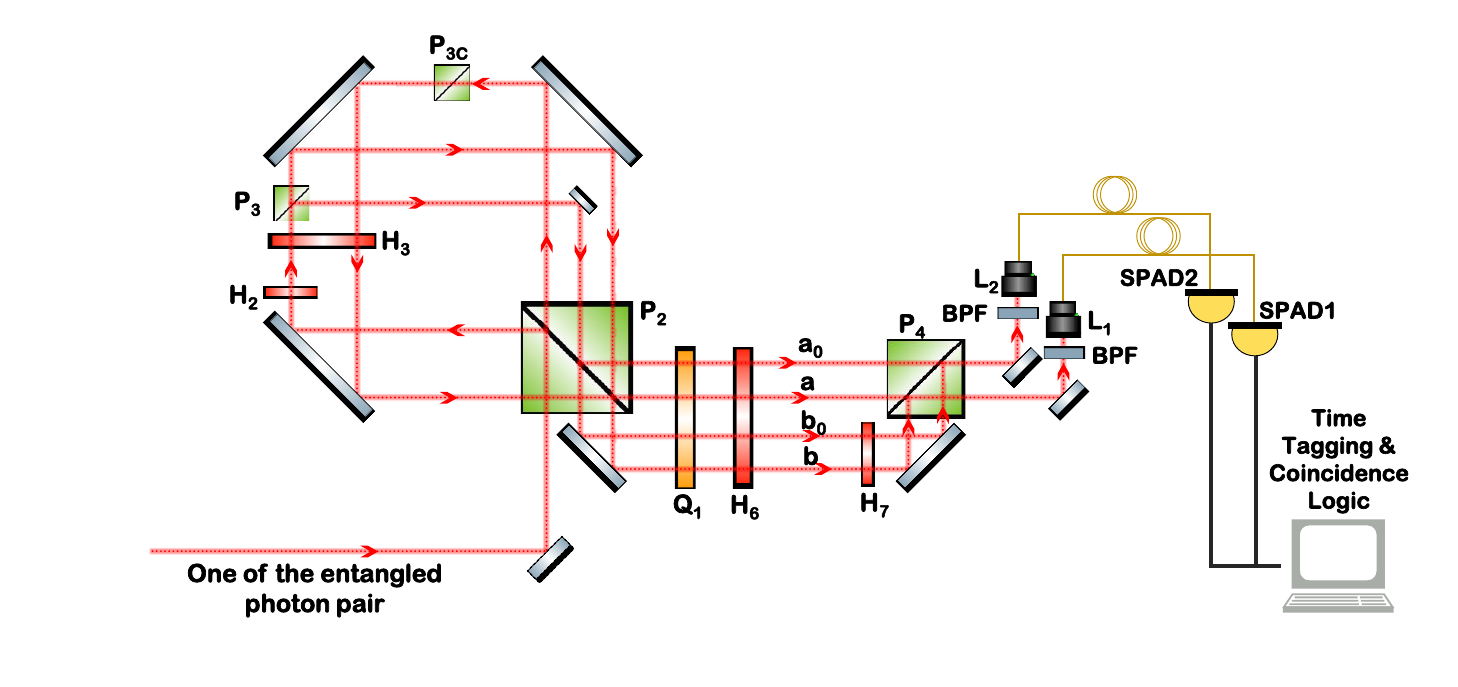}
    \caption{Schematic of the new correlated amplitude damping-like channel. $H_2$, $H_3$, $H_6$, $H_7$ are half-wave plates, $Q_1$ is a quarter-wave plate, $P_2$, $P_3$, $P_{3C}$ and $P_4$ are polarization beam splitters, SPAD: Single Photon Avalanche Diode detectors. $a$, $a_0$, $b$ and $b_0$ are environmental modes.}
    \label{fig:ESDExptADC1}
\end{figure}

In our protocol, we apply a controlled initial damping followed by a NOT operation and another amplitude damping channel (ADC), to manipulate ESD. In Fig. \ref{fig:ESDExptADC1}, we can observe the implementation of the first amplitude damping followed by a NOT operation on one of the subsystems. Within the displaced-Sagnac design PBS $P_2$ followed by HWP $H_2$, create the following mapping
\begin{equation}
\begin{aligned}
|H\rangle_S|a\rangle_R &\rightarrow|H\rangle_S|a\rangle_R,\\
|V\rangle_S|a\rangle_R &\rightarrow\sqrt{1-p}|V\rangle_S|a\rangle_R+\sqrt{p}|H\rangle_S|b\rangle_R;~~ p=\sin^2(2\theta),
\end{aligned}
\label{GenADC}
\end{equation}
where we regard $H$ and $V$ polarization as our ground and excited states, respectively. The channel parameter $p$ is directly related to the rotation angle of the HWP inside the Displaced Sagnac Interferometer (DSI) along the vertical polarization path. In spontaneous decay in a two-level system, the damping strength is given by $p=1-\exp(-\Gamma t)$, where $\Gamma$ is the decay rate from the excited state to the ground state. In a photonic system, the term $\sin^2(2\theta)$ from the HWP rotation represents the same analogous action mimicking this notion of time-dependent damping. The fast axis of the HWP is rotated by an angle $\theta$ with respect to the vertical polarization state of the photon. The spatial modes $|a\rangle$ and $|b\rangle$ serve as the modes of the reservoir.\\
In our experimental design (Fig. \ref{fig:ESDExptADC1}) beyond HWP $H_2$, we introduced HWP $H_3$, which is set at an angle $45^o$, implementing NOT operations on both the qubits. For this action, the waveplate has to act on both the spatial modes (clockwise and counterclockwise) within the DSI. Following this, our design involves PBSs $P_3$ introducing another mode with a path difference compared to the other two modes and $P_{3C}$ compensating for the path delay introduced by P3 to enable coherent recombination of interfering modes in the DSI within the coherence length of photons.\\

Due to the above design, we populate the modes $a_0$ and $b$ outside the interferometer. These modes serve as our environment. This unique design for the experiment lead to the quantum mapping of the bipartite system as,
\begin{equation}
\begin{aligned}
|HH\rangle_S|a\bar{a}\rangle_R &\rightarrow|VV\rangle_S|b\bar{b}\rangle_R,\\
|HV\rangle_S|a\bar{a}\rangle_R &\rightarrow\sqrt{1-p}|VH\rangle_S|b\bar{b}\rangle_R+\bar{X}\sqrt{p}|VV\rangle_s|b\bar{a_0}\rangle_R,\\
|VH\rangle_S|a\bar{a}\rangle_R &\rightarrow\sqrt{1-p}|HV\rangle_S|b\bar{b}\rangle_R+X\sqrt{p}|VV\rangle_s|a_0\bar{b}\rangle_R,\\
|VV\rangle_S|a\bar{a}\rangle_R &\rightarrow(1-p)|HH\rangle_S|b\bar{b}\rangle_R+X\bar{X}p|VV\rangle_S|a_0\bar{a_0}\rangle_R +\bar{X}\sqrt{p(1-p}|HV\rangle_s|b\bar{a_0}\rangle_R +X\sqrt{p(1-p)}|VH\rangle_s|a_0\bar{b}\rangle_R.
\end{aligned}
\label{ADC1Sup}
\end{equation}
Where $X$ is an operator we use to model the temporal mismatch along different spatial modes. $X$ introduces a temporal shift proportional to the path length (discussed in the previous paragraph) acting on the photon amplitudes: {\it i.e.} \begin{equation}
    X: X[x(t)]=x(t+\delta t)~~,
    \label{TempShift}
\end{equation} and similarly for $\bar{X}$ on $\bar{x}$. Consequently, $X$ acts on the ensemble of two-photon states given by
\begin{eqnarray}
|\Psi\rangle = \alpha|HH\rangle - \beta|VV\rangle,
\label{stateSup}
\end{eqnarray}
where, $|\alpha|^2+|\beta|^2=1$ and these state parameters can be tuned experimentally using HWP before the crystals. To explain our results, we generalize Eq. \ref{stateSup} in a time-dependent framework defining $\alpha = x(t)\bar{x}(t)$ and $\beta = y(t)\bar{y}(t)$, subject to $\sum_t \left( |x(t)\bar{x}(t)|^2 + |y(t)\bar{y}(t)|^2 \right) = 1$. $X=\exp(-i\chi)$ is the time-delay operator defined by its action on a function $x(t)\mapsto X x(t) X^\dagger=x(t+\delta t)$. $X$ acts on Eq. \ref{stateSup} as $X\sum_t x(t)\bar{x}(t)=\sum_t x(t+\delta t)\bar{x}(t)$ such that $\sum_t x(t+\delta t)\bar{x}(t)=\sum_t x(t)\bar{x}(t) \ \iff \delta t< \Delta t$ and zero otherwise, where $\Delta t$ is the coincidence window for measurement.
The Kraus operators for the state evolution include $X$ and are of the form,
\begin{equation}
    \begin{aligned}
        \mathbb{K}_{1}&=|VV\rangle\langle HH|+(1-p)|HH\rangle\langle VV| +\sqrt{1-p}|VH\rangle\langle HV|+\sqrt{1-p}|HV\rangle\langle VH|\,, \mathbb{K}_{4}=p|VV\rangle\langle VV|\,,\\
        \mathbb{K}_{2}&=\sqrt{X}\sqrt{p}|VV\rangle\langle HV|+\sqrt{X}\sqrt{p(1-p)}|HV\rangle\langle VV|\,,
        \mathbb{K}_{3}=\sqrt{X}\sqrt{p}|VV\rangle\langle VH|+\sqrt{X}\sqrt{p(1-p)}|VH\rangle\langle VV|\,.
    \end{aligned}
    \label{KrausSup}
\end{equation}

We tested this new ADC under different input conditions. We use highly entangled state as input to the experiment and observe the decay characteristics. (see Fig. \ref{fig:1stADC}) Also, we create a separable state and test the new damping channel with the separable state $|VV\rangle$ as input. In this case, we cannot consider Concurrence as a measure to validate our channel; instead, we observe the variation of purity and compare our experimental results with evolution through both the new ADC and the conventional ADC (see Fig. \ref{fig:Purity_Separable_State}).

\subsection{Second damping channel}\label{Supp_2ndADC}
\begin{figure}
    \centering
    \includegraphics[width=1\linewidth,trim=10 200 250 10,clip]{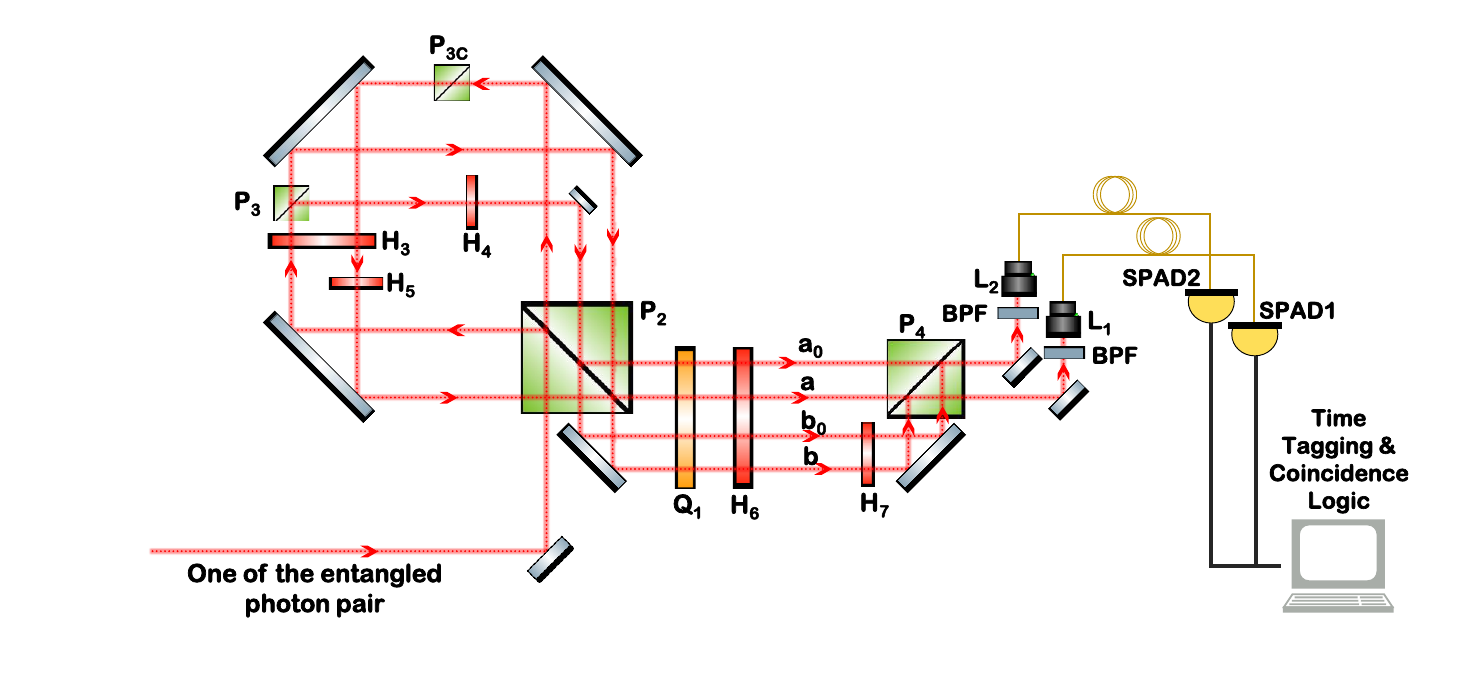}
    \caption{Schematic of the second amplitude damping channel. $H_3$, $H_4$, $H_5$, $H_6$, $H_7$ are half-wave plates, $Q_1$ is a quarter-wave plate, $P_2$, $P_3$, $P_{3C}$ and $P_4$ are polarization beam splitters, SPAD: Single Photon Avalanche Diode detectors. $a$, $a_0$, $b$ and $b_0$ are environmental modes.}
    \label{fig:ESDExptADC2}
\end{figure}

After the application of the NOT operation using the HWP $H_3$, we implement a controlled evolution of the state through a subsequent amplitude damping. We achieve this using the PBSs $P_3$, $P_{3C}$, $P_2$ and HWPs $H_4$, $H_5$. (See Fig. \ref{fig:ESDExptADC2}) To test the second damping channel, we keep the HWP $H_2$ at its fast axis throughout the run of the experiment and rotate $H_4$ and $H_5$ simultaneously by the same angles from zero to $45^o$ angle. From doing the optical evolution, it can be seen that this part of the experimental design exactly creates the quantum mapping described by Eq. \ref{ADCSup}. We can compare the experimental data with the evolution of the initially measured state through the standard amplitude-damping evolution. The Kraus operators for this channel can be derived from the unitary evolution of the system-environment.
The unitary transformation that dictates the system environment evolution through the second damping channel can be written as
\begin{equation}
\begin{aligned}
U_{ADC}=|H\rangle\langle H|_S|a\rangle\langle a|_R + \sqrt{1-p}|V\rangle\langle V|_S|a\rangle\langle a|_R+\sqrt{p}|H\rangle\langle V|_S|b\rangle\langle a|_R~.
\end{aligned}
\label{ADCSup}
\end{equation}
Tracing out the path from Eq. \ref{ADCSup}, we obtain the Kraus operators as,
\begin{equation}
\begin{aligned}
    \mathbb{K}_{1}&=_R\langle a|U_{ADC}|a\rangle_R =|H\rangle\langle H|_S+ \sqrt{1-p}|V\rangle\langle V|_S =\begin{pmatrix}
    1 & 0\\0 & 0
\end{pmatrix}+\sqrt{1-p}\begin{pmatrix}
    0 & 0\\0 & 1
\end{pmatrix} =\begin{pmatrix}
1 & 0\\
0 & \sqrt{1-p}
\end{pmatrix}\\
\mathbb{K}_{2}&=_R\langle b|U_{ADC}|a\rangle_R =\sqrt{p}|H\rangle\langle V|_S =\sqrt{p}\begin{pmatrix}
    0 & 1\\0 & 0
\end{pmatrix} =\begin{pmatrix}
0 & \sqrt{p}\\
0 & 0
\end{pmatrix}
\end{aligned}
\label{KrausADCSup}
\end{equation}
Using these Kraus operators, the density matrix evolution can be explained as
\begin{equation}
\begin{aligned}
\rho_{out} &=\sum_{i,j}\mathbb{K}_{ij}\rho_{in}\mathbb{K}^\dagger_{ij}~,
\label{Rho_ADC}
\end{aligned}
\end{equation}
The two-qubit Kraus operators are $\mathbb{K}_{ij}=\mathbb{K}_i\otimes \mathbb{K}_j~;~~i,j=1,2$, with $\mathbb{K}$ from Eq. \ref{KrausADCSup}.

\subsection{Additional Data}
\label{Sup: Data}
Here, we discuss additional experimental data validating the independent operation of the damping channels and the controlled manipulation of ESD via a local NOT operation. We implement the protocol for additional input states characterized by $\alpha=0.45$ and $\alpha=0.5$ to demonstrate the robustness of the protocol for various input states and entanglement regimes.

\subsection*{Characterization of the First ADC}

\begin{figure}[h]
    \centering
    \includegraphics[width=0.6\linewidth]{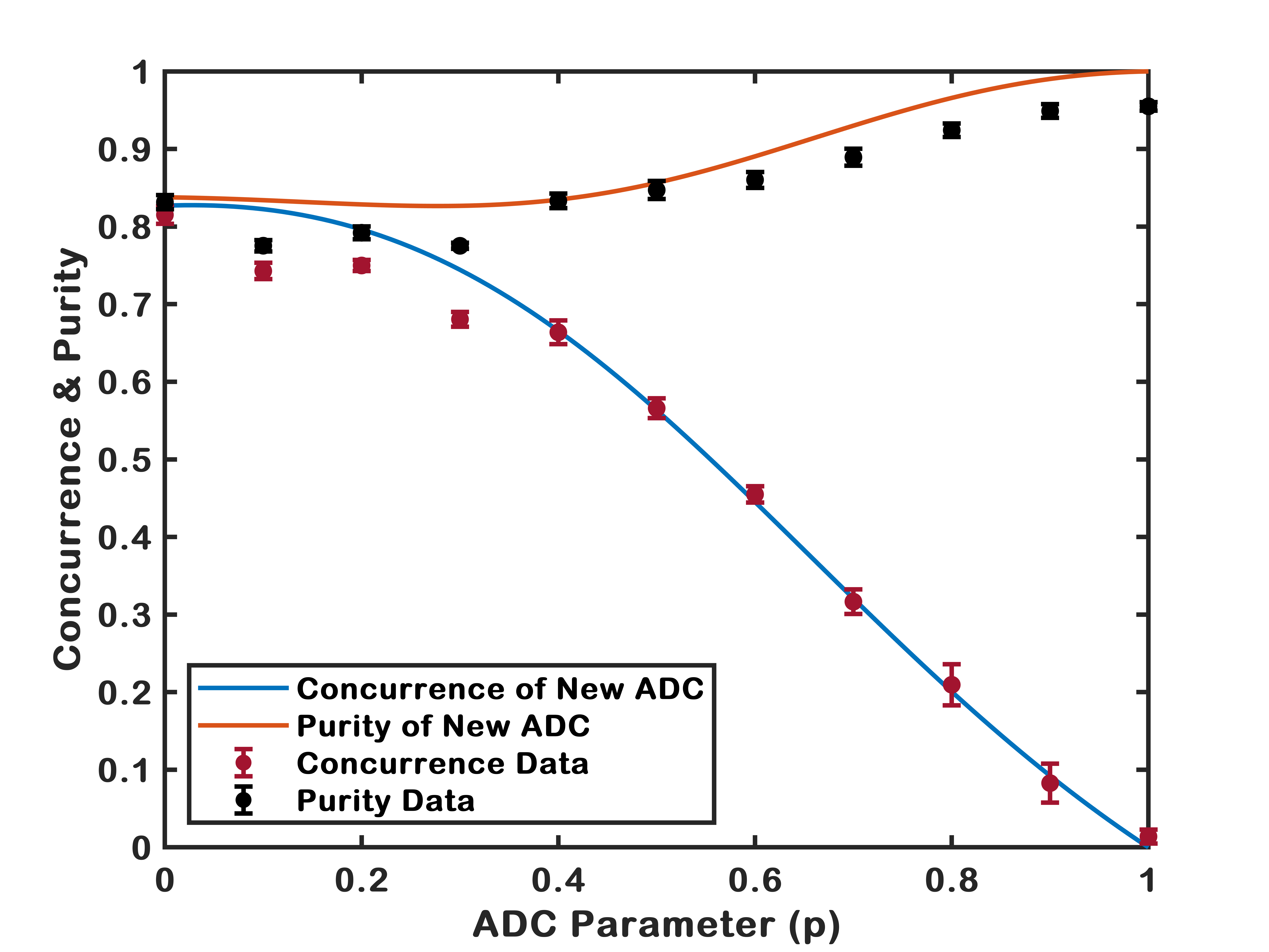}
    \caption{Characterization of the first amplitude damping channel with an entangled state as input. The initial state is measured at $p=0$ (HWP $H_2$ aligned vertically). HWPs $H_4$ and $H_5$ were kept at zero, disabling the second ADC throughout the experiment, while $p$ was varied from $0$ to $1$. The NOT operation was implemented with HWP $H_3$.}
    \label{fig:1stADC}
\end{figure}

Figure~\ref{fig:1stADC} shows the characterization of the first ADC using an entangled input state. The second damping channel was turned off by keeping HWPs $H_4$ and $H_5$ aligned with the vertical axis, ensuring that only the first ADC contributes to the evolution. The damping parameter $p$ was varied continuously from $0$ to $1$ by rotating HWP $H_2$.

The concurrence exhibits a decay with increasing $p$, consistent with the expected correlated amplitude-damping-like behavior. The experimental data closely follow the theoretical curve, with deviations remaining within the systematic uncertainty discussed in Sec.~\ref{errorc}. These deviations primarily arise from finite PBS extinction ratios and small waveplate misalignments.

\subsection*{First ADC with Separable Input}

\begin{figure}[h]
    \centering
    \includegraphics[width=0.6\linewidth]{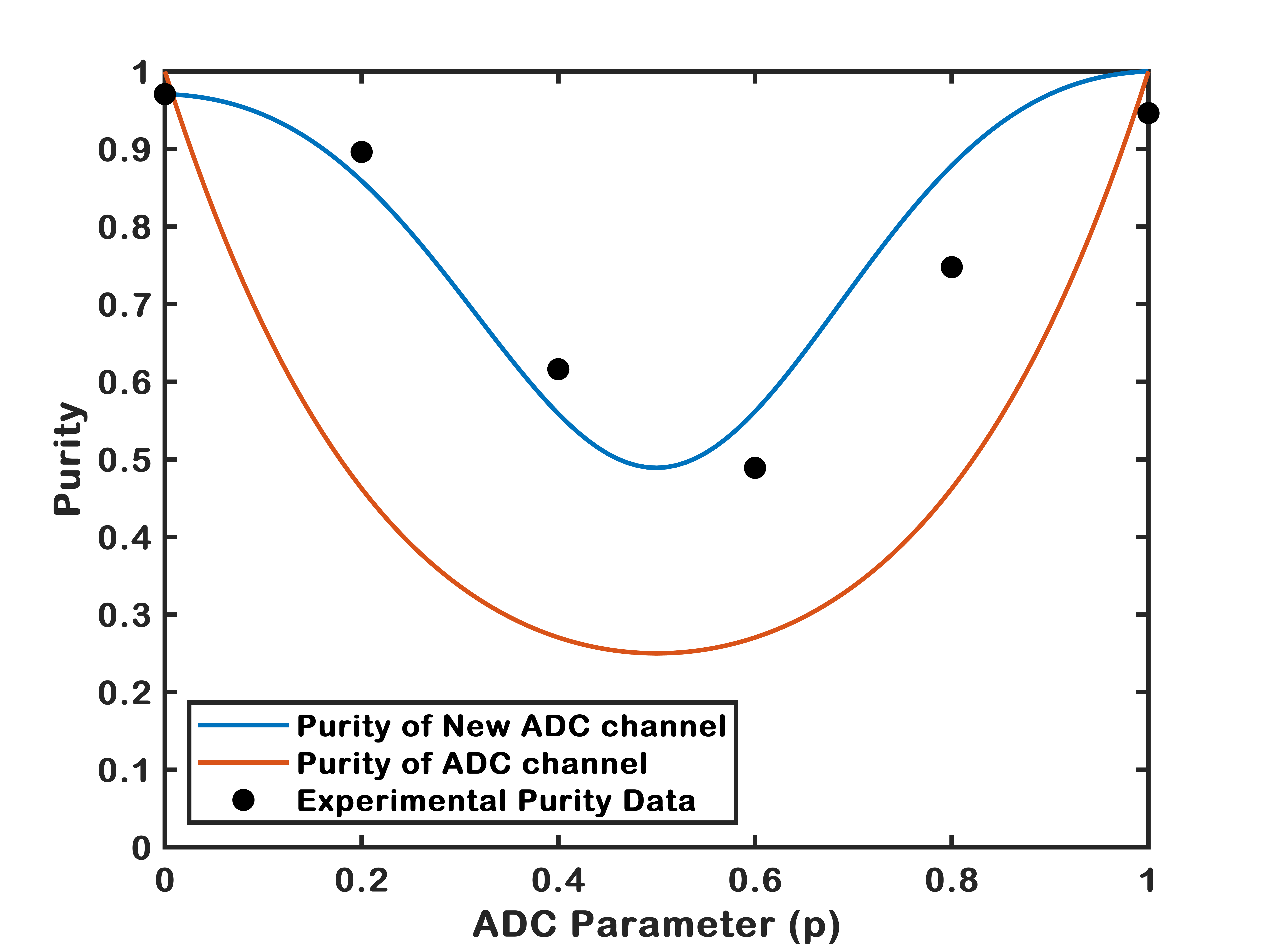}
    \caption{Characterization of the first amplitude damping channel with separable state $|VV\rangle$ as input. The initial state was measured at $p=0$ (HWP $H_2$ aligned vertically). HWPs $H_4$ and $H_5$ were kept at zero, disabling the second ADC throughout the experiment, while $p$ was varied from $0$ to $1$. The NOT operation was implemented with HWP $H_3$.}
    \label{fig:Purity_Separable_State}
\end{figure}

To further validate our first ADC, we repeated the measurement using the separable input state $|VV\rangle$, as shown in Fig.~\ref{fig:Purity_Separable_State}. Since $|VV\rangle$ is a separable state, no entanglement dynamics are expected.

Purity was taken as the measure in this case, and it followed the trajectory for evolution through a $\rm CADC-like$ environment, confirming and validating the implemented CADC-like damping process. This serves as an important consistency check for the physical implementation of the channel.

\subsection*{Characterization of the Second ADC}

\begin{figure}[h]
    \centering
    \includegraphics[width=0.6\linewidth]{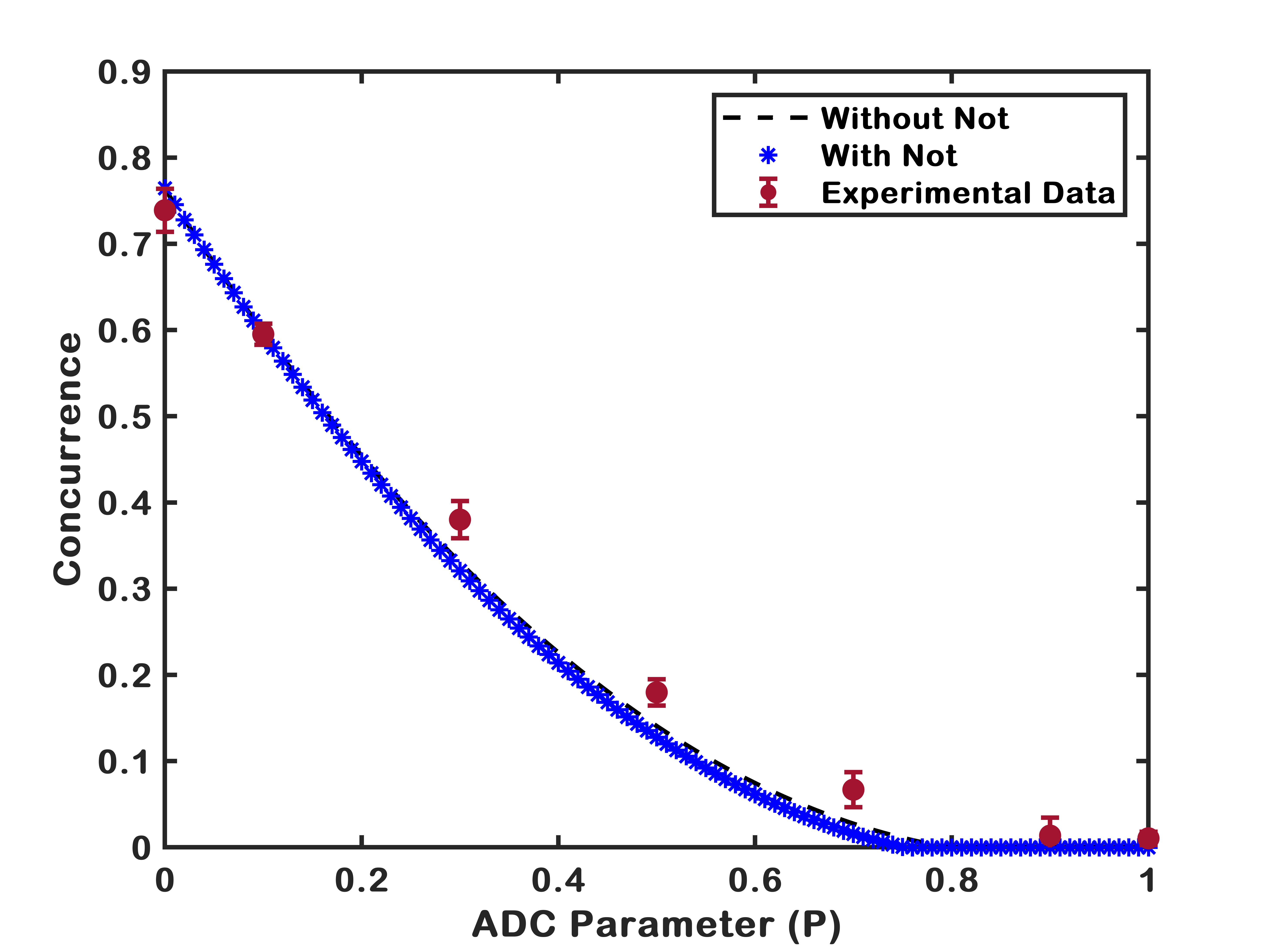}
    \caption{Characterization of the second ADC with a highly entangled state as input. The initial state was measured at $P=0$ (HWPs $H_4$ and $H_5$ aligned vertically). HWP $H_2$ was kept at zero, disabling the first damping channel throughout the experiment, while $P$ was varied from $0$ to $1$. The NOT operation was implemented with HWP $H_3$.}
    \label{fig:ADE_Old}
\end{figure}

The second ADC was independently characterized by disabling the first damping module (keeping $H_2$ at zero) and varying the damping parameter $P$ through HWPs $H_4$ and $H_5$. The results are shown in Fig.~\ref{fig:ADE_Old}.

A controlled reduction of the concurrence is observed as $P$ increases, confirming that the second ADC faithfully implements the amplitude damping map. Thus, independently both channels are experimentally verified.

\subsection*{Avoidance of ESD}

\begin{figure}[h]
    \centering
    \includegraphics[width=0.6\linewidth]{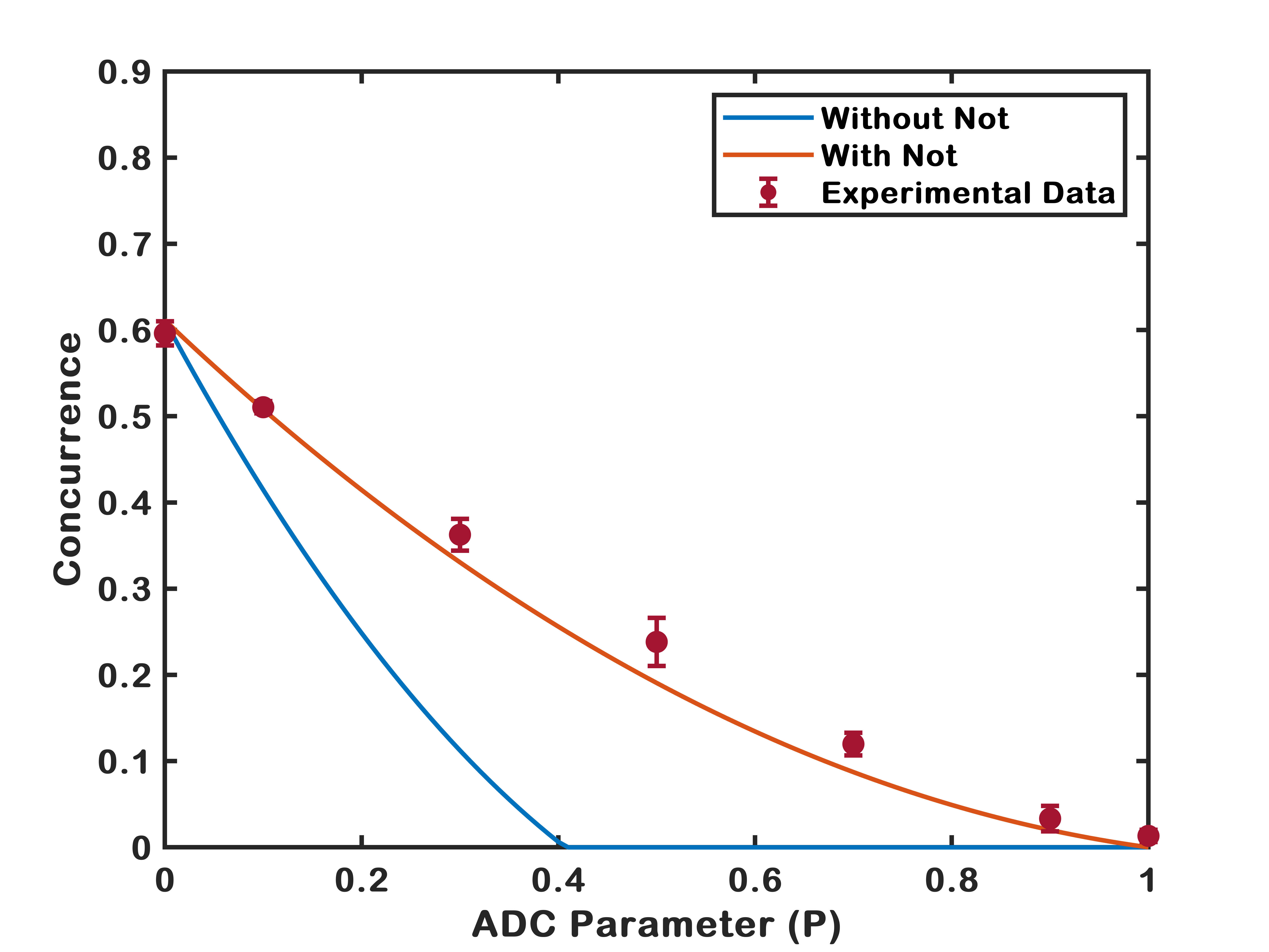}
    \caption{Avoidance of ESD for $\alpha=0.45$. Without the NOT operation, ESD is expected near $P=0.4$. With the NOT implemented (red solid line), Concurrence remains finite. Red data points with error bars represent experimental measurements.}
    \label{fig:Avoidance_Old}
\end{figure}

Figure~\ref{fig:Avoidance_Old} demonstrates the avoidance of entanglement sudden death for $\alpha=0.45$. Due to the relatively low initial concurrence, theory predicts ESD at $P\approx 0.4$ without intervention.

When the local NOT operation (HWP $H_3$) is applied, the concurrence remains finite throughout the damping range explored. The red solid curve represents the theoretical prediction with the NOT operation, and the experimental data (red points with error bars) show strong agreement. This confirms that the local unitary redistributes the populations and prevents complete entanglement collapse at a finite time leading to an avoidance of ESD.

\subsection*{Delay of ESD}

\begin{figure}[h]
    \centering
    \includegraphics[width=0.6\linewidth]{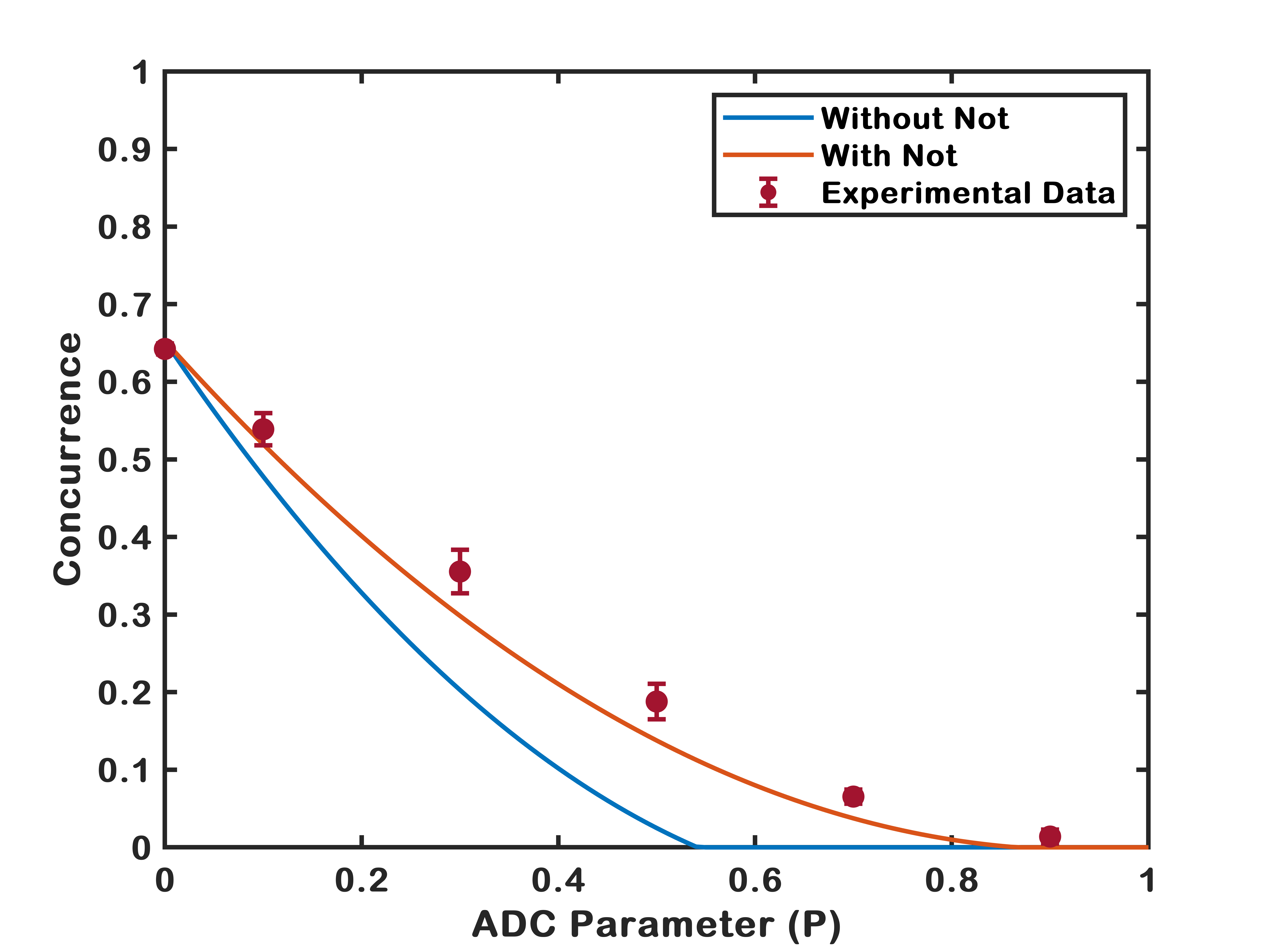}
    \caption{Delay of ESD for $\alpha=0.5$. Without the NOT operation, ESD is expected near $P=0.54$. With the NOT implemented (red solid line), Concurrence decays to zero near $P=0.82$. Experimental data (red points with error bars) show zero Concurrence around $P\approx 0.9$.}
    \label{fig:Delay_Old}
\end{figure}

Figure~\ref{fig:Delay_Old} illustrates the delay of ESD for $\alpha=0.5$. With an initial measured Concurrence of $0.65$, ESD is predicted to occur near $P\approx 0.54$ without control.

With the NOT operation implemented, the theoretical prediction (red curve) shows the concurrence vanishing only near $P=0.82$. Experimentally, the concurrence approaches zero around $P\approx 0.9$, demonstrating a substantial extension of entanglement lifetime.

These supplementary measurements collectively confirm:
\begin{itemize}
    \item Validation for the $\rm CADC
    $ evolution,
    \item Faithful implementation of the amplitude damping map,
    \item Independent characterization of both channels,
    \item Controlled avoidance and delay of ESD via a local NOT operation.
\end{itemize}

All deviations from theoretical predictions remain within the estimated systematic uncertainty of $0.06$--$0.10$ in Concurrence, demonstrating the robustness of the protocol against realistic optical imperfections.

\section{Systematic Error Analysis}\label{errorc}

We now analyze the impact of systematic imperfections arising from non-ideal optical components. The essential computation is reported in the appendix of the main text. We present the additional mathematical details in this section for completion. The total unitary evolution in the presence of such imperfections is

\begin{equation}
U_T = U^A \otimes U^B ,
\end{equation}

where $U^A$ and $U^B$ describe the unitary evolution of the two qubits through the respective DSIs $A$ and $B$. 
For subsystem $A$,

\begin{equation}
U^A = P_1^\dagger \cdot H_2(q)\cdot P_2 \cdot U_{NOT}\cdot H_1(p)\cdot P_1 .
\end{equation}

Here $H_1$ and $H_2$ implement successive amplitude damping channels (ADCs), while $P_1$ and $P_2$ represent multi-port PBS transformations including finite extinction ratios.

The composite transformation
\begin{align}
\begin{split}
H_2(q)\cdot P_2\cdot U_{NOT}\cdot H_1(p)\cdot P_1 =& (1-\delta')\Big[f_+\big((1-\delta_2)|H_4\rangle + X\delta_2 H(\phi)|H_5\rangle\big)+ g_-\big(X(1-\delta_2')H(\phi)|V_5\rangle + \delta_2'|V_4\rangle\big)\Big]\langle V_0| \\
&+ \delta \Big[g_+\big((1-\delta_2)|H_4\rangle + X\delta_2 H(\phi)|H_5\rangle\big) + f_-\big(X(1-\delta_2')H(\phi)|V_5\rangle + \delta_2'|V_4\rangle\big)\Big]\langle H_0| \\
&+ (1-\delta)H(\phi)(|V_2\rangle - \mu |H_2\rangle)\langle H_0|+ \delta'H(\phi)(|H_2\rangle + \mu |V_2\rangle)\langle V_0| ,
\end{split}
\end{align}
with
\begin{equation}
f_\pm = \cos 2\theta \pm \mu \sin 2\theta , 
\qquad 
g_\pm = \sin 2\theta \pm \mu \cos 2\theta .
\end{equation}
The waveplate $H(\phi)$ acts on polarization in any path $a$ as
\begin{equation}
H(\phi)|H_a\rangle = -\cos2\phi |H_a\rangle + \sin2\phi |V_a\rangle ,
\end{equation}
\begin{equation}
H(\phi)|V_a\rangle = \sin2\phi |H_a\rangle + \cos2\phi |V_a\rangle .
\end{equation}
We define
\begin{equation}
U_3 = H_2(q)\cdot P_2 \cdot U_{NOT}\cdot H_1(p)\cdot P_1 
= \big(M_2|2\rangle + M_4|4\rangle + M_5|5\rangle\big)\langle H_0| 
+ \big(N_2|2\rangle + N_4|4\rangle + N_5|5\rangle\big)\langle V_0| ,
\end{equation}
where
\begin{align}
\begin{split}
M_2 &= (1-\delta)\Big[(\cos2\phi - \mu\sin2\phi)|V\rangle 
+ (\sin2\phi + \mu\cos2\phi)|H\rangle\Big]\,,
M_4 = \delta\left(g_+(1-\delta_2)|H\rangle + f_- \delta_2'|V\rangle\right), \\
M_5 &= X\delta\Big[(f_-(1-\delta_2')\sin2\phi - g_+\delta_2\cos2\phi)|H\rangle+ (f_-(1-\delta_2')\cos2\phi + g_+\delta_2\sin2\phi)|V\rangle\Big], \\
N_2 &= \delta'\Big[(\sin2\phi + \mu\cos2\phi)|V\rangle 
- (\cos2\phi - \mu\sin2\phi)|H\rangle\Big]\,,
N_4= (1-\delta')\left(f_+(1-\delta_2)|H\rangle + g_- \delta_2'|V\rangle\right), \\
N_5 &= X(1-\delta')\Big[(g_-(1-\delta_2')\sin2\phi - f_+\delta_2\cos2\phi)|H\rangle + (g_-(1-\delta_2')\cos2\phi + f_+\delta_2\sin2\phi)|V\rangle\Big].
\end{split}
\end{align}

Using
\begin{equation}
P_1^\dagger = F^2_{ab}\langle 2| + F^4_{ab}\langle 4| + F^5_{a'b'}\langle 5| ,
\end{equation}

we obtain $U^A = |F\rangle\langle H_0| + |G\rangle\langle V_0|$, with
\begin{equation}
|F\rangle = F^2_{ab}M_2 + F^4_{ab}M_4 + F^5_{a'b'}M_5 ,
\qquad
|G\rangle = F^2_{ab}N_2 + F^4_{ab}N_4 + F^5_{a'b'}N_5 .
\end{equation}

The explicit forms are

\begin{align}
\begin{split}
|F\rangle &= \Big((1-\delta)(1-\delta_1)(\sin2\phi+\mu\cos2\phi) 
+ \delta^2(1-\delta_2)g_+\Big)|H_a\rangle \\
&+ \Big((1-\delta)\delta_1(\sin2\phi+\mu\cos2\phi) 
+ \delta(1-\delta)(1-\delta_2)g_+\Big)|H_b\rangle \\
&+ \Big((1-\delta)\delta_1'(\cos2\phi+\mu\sin2\phi) 
+ \delta\delta_2'(1-\delta')f_-\Big)|V_a\rangle \\
&+ \Big((1-\delta)(1-\delta_1')(\cos2\phi-\mu\sin2\phi) 
+ \delta\delta'\delta_2'f_-\Big)|V_b\rangle \\
&+ X\delta^2\big(f_-(1-\delta_2')\sin2\phi - g_+\delta_2\cos2\phi\big)|H_{a'}\rangle \\
&+ X\delta(1-\delta)\big(f_-(1-\delta_2')\sin2\phi - g_+\delta_2\cos2\phi\big)|H_{b'}\rangle \\
&+ X\delta(1-\delta')\big(f_-(1-\delta_2')\cos2\phi + g_+\delta_2\sin2\phi\big)|V_{a'}\rangle \\
&+ X\delta\delta'\big(f_-(1-\delta_2')\cos2\phi + g_+\delta_2\sin2\phi\big)|V_{b'}\rangle ,
\end{split}
\end{align}

and

\begin{align}
\begin{split}
|G\rangle &= \delta'(\sin2\phi+\mu\cos2\phi)
\big(\delta_1'|V_a\rangle + (1-\delta_1')|V_b\rangle\big) \\
&- \delta'(\cos2\phi-\mu\sin2\phi)
\big((1-\delta_1)|H_a\rangle + \delta_1|H_b\rangle\big) \\
&+ (1-\delta')f_+(1-\delta_2)
\big(\delta|H_a\rangle + (1-\delta)|H_b\rangle\big) \\
&+ (1-\delta')g_-\delta_2'
\big((1-\delta')|V_a\rangle + \delta'|V_b\rangle\big) \\
&+ X(1-\delta')
\big(g_-(1-\delta_2')\sin2\phi - f_+\delta_2\cos2\phi\big)
\big(\delta|H_{a'}\rangle + (1-\delta)|H_{b'}\rangle\big) \\
&+ X(1-\delta')
\big(g_-(1-\delta_2')\cos2\phi + f_+\delta_2\sin2\phi\big)
\big((1-\delta')|V_{a'}\rangle + \delta'|V_{b'}\rangle\big) .
\end{split}
\end{align}

Neglecting contributions from input path $|1\rangle$, the total unitary becomes

\begin{equation}
U_T = U^A \otimes U^B 
= |GG\rangle\langle V_0V_0| 
+ |FF\rangle\langle H_0H_0| 
+ |FG\rangle\langle H_0V_0| 
+ |GF\rangle\langle V_0H_0| .
\end{equation}

Applying this to the input path $|00\rangle$ yields

\begin{equation}
\widetilde{U} = |FF\rangle\langle HH| 
+ |GG\rangle\langle VV| 
+ |GF\rangle\langle VH| 
+ |FG\rangle\langle HV| ,
\end{equation}

which can be expanded in terms of Kraus operators as

\begin{equation}
\widetilde{U} = \sum_{ij} \mathbb{K}_{ij} |ij\rangle ,
\qquad |ij\rangle \in \text{output paths}, 
\quad (i,j)\in\{a,b,a',b'\}.
\end{equation}

\subsection*{Dominant Systematic Error Contributions}

1. PBS extinction ratio:  
Finite extinction ratios introduce leakage of the wrong polarization:

\begin{align}
P_{\text{real}} &= (1-\delta)|H_{\text{trans}}\rangle\langle H| 
+ \delta|H_{\text{refl}}\rangle\langle H| \nonumber \\
&+ \delta'|V_{\text{trans}}\rangle\langle V| 
+ (1-\delta')|V_{\text{refl}}\rangle\langle V|.
\end{align}






2. HWP angle uncertainty:
For mount precision $\delta\theta$ and 
$p=\sin^2(2\theta)$,

\begin{equation}
\delta p = 2\sin(4\theta)\delta\theta.
\end{equation}



3. Path mismatch:
Partial distinguishability reduces interference visibility

\begin{equation}
V = \frac{I_{\max}-I_{\min}}{I_{\max}+I_{\min}},
\end{equation}

leading to a loss in purity of the reconstruted state and hence lower concurrence.




Total Systematic Deviation: Combining all contributions produces an experimentally observed $\sim 1$--$14\%$ deviation from the ideal theoretical prediction. 

The origin of this discrepancy, as mentioned earlier in this section, can be traced to the non-ideal characteristics of the optical components—such as finite PBS extinction ratios, waveplate rotation uncertainties, and residual path distinguishability, which have been carefully characterized and analyzed in the following. 

\begin{figure}[h]
\centering

\subfloat[$\delta_i \sim 10^{-3}$]{
\includegraphics[scale=0.65]{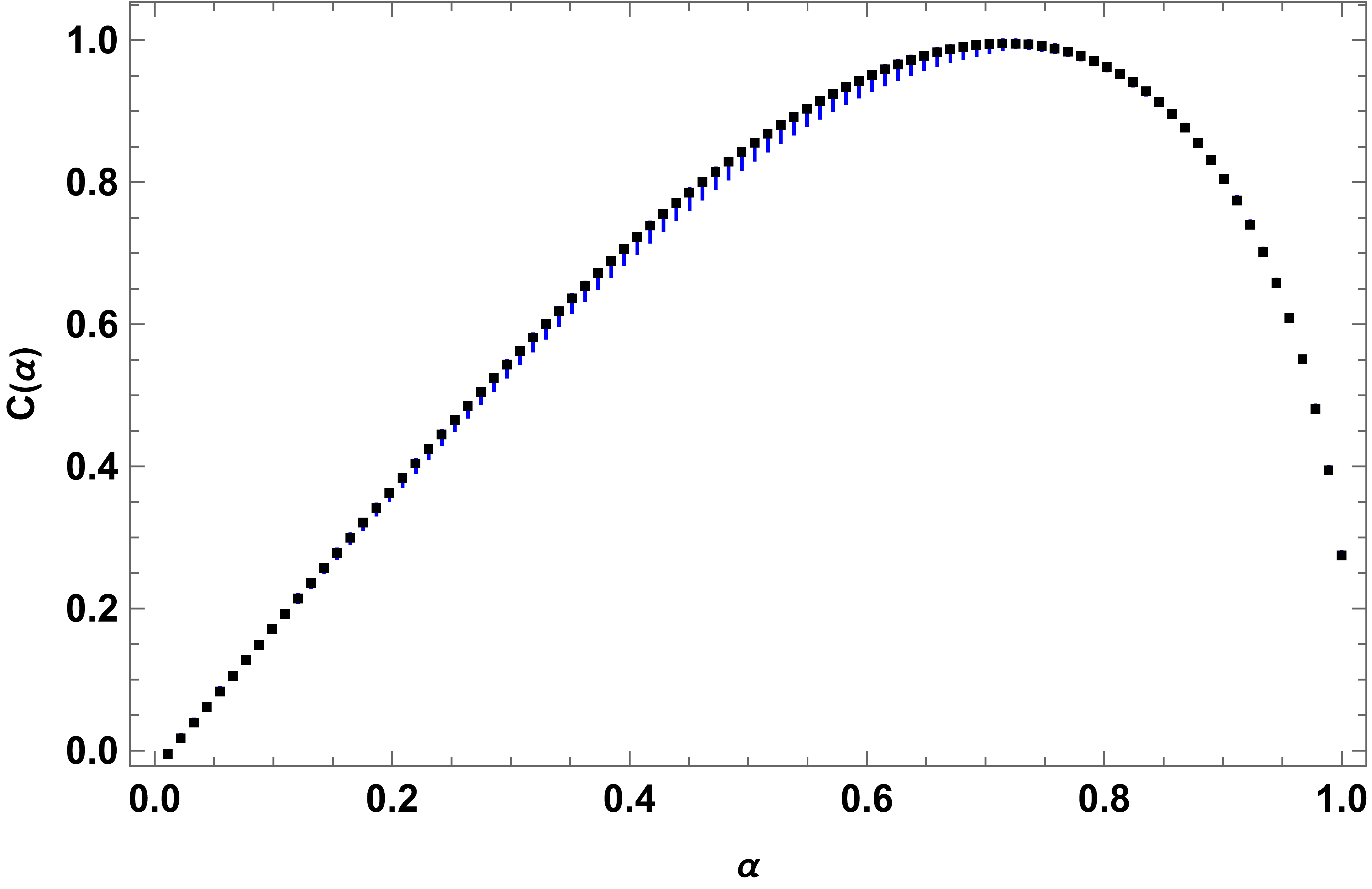}
\label{fig:image2a}
}
\hfill
\subfloat[$\delta_i \sim 10^{-2}$]{
\includegraphics[scale=0.65]{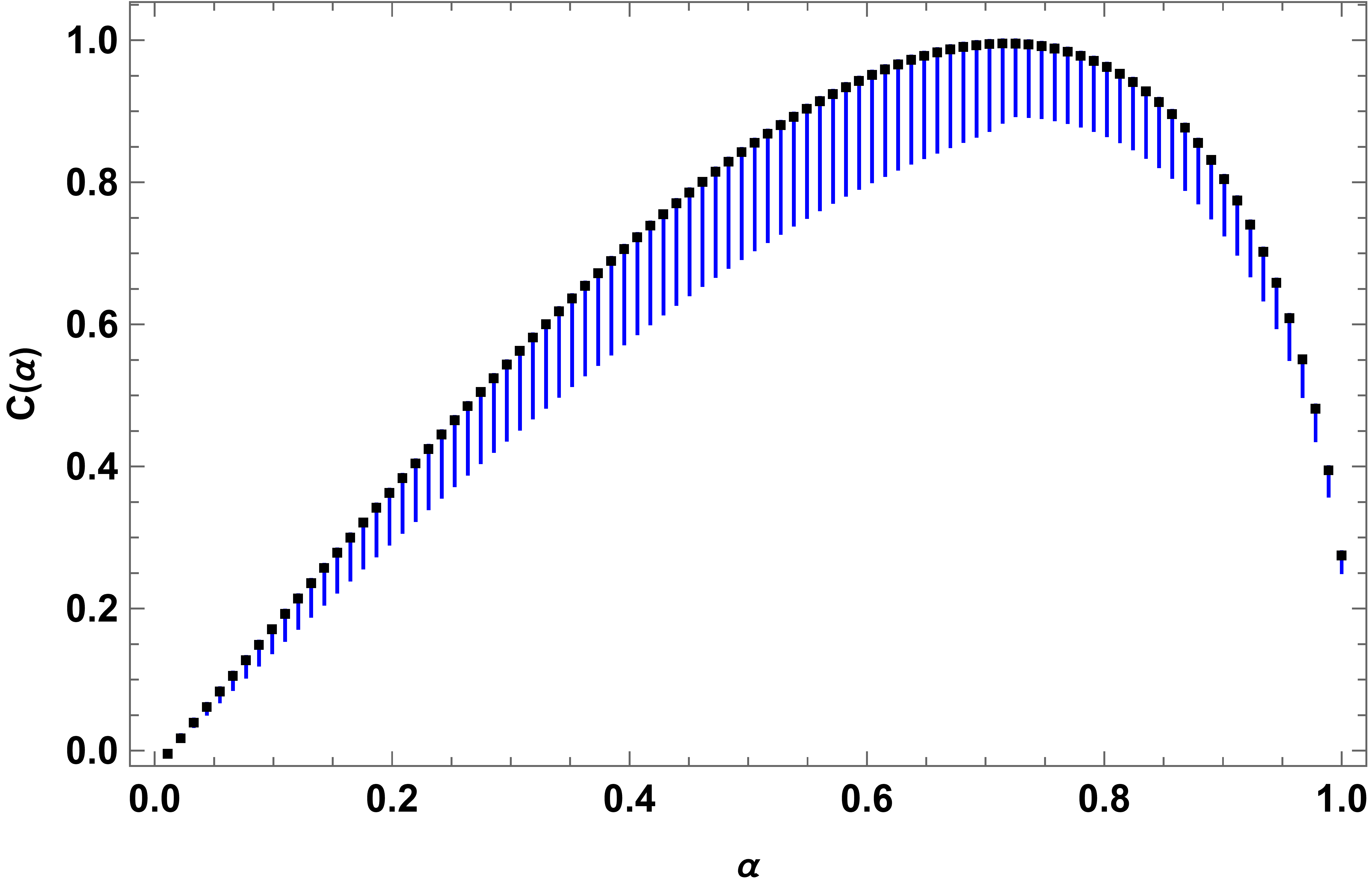}
\label{fig:image2b}
}

\caption{Concurrence of the initial state $|\psi\rangle=\alpha|HH\rangle+\sqrt{1-\alpha^2}|VV\rangle$ as a function of $\alpha$ with varying inaccuracies of the components of the setup.}
\label{fig:image2}

\end{figure}

In simulations shown in the figures. \ref{fig:ADE_Ideal}, \ref{fig:Avoidance_Ideal},\ref{fig:Delay_Ideal},\ref{fig:Hastening_Ideal} we have considered ideal instruments, meaning that there is no imperfection or loss. Hence, we see very high initial entanglement and ideal evolution of the state as we assume a lossless system. However, it is important to incorporate device imperfections in order to illustrate practical scenarios. For example, in our experimental data explained in figures \ref{fig:ADE_Old},\ref{fig:Avoidance_Old},\ref{fig:Delay_Old}, we can clearly observe the initially measured entanglement to be low and the concurrence measurement having error bars. These errors come from non-ideal optical components used in the experiments, such as waveplates, polarization beam splitters, mirrors, etc. For example, the PBS can have a slight error in the transmission of polarizations at the output ports. Similarly, the wave plates can introduce systematic errors depending on the resolution of the rotation mounts they are mounted on. In the present section, we will incorporate these aspects to analyze the error in Concurrence computation. To begin with, an imperfect PBS will have the following form
\begin{equation}
P_1 = \left(\begin{array}{cccc}
   1-\delta  & 0 & \delta & 0  \\
    0 & \delta' & 0 & 1-\delta' \\
    \delta_1 & 0 & 1-\delta_1 & 0 \\
    0 & 1-\delta'_1 & 0 & \delta'_1
\end{array}\right)\,.
\end{equation}
Note that the errors are port-dependent, which is a manifestation of uneven coating in the material. This could also be generalized to polarization dependence; however, for illustration purposes, the above form suffices. We will not be assuming the errors introduced by the waveplates separately, since,
\begin{equation}\label{delpq}
\Delta C(p,q)= \frac{\partial C}{\partial p}\delta p +\frac{\partial C}{\partial q}\delta q\,,
\end{equation}
where $p$ and $q$ are two damping parameters directly related to the waveplate rotation angles.
Going back to the PBS, the contribution of the imperfections, with input at port $|0\rangle$ or $|0'\rangle$ for the setup, is then given by the matrix,
\begin{equation}
P_1= (1-\delta)|H_2\rangle\langle H_0|+(1-\delta')|V_3\rangle\langle V_0|+\delta|H_3\rangle\langle H_0|+\delta'|V_2\rangle\langle V_0|+\dots\,.
\end{equation}
where we are assuming $P_1$ to be the first PBS of the DSI. The modes $|0\rangle$, $|2\rangle$ and $|3\rangle$ are the input mode to the interferometer, counterclockwise and clockwise rotating spatial modes within the interferometer, respectively.
The error coming from the NOT plate is,
\begin{equation}
\text{Err}_{NOT} = \frac{\partial}{\partial\theta}\left(\begin{array}{cc}
  \cos2\theta   & \sin2\theta  \\
    \sin2\theta & -\cos2\theta 
\end{array}\right)_{\theta=\pi/4}\delta\theta=-2\sigma_z \delta\theta\,.
\end{equation}
The NOT operator, including the error, is given by,
\begin{equation}
U_{NOT}=\left(\sigma_x-\mu \sigma_z\right)\otimes \left(|2\rangle\langle 2|+|3\rangle\langle 3|\right)\,, \ \ \mu=\delta\theta\,.
\end{equation}
Similarly for $P_2$, we can write,
\begin{equation}
P_2= (1-\delta_2)|H_4\rangle\langle H_3|+X(1-\delta_2')|V_5\rangle\langle V_3|+X\delta_2 |H_5\rangle\langle H_3|+ \delta'_2 |V_4\rangle\langle V_3|+\dots\,,
\end{equation}
where $P_2$ is the second PBS within the interferometer, and subscripts $4$ and $5$ correspond to the two output modes of PBS $P_2$. The operator $X$ is explained in Eq. \ref{TempShift}. $\delta_2$ and $\delta_2'$ are the extinction ratios of PBS $P_2$.
Further, the operator $P_1^\dagger$ is given by,
\begin{align}
\begin{split}
P_1^\dagger = &(1-\delta_1)|H_a\rangle\langle H_2|+\delta_1|H_b\rangle\langle H_2|+(1-\delta_1')|V_b\rangle\langle V_2|+\delta_1'|V_a\rangle\langle V_2|+(1-\delta)|H_b\rangle\langle H_4|+\delta|H_a\rangle\langle H_4|\\
&+(1-\delta')|V_{a'}\rangle\langle V_5|+\delta'|V_{b'}\rangle\langle V_5|+ (1-\delta)|H_{b'}\rangle\langle H_5|+\delta |H_{a'}\rangle\langle H_5|+(1-\delta')|V_a\rangle\langle V_4|+\delta'|V_b\rangle\langle V_4|\,,
\end{split}
\end{align}
where $a$, $a'$, $b$ and $b'$ are the output modes of the DSI and also represent the environmental modes. Using the Kraus operator derivation in Section \ref{errorc}, we can expand the output state as
\begin{equation}\label{rho_evolution}
\rho_{out}=\sum_i \mathbb{K}_i\rho_{in}\mathbb{K}_i^\dagger\,, \ \ \text{subject to :}\ \ \sum_i \mathbb{K}_i^\dagger\mathbb{K}_i=\mathbb{I}\,.
\end{equation}
to linear order in the fluctuations, $\mathbb{K}_i=\mathbb{K}_i^0+\delta \mathbb{K}_i$, so that,
\begin{equation}
\rho_{out}=\rho_{out}^0+\delta\rho_{out}\,,\ \delta\rho_{out}=\sum_i\left(\mathbb{K}_i^0\rho_{in}\delta\mathbb{K}_i^\dagger+\delta\mathbb{K}_i\rho_{in}\mathbb{K}_i^{0\dagger}\right)\,.
\end{equation}
Enforcing the eigenvalue equation on $\widetilde{\rho_{out}}=\rho^0_{out}(\sigma_y\otimes\sigma_y)\rho^0_{out}(\sigma_y\otimes\sigma_y)$,
\begin{equation}
\widetilde{\rho_{out}} V=\lambda V\,,  \ \text{so that}\ \Delta\lambda_i= \langle V_i|\widetilde{\delta\rho_{out}}|V_i\rangle\,,
\end{equation}
we finally arrive at the first order error introduced in the concurrence ($C$),

\begin{equation}
\Delta C= \frac{1}{2}\left(\frac{\Delta\lambda_1}{\sqrt{\lambda_1}}-\frac{\Delta\lambda_2}{\sqrt{\lambda_2}}-\frac{\Delta\lambda_3}{\sqrt{\lambda_3}}-\frac{\Delta\lambda_4}{\sqrt{\lambda_4}}\right)\,, \ \ \lambda_1>\lambda_2>\lambda_3>\lambda_4\,.
\end{equation}    
For our case, as demonstrated in Fig. \ref{fig:image2}, we have taken the least counts of the HWPs to be $\delta p$ and $\delta q$ and $\mu$ is the least count of the NOT plate given by,
\begin{equation}
\mu = \frac{\pi}{180}\,,\ \delta p=\sqrt{p(1-p)}\frac{\pi}{90}\,,\ \delta q=\sqrt{q(1-q)}\frac{\pi}{90}\,,
\end{equation}
The extinction ratios $\delta_i$ for the PBSs in the setup are taken to be in the range $\delta_i\sim 10^{-3}-10^{-2}$. We find that,
\begin{align}
    \begin{split}
        \delta_i\sim 10^{-3}&: \ \ \Delta C = 1.57\%\pm 0.91\% \\
        \delta_i\sim 10^{-2}&: \ \ \Delta C = 9.8\%\pm 4.5\%\,. 
    \end{split}
    \label{ErrorC}
\end{align}
Here, the error in Concurrence computation with varying state parameters is discussed. As the state parameter also has an error in preparation, we find the final error in Concurrence measurement as shown in Eq. \ref{ErrorC}.
For a maximally entangled state {\it i.e.} $\alpha=1/\sqrt{2}$, $\Delta C \sim 1\%-14\%$ for the above ranges of $\delta_i$. 

\newpage
\bibliography{Reference}